\documentclass[english]{article}
\usepackage[T1]{fontenc}
\usepackage[applemac]{inputenc}
\usepackage{geometry}
\usepackage{color}
\usepackage{babel}
\usepackage{mathtools}
\usepackage{amsmath}
\usepackage{amssymb}
\usepackage{esint}
\usepackage[unicode=true,pdfusetitle,
 bookmarks=true,bookmarksnumbered=false,bookmarksopen=false,
 breaklinks=false,pdfborder={0 0 1},backref=false,colorlinks=false]
 {hyperref}

\makeatletter

\usepackage{babel}

\makeatother

\begin{document}
\title{\textcolor{blue}{\huge{}A Suggested Answer To Wallstrom's Criticism:
Zitterbewegung Stochastic Mechanics II}}
\author{Maaneli Derakhshani\thanks{Email: m.derakhshani@uu.nl and maanelid@yahoo.com}\medskip{}
\medskip{}
\\
\emph{Department of Mathematics }\\
\emph{Utrecht University, Utrecht, The Netherlands}\thanks{Address when this work was initiated: Department of Physics \& Astronomy,
University of Nebraska-Lincoln, Lincoln, NE 68588, USA.} }
\maketitle
\begin{abstract}
The ``zitterbewegung stochastic mechanics'' (ZSM) answer to Wallstrom's
criticism, introduced in Part I \cite{Derakhshani2016}, is extended
to many particles. We first formulate the many-particle generalization
of Nelson-Yasue stochastic mechanics (NYSM), incorporating external
and classical interaction potentials. Then we formulate the many-particle
generalization of the classical zitterbewegung \emph{zbw} model introduced
in Part I, for the cases of free particles, particles interacting
with external fields, and classically interacting particles. On the
basis of these developments, ZSM is constructed for classically free
particles, as well as for particles interacting both with external
fields and through inter-particle scalar potentials. Throughout, the
beables of ZSM (based on the many-particle formulation) are made explicit.
Subsequently, we assess the plausibility and generalizability of the\emph{
zbw} hypothesis. We close with an appraisal of other proposed answers,
and compare them to ZSM.

\pagebreak{}

\tableofcontents{} 
\end{abstract}

\section{Introduction}

This paper is a direct continuation of the preceding paper, Part I
\cite{Derakhshani2016}. There we proposed an answer to the Wallstrom
criticism of stochastic mechanical theories by modifying Nelson-Yasue
stochastic mechanics (NYSM) for a single non-relativistic particle
with the following hypothesis: Nelson's hypothetical stochastic ether
medium that drives the conservative diffusions of the particle, also
induces steady-state harmonic oscillations of zitterbewegung ($\emph{zbw}$)
frequency in the particle's instantaneous mean forward/backward translational
rest frame. We then showed that, in the lab frame, the function \emph{S}
arises from imposing the constraint of conservative diffusions on
the time-symmetrized steady-state phase of the \emph{zbw} particle,
satisfies the required quantization condition, and evolves in time
by the Hamilton-Jacobi-Madelung equations (when generalized to describe
a statistical ensemble of \emph{zbw} particles). This allowed us to
recover the Schrödinger equation for single-valued wave functions
with (potentially) multi-valued phases, for the cases of a free particle
and a particle interacting with external fields (the latter of which
we illustrated with the two-dimensional central potential problem).
We termed this modification of NYSM ``zitterbewegung stochastic mechanics''
or ZSM.

The approach of this paper is similar to that of Part I. In section
2, we formulate the many-particle generalization of NYSM and point
out where in the derivation of the many-particle Schrödinger equation
the Wallstrom criticism applies. Section 3 discusses how to properly
physically interpret the wave function in NYSM. Section 4 formulates
the classical model of constrained zitterbewegung motion for the cases
of many free particles, many particles interacting with external fields,
and classically interacting particles. Section 5 generalizes ZSM to
the cases of many free particles, many particles interacting with
external fields, and classically interacting particles; throughout,
the beables \footnote{This term was coined by J.S. Bell \cite{BellTLB} as a play on ``observables”
in standard quantum mechanics. It refers to ``those elements which
might correspond to elements of reality, to things which exist. Their
existence does not depend on `observation.' Indeed observation and
observers must be made out of beables” \cite{BellBQFT}.} of ZSM are made explicit. Section 6 assesses the plausibility and
generalizability of the $\emph{zbw}$ hypothesis through multiple
considerations. Finally, section 7 appraises other proposed answers
to Wallstrom's criticism, and compares them to ZSM.

\section{Nelson-Yasue Stochastic Mechanics for Many Particles}

The first non-relativistic, $N$-particle, stochastic mechanical reconstruction
of the \emph{N}-particle Schrödinger equation was given by Loffredo
and Morato \cite{M.I.Loffredo2007}, who used the Guerra-Morato variational
formulation. \footnote{Prior to Loffredo-Morato, Nelson \cite{Nelson1985} and Bacciagaluppi
\cite{Bacciagaluppi2003} employed \emph{N}-particle extensions of
stochastic mechanics for scalar particles. However, they did so by
assuming (rather than reconstructing) the \emph{N}-particle Schrödinger
equation, and constructing \emph{N}-particle extensions of the stochastic
mechanical equations of motion from solutions of the \emph{N}-particle
Schrödinger equation. The \emph{N}-particle stochastic mechanical
equations obtained by Nelson and Bacciagaluppi are formally the same
as those obtained by Loffredo-Morato.} However, as noted in footnote 8 of Part I \cite{Derakhshani2016},
the Guerra-Morato formulation is not applicable to ZSM because the
Guerra-Morato variational principle entails a globally single-valued
$S$ function, and this excludes the possibility of single-valued
wave functions with multi-valued phases (as in systems with angular
momentum \cite{Wallstrom1989,Wallstrom1994}. Koide \cite{Koide2015}
has given a brief two-particle extension of the non-relativistic Nelson-Yasue
formulation, for the case of a classical interaction potential, but
otherwise no comprehensive \emph{N}-particle Nelson-Yasue reconstruction
of the \emph{N}-particle Schrödinger equation has been given (to the
best of our knowledge). Accordingly, we shall develop the \emph{N}-particle
extension of NYSM before extending ZSM to the many-particle case.
This will also be useful for identifying the various differences between
NYSM and ZSM in the many-particle formulation. For completeness, we
will incorporate coupling of the particles to external (scalar and
vector) potentials and to each other through scalar interaction potentials.

As in the single-particle formulation of NYSM \cite{Nelson1966,Nelson1967,Nelson1985},
we hypothesize that the vacuum of 3-D space is pervaded by a homogeneous
and isotropic ether with classical stochastic fluctuations that impart
a frictionless, conservative diffusion process to a point particle
of mass $m$ and charge $e$ immersed within the ether. Accordingly,
for $N$ point particles of masses $m_{i}$ and charges $e_{i}$ immersed
in the ether, each particle will in general have its position 3-vector
$\mathbf{q}_{i}(t)$ constantly undergoing diffusive motion with drift,
as modeled by the first-order forward stochastic differential equations
\begin{equation}
d\mathbf{q}_{i}(t)=\mathbf{b}_{i}(q(t),t)dt+d\mathbf{W}_{i}(t).
\end{equation}
Here $q(t)=\{\mathbf{q}_{1}(t),\mathbf{q}_{2}(t),...,\mathbf{q}_{N}(t)\}$
$\in$ $\mathbb{R}^{3N}$, $\mathbf{b}_{i}(q(t),t)$ is the deterministic
mean forward drift velocity of the $i$-th particle (which in general
may be a function of the positions of all the other particles, such
as in the case of particles interacting with each other gravitationally
and/or electrostatically), and $\mathbf{W}_{i}(t)$ is the Wiener
process modeling the $i$-th particle's interaction with the ether
fluctuations. (Recall that ``mean'', here, refers to averaging over
the Wiener process in the sense of the conditional expectation at
time \emph{t}.)

The Wiener increments $d\mathbf{W}_{i}(t)$ are assumed to be Gaussian
with zero mean, independent of $d\mathbf{q}_{i}(s)$ for $s\leq t$,
and with variance 
\begin{equation}
\mathrm{E}_{t}\left[d\mathbf{W}_{in}(t)d\mathbf{W}_{im}(t)\right]=2\nu_{i}\delta_{nm}dt,
\end{equation}
where $\mathrm{E}_{t}$ denotes the conditional expectation at time
\emph{t}. We then hypothesize that the magnitudes of the diffusion
coefficients $\nu_{i}$ are given by 
\begin{equation}
\nu_{i}=\frac{\hbar}{2m_{i}}.
\end{equation}

In addition to (1), we also have the backward stochastic differential
equations 
\begin{equation}
d\mathbf{q}_{i}(t)=\mathbf{b}_{i*}(q(t),t)dt+d\mathbf{W}_{i*}(t),
\end{equation}
where $\mathbf{b}_{i*}(q(t),t)$ are the mean backward drift velocities,
and $d\mathbf{W}_{i*}(t))$ are the backward Wiener processes. As
in the single-particle case, the $d\mathbf{W}_{i*}(t)$ have all the
properties of $d\mathbf{W}_{i}(t)$ except that they are independent
of the $d\mathbf{q}_{i}(s)$ for $s\geq t$. With these conditions
on $d\mathbf{W}_{i}(t)$ and $d\mathbf{W}_{i*}(t)$, Eqs. (1) and
(4) respectively define forward and backward Markov processes for
$N$ particles on $\mathbb{R}^{3}$ (or, mathematically equivalently,
for a single particle on $\mathbb{R}^{3N}$).

The forwards and backwards transition probabilities defined by (1)
and (4), respectively, should be understood, in some sense, as ontic
probabilities \cite{Uffink2006,BacciagaluppiProbab2011}. (Broadly
speaking, `ontic probabilities' can be understood as probabilities
about objective physical properties of the \emph{N}-particle system,
as opposed to `epistemic probabilities' \cite{Arntzenius1995} which
are about our ignorance of objective physical properties of the \emph{N}-particle
system.) Just how `ontic' these transition probabilities should be
is an open question. One possibility is that these transition probabilities
should be viewed as phenomenologically modeling the complicated deterministic
interactions of a massive particle (or particles) with the fluctuating
ether, in analogy with how equations such as (1) and (4) are used
in the Einstein-Smoluchowski theory \cite{Nelson1967} to phenomenologically
model the complicated deterministic interactions of a macroscopic
particle immersed in a fluctuating classical fluid of finite temperature.
Another possibility is that the fluctuations of the ether are irreducibly
stochastic, and this irreducible stochasticity is `transferred' to
a particle immersed in and interacting with the ether. We prefer the
former possibility, but acknowledge that the latter possibility is
also viable. \footnote{Concerning whether or not the forward and backwards transition probabilities
should be understood as `objective' (i.e., as chances governed by
natural law) versus `subjective' (i.e., encoding our expectations
or degrees of belief) \cite{FriggHoefer2010,Glynn2010,Emery2015},
this seems to depend on whether the transition probabilities are merely
phenomenological (in which case they would seem to be subjective)
or reflect irreducible stochasticity in the ether (in which case they
would seem to be objective). Our preference for viewing the transition
probabilities as phenomenological seems to commit us to the subjective
view, but the objective view also seems viable (the objective view
is taken by Bacciagaluppi in \cite{Bacciagaluppi2005,Bacciagaluppi2012}).
It is worth noting that, under the objective view, the backwards transition
probabilities can be regarded as being just as objective/law-like
as the forwards transition probabilities (but see \cite{Arntzenius1995}
for a different view).}

Associated to the trajectories $\mathbf{q}_{i}(t)$ is the $N$-particle
probability density $\rho(q,t)=n(q,t)/N$ where $n(q,t)$ is the number
of particles per unit volume. Corresponding to (1) and (4), then,
are the $N$-particle forward and backward Fokker-Planck equations
\begin{equation}
\frac{\partial\rho(q,t)}{\partial t}=-\sum_{i=1}^{N}\nabla_{i}\cdot\left[\mathbf{b}_{i}(q,t)\rho(q,t)\right]+\sum_{i=1}^{N}\frac{\hbar}{2m_{i}}\nabla_{i}^{2}\rho(q,t),
\end{equation}
and 
\begin{equation}
\frac{\partial\rho(q,t)}{\partial t}=-\sum_{i=1}^{N}\nabla_{i}\cdot\left[\mathbf{b}_{i*}(q,t)\rho(q,t)\right]-\sum_{i=1}^{N}\frac{\hbar}{2m_{i}}\nabla_{i}^{2}\rho(q,t),
\end{equation}
where we assume that the solutions $\rho(q,t)$ in each time direction
satisfy the normalization condition 
\begin{equation}
\int_{\mathbb{R}^{3N}}\rho_{0}(q)d^{3N}q=1.
\end{equation}
In contrast to the transition probabilities defined by (1) and (4),
the probability distributions satisfying (5) and (6) are epistemic
distributions in the sense that they are distributions over a Gibbsian
ensemble of identical systems (i.e., the distributions reflect our
ignorance of the actual positions of the particles). Nevertheless,
for an epistemic distribution satisfying (5) or (6) at time $t$,
its subsequent evolution will be determined by the ontic transition
probabilities so that the distribution at later times will partly
come to reflect ontic features of the \emph{N}-particle system, and
may asymptotically become independent of the initial distribution.
\footnote{I thank Guido Bacciagaluppi for emphasizing this point.}
Of course, the asymptotic distribution would still be epistemic in
the sense of encoding our ignorance of the actual particle positions,
even though it would be determined by the ontic features of the system.

Up to this point, (5) and (6) correspond to independent diffusion
processes in opposite time directions. \footnote{In fact, given all possible solutions to (1), one can define as many
forward processes as there are possible initial distributions satisfying
(5); likewise, given all possible solutions to (4), one can define
as many backward processes as there are possible `initial' distributions
satisfying (6). Consequently, the forward and backward processes are
both underdetermined, and neither (1) nor (4) has a well-defined time-reversal.} To fix the diffusion process uniquely for both time directions, we
must constrain the diffusion process to simultaneous solutions of
(5) and (6).

Note that the sum of (5) and (6) yields the $N$-particle continuity
equation 
\begin{equation}
\frac{\partial\rho({\normalcolor q},t)}{\partial t}=-\sum_{i=1}^{N}\nabla_{i}\cdot\left[\mathbf{v}_{i}(q.t)\rho(q,t)\right],
\end{equation}
where 
\begin{equation}
\mathbf{v}_{i}(q.t)\coloneqq\frac{1}{2}\left(\mathbf{b}_{i}(q,t)+\mathbf{b}_{i*}(q,t)\right)
\end{equation}
is the current velocity field of the $i$-th particle. We shall also
require that $\mathbf{v}_{i}(q.t)$ is equal to the gradient of a
scalar potential $S(q,t)$ (since, if we allowed $\mathbf{v}_{i}(q.t)$
a non-zero curl, then the time-reversal operation would change the
orientation of the curl, thus distinguishing time directions \cite{PenaCetto1982,Bacciagaluppi2012}).
And for particles classically interacting with an external vector
potential $\mathbf{A}_{i}^{ext}(\mathbf{q}_{i},t)$, the current velocities
get modified by the usual expression 
\begin{equation}
\mathbf{v}_{i}(q.t)=\frac{\nabla_{i}S(q,t)}{m_{i}}-\frac{e_{i}}{m_{i}c}\mathbf{A}_{i}^{ext}(\mathbf{q}_{i},t).
\end{equation}
So (8) becomes 
\begin{equation}
\frac{\partial\rho({\normalcolor q},t)}{\partial t}=-\sum_{i=1}^{N}\nabla_{i}\cdot\left[\left(\mathbf{\frac{\nabla_{\mathit{i}}\mathrm{\mathit{S\mathrm{(\mathit{q},\mathit{t})}}}}{\mathit{m}_{\mathit{i}}}}-\frac{e_{i}}{m_{i}c}\mathbf{A}_{i}^{ext}(\mathbf{q}_{i},t)\right)\rho(q,t)\right],
\end{equation}
which is now a time-reversal invariant evolution equation for $\rho$.

The function \textit{S} is an $N$-particle velocity potential, defined
here as a field over the possible positions of the particles (hence
the dependence of $S$ on the generalized coordinates $\mathbf{q}_{i}$),
and generates different possible initial irrotational velocities for
the particles via (10). We make no assumptions at this level as to
whether or not $S$ can be written as a sum of single-particle velocity
potentials. Rather, this will depend on the initial conditions and
constraints specified for a system of $N$ Nelsonian particles, as
well as the dynamics we obtain for $S$. For example, for $N$ particles
constrained to interact with each other through a classical Newtonian
gravitational and/or electrostatic potential, and $S$ evolving by
the $N$-particle generalization of the quantum Hamilton-Jacobi equation
(which will turn out to be the case), we will find that $S$ won't
be decomposable into a sum as long as the interactions are appreciable.
On the other hand, for $N$ non-interacting particles, we will find
that $S$ evolving by the quantum Hamilton-Jacobi equation can (in
certain cases) be written as $\sum_{i=1}^{N}S_{i}(\mathbf{q}_{i},t)$.

Note also that subtracting (5) from (6) yields the equality on the
right hand side of 
\begin{equation}
\mathbf{u}_{i}(q,t)\coloneqq\frac{1}{2}\left[\mathbf{b}_{i}(q,t)-\mathbf{b}_{i*}(q,t)\right]=\frac{\hbar}{2m_{i}}\frac{\nabla_{i}\rho(q,t)}{\rho(q,t)},
\end{equation}
where $\mathbf{u}_{i}(q,t)$ is the osmotic velocity field of the
$i$-th particle. From (10) and (12), we then have $\mathbf{b}_{i}=\mathbf{v}_{i}+\mathbf{u}_{i}$
and $\mathbf{b}_{i*}=\mathbf{v}_{i}-\mathbf{u}_{i}$, which when inserted
back into (5) and (6), respectively, returns (11). Thus $\rho$ is
fixed as the unique, single-time, `quantum equilibrium' distribution
for the solutions of (1) and (4), and evolves by (11). Moreover, the
epistemic probabilities associated with $\rho$ are now fully determined
by the ontic transition probabilities corresponding to solutions of
(1) and (4).

As in the single-particle case, we can give physical meaning to the
osmotic velocities by analogy with the Einstein-Smoluchowski theory:
We postulate the presence of an external ``osmotic'' potential (which
we will formally write as a field on the $N$-particle configuration
space, in analogy with a classical $N$-particle external potential),
$U(q,t)$, which couples to the $i$-th particle as $R(q(t),t)=\mu U(q(t),t)$
(we assume that the coupling constant $\mu$ is identical for particles
of the same species), and imparts to the $i$-th particle a momentum,
$\nabla_{i}R(q,t)|_{\mathbf{q}_{j}=\mathbf{q}_{j}(t)}$. This momentum
then gets counter-balanced by the ether fluid's osmotic impulse pressure,
$\left(\hbar/2m_{i}\right)\nabla_{i}\ln[n(q,t)]|_{\mathbf{q}_{j}=\mathbf{q}_{j}(t)}$.
This leads to the equilibrium condition $\nabla_{i}R/m_{i}=\left(\hbar/2m_{i}\right)\nabla_{i}\rho/\rho$
(using $\rho=n/N$), which implies that $\rho$ depends on $R$ as
$\rho=e^{2R/\hbar}$ for all times. So the osmotic velocity of the
\emph{i}-th particle is the `equilibrium velocity' that the $i$-th
particle would acquire in the absence of any current velocity $\nabla_{i}S/m_{i}$.
(Note that the sense here in which the osmotic velocity is an equilibrium
velocity is different from the sense in which $\nabla_{i}S$ is an
equilibrium velocity; the latter is an equilibrium velocity in the
sense that it's the velocity that transports the quantum equilibrium
distribution $\rho$ on configuration space via the continuity equation
(11).)

It might be thought that, as an external potential (in the sense of
a potential not sourced by the particle), it should be reasonable
to assume that $R$ is a separable function of the $N$ coordinates
so that we can write $R(q,t)=\sum_{i=1}^{N}R_{i}(\mathbf{q}_{i},t)$.
However, we know from the single-particle case that the evolution
of $R$ depends on the evolution of $S$ (through the continuity equation
for $\rho$), and that the evolution of $S$ depends on the classical
potential $V$. Since, for many particles, $V$ can be an interaction
potential (such as an \emph{N}-particle Coulomb potential), and since
we expect to find that the \emph{N}-particle evolution equations for
$R$ and $S$ are the \emph{N}-particle generalizations of the HJM
equations, we should expect $R$ to possibly depend on the positions
of all the other particle coordinates as a consequence of its nonlinear
coupling to $S$.

From a more physical point of view, it would be reasonable to expect
that $R$ functionally depends on the coordinates of all the other
particles if either (i) the source of the potential $U$ dynamically
couples to all the particles in such a way that the functional dependence
of $U$ is determined by the magnitude of inter-particle physical
interactions, or (ii) $U$ is an independently existing field in space-time
that directly exchanges energy-momentum with the particles. Since,
by Nelson's hypothesis, each particle undergoes a conservative diffusion
process through the ether, on the (ensemble) average, the energy-momentum
of each particle is a constant (assuming no time-dependent classical
external potentials are present). This suggests that the source of
$U$ should be Nelson's ether \footnote{So the idea would be that the ether fluid produces a potential field
$U$ that imparts a momentum of $\nabla_{i}(\mu U)$ to each particle,
causing the particles to scatter through the ether constituents and
thereby experience a counter-balancing osmotic impulse pressure of
magnitude $\left(\hbar/2m_{i}\right)\nabla_{i}ln[n]$.} (otherwise the diffusions would not be conservative). So the functional
dependence of $U$ must be determined by the (hypothetical) dynamical
coupling of the ether to the particles, and whether or not the particles
classically interact with one another. In this way, it is conceivable
how $U$ could have a non-separable functional dependence on the coordinates
associated with all the particles. Moreover, we should expect the
`strength' of the non-separability (i.e., the inter-particle correlations)
of $U$ to be proportional to the strength of the classical interactions
between the particles. (As it turns out, a dust grain undergoing Brownian
motion in a nonequilibrium plasma induces an electrostatic osmotic
potential from the plasma through an analogous mechanism to what we've
sketched here \textcolor{black}{\cite{Lev2009}}; moreover, the corresponding
Fokker-Planck equation for the stationary probability distribution
in velocity space is formally equivalent to Eq. (5) here.)

Since we do not at present have a physical model for Nelson's ether
and its dynamical interactions with the particles, in practice, hypothesis
(i) in the previous paragraph gets implemented via Eq. (11) (which,
as we've noted, equivalently describes the time-evolution of $R$
and thereby the time-evolution of the coupling of the particles to
$U$) and Yasue's stochastic variational principle for the particles.
Thus, for $N$ particles constrained to interact with each other through
a classical Newtonian gravitational and/or electrostatic potential,
and $R$ coupled to $S$ by the $N$-particle HJM equations, we will
indeed see that $R$ (and hence $\rho$) is not separable, from which
we can deduce that $U$ will also be non-separable. On the other hand,
in the case of non-interacting particles, we will find that it is
possible to have $R(q,t)=\sum_{i=1}^{N}R_{i}(\mathbf{q}_{i},t)$ (hence
$\rho(q,t)=\prod_{i=1}^{N}\rho_{i}(\mathbf{q}_{i},t)$). So, for now,
we will keep writing the general form $R=R(q,t)$.

In order to formulate the second-order dynamics of the particles,
we need to construct the $N$-particle generalizations of Nelson's
mean forward and backward derivatives. This generalization is straightforwardly
given by 
\begin{equation}
D\mathbf{q}_{i}(t)=\underset{_{\Delta t\rightarrow0^{+}}}{lim}\mathrm{E}_{t}\left[\frac{q_{i}(t+\Delta t)-q_{i}(t)}{\Delta t}\right],
\end{equation}
and 
\begin{equation}
D_{*}\mathbf{q}_{i}(t)=\underset{_{\Delta t\rightarrow0^{+}}}{lim}\mathrm{E}_{t}\left[\frac{q_{i}(t)-q_{i}(t-\Delta t)}{\Delta t}\right].
\end{equation}
By the Gaussianity of $d\mathbf{W}_{i}(t)$ and $d\mathbf{W}_{i*}(t)$,
we obtain $D\mathbf{q}_{i}(t)=\mathbf{b}_{i}(q(t),t)$ and $D_{*}\mathbf{q}_{i}(t)=\mathbf{b}_{i*}(q(t),t)$.
To compute $D\mathbf{b}_{i}(q(t),t)$ (or $D_{*}\mathbf{b}_{i}(q(t),t)$),
we expand $\mathbf{b}_{i}$ in a Taylor series up to terms of order
two in $d\mathbf{q}_{i}(t)$: 
\begin{equation}
\begin{aligned}d\mathbf{b}_{i}(q(t),t) & =\frac{\partial\mathbf{b}_{i}(q(t),t)}{\partial t}dt+\sum_{i=1}^{N}d\mathbf{q}_{i}(t)\cdot\nabla_{i}\mathbf{b}_{i}(q,t)|_{\mathbf{q}_{j}=\mathbf{q}_{j}(t)}\\
 & +\sum_{i=1}^{N}\frac{1}{2}\underset{n,m}{\sum}d\mathbf{\mathit{q}}_{in}(t)d\mathbf{\mathit{q}}_{im}(t)\frac{\partial^{2}\mathbf{b}_{i}(q,t)}{\partial\mathbf{\mathit{q}}_{in}\partial\mathit{q}_{im}}|_{\mathbf{q}_{j}=\mathbf{q}_{j}(t)}+\ldots.
\end{aligned}
\end{equation}
From (1), we can replace $dq_{i}(t)$ by $dW_{i}(t)$ in the last
term, and when taking the conditional expectation at time \emph{t}
in (13), we can replace $d\mathbf{q}_{i}(t)\cdot\nabla_{i}\mathbf{b}_{i}|_{\mathbf{q}_{j}=\mathbf{q}_{j}(t)}$
by $\mathbf{b}_{i}(\mathbf{q}(t),t)\cdot\nabla_{i}\mathbf{b}_{i}|_{\mathbf{q}_{j}=\mathbf{q}_{j}(t)}$
since $d\mathbf{W}_{i}(t)$ is independent of $\mathbf{q}_{i}(t)$
and has mean 0. From (2), we then obtain 
\begin{equation}
D\mathbf{b}_{i}(q(t),t)=\left[\frac{\partial}{\partial t}+\sum_{i=1}^{N}\mathbf{b}_{i}(q(t),t)\cdot\nabla_{i}+\sum_{i=1}^{N}\frac{\hbar}{2m_{i}}\nabla_{i}^{2}\right]\mathbf{b}_{i}(q(t),t),
\end{equation}
and likewise 
\begin{equation}
D_{*}\mathbf{b}_{i*}(q(t),t)=\left[\frac{\partial}{\partial t}+\sum_{i=1}^{N}\mathbf{b}_{i*}(q(t),t)\cdot\nabla_{i}-\sum_{i=1}^{N}\frac{\hbar}{2m_{i}}\nabla_{i}^{2}\right]\mathbf{b}_{i*}(q(t),t).
\end{equation}
Using (16-17), and assuming the particles also couple to an external
electric potential, $\Phi_{i}^{ext}(\mathbf{q}_{i}(t),t)$, as well
as to each other by the Coulomb interaction potential $\Phi_{c}^{int}(\mathbf{q}_{i}(t),\mathbf{q}_{j}(t))=\frac{1}{2}\sum_{j=1}^{N(j\neq i)}\frac{e_{j}}{|\mathbf{q}_{i}(t)-\mathbf{q}_{j}(t)|}$
we can then construct the \emph{N}-particle generalization of Yasue's
ensemble-averaged, time-symmetric mean action: 
\begin{equation}
\begin{aligned}J & =\mathrm{E}\left[\int_{t_{I}}^{t_{F}}\sum_{i=1}^{N}\left\{ \frac{1}{2}\left[\frac{1}{2}m_{i}\mathbf{b}_{i}^{2}+\frac{1}{2}m_{i}\mathbf{b}_{i*}^{2}\right]+\frac{e_{i}}{c}\mathbf{A}_{i}^{ext}\cdot\frac{1}{2}\left(D+D_{*}\right)\mathbf{q}_{i}(t)-e_{i}\left[\Phi_{i}^{ext}+\Phi_{c}^{int}\right]\right\} dt\right]\\
 & =\mathrm{E}\left[\int_{t_{I}}^{t_{F}}\sum_{i=1}^{N}\left\{ \frac{1}{2}m_{i}\mathbf{v}_{i}^{2}+\frac{1}{2}m_{i}\mathbf{u}_{i}^{2}+\frac{e_{i}}{c}\mathbf{A}_{i}^{ext}\cdot\mathbf{v}_{i}-e_{i}\left[\Phi_{i}^{ext}+\Phi_{c}^{int}\right]\right\} dt\right],
\end{aligned}
\end{equation}
where $\mathrm{E}\left[...\right]$ denotes the absolute expectation,
and we note that $\mathbf{v}_{i}(q(t),t)=\frac{1}{2}\left(D+D_{*}\right)\mathbf{q}_{i}(t)$.

Upon imposing the conservative diffusion constraint through the \emph{N}-particle
generalization of Yasue's variational principle 
\begin{equation}
J=extremal,
\end{equation}
a straightforward computation (see the Appendix) shows that (19) implies
\begin{equation}
\sum_{i=1}^{N}\frac{m_{i}}{2}\left[D_{*}D+DD_{*}\right]\mathbf{q}_{i}(t)=\sum_{i=1}^{N}e_{i}\left[-\frac{1}{c}\partial_{t}\mathbf{A}_{i}^{ext}-\nabla_{i}\left(\Phi_{i}^{ext}+\Phi_{c}^{int}\right)+\frac{\mathbf{v}_{i}}{c}\times\left(\nabla_{i}\times\mathbf{A}_{i}^{ext}\right)\right]|_{\mathbf{q}_{j}=\mathbf{q}_{j}(t)}.
\end{equation}
Moreover, since the $\delta\mathbf{q}_{i}(t)$ are independent (as
we show in the Appendix), it follows from (20) that we have the equations
of motion 
\begin{equation}
\begin{aligned}m_{i}\mathbf{a}_{i}(q(t),t) & =\frac{m_{i}}{2}\left[D_{*}D+DD_{*}\right]\mathbf{q}_{i}(t)\\
 & =\left[-\frac{e_{i}}{c}\partial_{t}\mathbf{A}_{i}^{ext}-e_{i}\nabla_{i}\left(\Phi_{i}^{ext}+\Phi_{c}^{int}\right)+\frac{e_{i}}{c}\mathbf{v}_{i}\times\left(\nabla_{i}\times\mathbf{A}_{i}^{ext}\right)\right]|_{\mathbf{q}_{j}=\mathbf{q}_{j}(t)},
\end{aligned}
\end{equation}
for $i=1,...,N$. Applying the mean derivatives in (20), using that
$\mathbf{b}_{i}=\mathbf{v}_{i}+\mathbf{u}_{i}$ and $\mathbf{b}_{i*}=\mathbf{v}_{i}-\mathbf{u}_{i}$,
and replacing $q(t)$ with $q$ in the functions on both sides, straightforward
manipulations show that (20) turns into 
\begin{equation}
\begin{aligned} & \sum_{i=1}^{N}m_{i}\left[\partial_{t}\mathbf{v}_{i}+\mathbf{v}_{i}\cdot\nabla_{i}\mathbf{v}_{i}-\mathbf{u}_{i}\cdot\nabla_{i}\mathbf{u}_{i}-\frac{\hbar}{2m_{i}}\nabla_{i}^{2}\mathbf{u}_{i}\right]\\
 & =\sum_{i=1}^{N}\left[-\frac{e_{i}}{c}\partial_{t}\mathbf{A}_{i}^{ext}-e_{i}\nabla_{i}\left(\Phi_{i}^{ext}+\Phi_{c}^{int}\right)+\frac{e_{i}}{c}\mathbf{v}_{i}\times\left(\nabla_{i}\times\mathbf{A}_{i}^{ext}\right)\right].
\end{aligned}
\end{equation}
Using (10) and (12), integrating both sides of (22), and setting the
arbitrary integration constants equal to zero, we then obtain the
\emph{N}-particle quantum Hamilton-Jacobi equation 
\begin{equation}
\begin{aligned}-\partial_{t}S(q,t) & =\sum_{i=1}^{N}\frac{\left[\nabla_{i}S(q,t)-\frac{e_{i}}{c}\mathbf{A}_{i}^{ext}(\mathbf{q}_{i},t)\right]^{2}}{2m_{i}}\\
 & +\sum_{i=1}^{N}e_{i}\left[\Phi_{i}^{ext}(\mathbf{q}_{i},t)+\Phi_{c}^{int}(\mathbf{q}_{i},\mathbf{q}_{j})\right]-\sum_{i=1}^{N}\frac{\hbar^{2}}{2m_{i}}\frac{\nabla_{i}^{2}\sqrt{\rho(q,t)}}{\sqrt{\rho(q,t)}},
\end{aligned}
\end{equation}
which describes the total energy of the possible mean trajectories
of the \emph{zbw} particles, and, upon evaluation at $q=q(t),$ the
total energy of the actual particles along their mean trajectories.
So (11) and (23) together define the \emph{N}-particle HJM equations.

Note that, as a consequence of the non-separability of $\Phi_{c}^{int}(\mathbf{q}_{i},\mathbf{q}_{j})$,
we will not be able to write (23) as a sum of total energies for each
particle (unless the particles are sufficiently spatially separated
from each other that we can effectively neglect this interaction term),
which means $S(q,t)\neq\sum_{i=1}^{N}S_{i}(\mathbf{q}_{i},t)$. Indeed,
as a consequence of this non-separability, we can now see from the
coupling of (11) and (23) that $R$ (and hence $U$) will also be
non-separable since its evolution depends on $\nabla_{i}S$ through
(11). We can make this more explicit by writing the general solutions,
$S$ and $R$, to (23) and the differentiated form of (11), respectively.
For (23), the general solution takes the form 
\begin{equation}
\begin{aligned}S(q,t) & =\sum_{i=1}^{N}\int\mathbf{p}_{i}(q,t)\cdot d\mathbf{q}_{i}\\
 & -\sum_{i=1}^{N}\int\left[\frac{\left[\mathbf{p}_{i}(q,t)-\frac{e_{i}}{c}\mathbf{A}_{i}^{ext}(\mathbf{q}_{i},t)\right]^{2}}{2m_{i}}+e_{i}\left[\Phi_{i}^{ext}(\mathbf{q}_{i},t)+\Phi_{c}^{int}(\mathbf{q}_{i},\mathbf{q}_{j})\right]-\frac{\hbar^{2}}{2m_{i}}\frac{\nabla_{i}^{2}\sqrt{\rho(q,t)}}{\sqrt{\rho(q,t)}}\right]dt.
\end{aligned}
\end{equation}

For the differentiated form of (11), the general solution $R$ can
be found most easily by first solving (11) directly in terms of $\rho$
and then using the relation $\rho=e^{2R/\hbar}$. Rewriting (11) as
$\left(\partial_{t}+\sum_{i}^{N}\mathbf{v}_{i}\cdot\nabla_{i}\right)\rho=-\rho\sum_{i}^{N}\nabla_{i}\cdot\mathbf{v}_{i}$,
we have $(d/dt)ln[\rho]=-\sum_{i}^{N}\nabla_{i}\cdot\mathbf{v}_{i}$.
Solving this last expression yields 
\begin{equation}
\rho(q,t)=\rho_{0}(q_{0})exp[-\int_{0}^{t}\left(\sum_{i=1}^{N}\nabla_{i}\cdot\mathbf{v}_{i}\right)dt'.
\end{equation}
The osmotic potential obtained from $\rho$ then takes the form 
\begin{equation}
R(q,t)=R_{0}(q_{0})-(\hbar/2)\int_{0}^{t}\left(\sum_{i=1}^{N}\nabla_{i}\cdot\mathbf{v}_{i}\right)dt'.
\end{equation}
Accordingly, we see clearly that $R$ depends on $S$ through $\mathbf{v}_{i}$,
and that $S$ depends on $R$ through the quantum kinetic. So the
non-separability of $\Phi_{c}^{int}$ alone entails non-factorizability
of $S(q,t)$, which entails non-factorizability of $R(q,t)$, which
entails non-factorizability of the quantum kinetic. \footnote{In Part I, we explained that we prefer to call the ``quantum potential''
the ``quantum kinetic'' in order to emphasize its physical origin
in the kinetic energy term associated with the osmotic velocity of
a Nelsonian particle. } That is, the nonlinear coupling between (24) and (26) entails that
$S$ is actually non-separable by virtue of the non-separability of
$\Phi_{c}^{int}$ \emph{and} (as a consequence thereof) that the quantum
kinetic is non-separable. Thus we've explicitly shown, from the \emph{N}-particle
HJM equations, that the presence of classical interactions between
Nelsonian particles means that the \emph{N}-particle osmotic potential
cannot be written as a separable sum of \emph{N} osmotic potentials
associated to each particle.

Let us now combine (11) and (23) into an \emph{N}-particle Schrödinger
equation and write down the most general form of the \emph{N}-particle
wave function. To do this, we first need to impose the \emph{N}-particle
generalization of the quantization condition 
\begin{equation}
\sum_{i=1}^{N}\oint_{L}\nabla_{i}S(q,t)\cdot d\mathbf{q}_{i}=nh,
\end{equation}
which, by (26), also constrains the osmotic potential sourced by the
ether. Then we can combine (11) and (23) into 
\begin{equation}
i\hbar\frac{\partial\psi(q,t)}{\partial t}=\sum_{i=1}^{N}\left[\frac{\left[-i\hbar\nabla_{i}-\frac{e_{i}}{c}\mathbf{A}_{i}^{ext}(\mathbf{q}_{i},t)\right]^{2}}{2m_{i}}+e_{i}\left(\Phi_{i}^{ext}(\mathbf{q}_{i},t)+\Phi_{c}^{int}(\mathbf{q}_{i},\mathbf{q}_{j})\right)\right]\psi(q,t),
\end{equation}
where the single-valued \emph{N}-particle wave function in polar form
is $\psi(q,t)=\sqrt{\rho(q,t)}e^{iS(q,t)/\hbar}$.

\section{Interpreting the Nelson-Yasue wave function}

How should we understand the NYSM-derived wave function satisfying
(28)? Is it part of NYSM's physical ontology, i.e., is it a beable?
Or should it be viewed as strictly epistemic, i.e., strictly reflecting
our ignorance about ontic aspects of an \emph{N}-particle NYSM system?

Straight off the bat, we can see that $\psi(q,t)$ is defined in terms
of $\rho(q,t)$ and $S(q,t)$. As noted in section 2, $\rho(q,t)$
is an epistemic distribution in that it reflects our ignorance of
the actual positions of the particles; hence $\rho(q,t)$ is not a
beable (in Bell's sense, see footnote 1). As also noted in section
2, $S(q,t)$ is a field over the possible positions of the actual
particles and describes the possible current velocities that the actual
particles can have at each possible point in 3-space they can occupy
at time $t$; hence $S(q,t)$ is also not a beable. Since $\psi(q,t)$
is defined in terms of $\rho(q,t)$ and $S(q,t)$, we must conclude
that $\psi(q,t)$ is also not a beable in NYSM. Rather, $\psi(q,t)$
can be said to be epistemic in the precise sense that it's defined
in terms of $\rho(q,t)$ and $S(q,t)$, and these latter two variables
reflect our ignorance about ontic properties of the actual particles
(their actual positions and velocities). In other words, $\psi(q,t)$
``represents our knowledge of the underlying reality'' \cite{Leifer2014},
rather than being an element of the underlying reality.

However, even though $\psi(q,t)$ is not a beable, it does indirectly
reflect certain ontic aspects of the \emph{N}-particle system in NYSM.
In particular, the evolution of $\rho(q,t)$ depends on the evolution
of $R(q,t)$ via $\rho=e^{2R/\hbar}$, where $R(q,t)=\mu U(q,t)$
and where $U(q,t)$ is a beable. So $\rho(q,t)$ reflects an ontic
aspect of the system, namely the system's osmotic potential field
$U(q,t)$, and by extension so does $\psi(q,t)$ through its modulus.
Additionally, recall from section 2 that the introduction of $S(q,t)$
through the constraint $\mathbf{v}_{i}=m_{i}^{-1}\nabla_{i}S$ implies
that the ether, which is a beable of NYSM, is irrotational, and this
irrotationality is an ontic property of the ether. $S(q,t)$ also
encodes the presence of classical fields in the system (which can
be reasonably regarded as beables, in the sense that the electromagnetic
field is typically regarded as a beable) via the quantum Hamilton-Jacobi
equation (23-24), while also satisfying an ontic (law-like) constraint
via the quantization condition (27). So insofar as $S(q,t)$ reflects
ontic aspects of the system, namely the irrotationality of the ether,
the presence of classical fields in the system, and the quantization
constraint on the current velocities of the particles, so does $\psi(q,t)$
through its complex phase.

It is worth emphasizing the significant conceptual differences between
$S(q,t)$ and $U(q,t)$, despite their formal mathematical similarities:
Even though both are fields on configuration space $\mathbb{R}^{3N}$,
and even though both enter into the stochastic differential equations
of motion (1) and (4) - $S(q,t)$ generating the current velocities,
and $U(q,t)$ generating the osmotic velocities - one field (the $U(q,t)$
field) is a beable and the other (the $S(q,t)$ field) isn't (though
it reflects ontic properties of a beable, the ether). Additionally,
$S(q,t)$ is subject to the quantization condition (27), which indirectly
constrains the evolution of $U(q,t)$ via (26).

It is also worth emphasizing that the epistemic features of the \emph{N}-particle
NYSM wave function are not in logical contradiction with the Pusey-Barrett-Rudolph
(PBR) theorem \cite{Pusey2012}: One of the assumptions of the PBR
theorem is that it is possible to prepare \emph{N} systems independently,
with pure quantum states $\psi_{q_{1},...,}\psi_{q_{N}}$, which results
in ontic states $\lambda_{1},...,\lambda_{N}$ distributed according
to the product distribution $\mu_{q_{1}}(\lambda_{1})\mu_{q_{2}}(\lambda_{2})...\mu_{q_{N}}(\lambda_{N})$.
(More simply, composite systems prepared in a product state will be
independent of one another, i.e., if Alice prepares her system in
a quantum state $\psi_{A}$, and Bob prepares his system in a quantum
state $\psi_{B}$, their joint state is a product of the two states.)
However, the complete set of ontic states of an \emph{N}-particle
NYSM universe, which include the \emph{N}-particle osmotic potential
$R(q,t)$ and the ether properties represented by the \emph{N}-particle
velocity potential $S(q,t)$, do not conform to this ``preparation
independence postulate (PIP)'' \cite{Pusey2012,Leifer2014} (the
same goes for the universal wave function in the de Broglie-Bohm pilot-wave
theory). The reason is that the complete set of ontic states of an
\emph{N}-particle NYSM universe are those of the \emph{universe},
and the universe is only given once, not in multiple independent `preparations'
(nor is it clear what a `preparation' of the universe could mean).
The kinds of pure quantum states to which the PIP applies (pure quantum
states of isolated \emph{sub}systems of the universe) correspond to
the ``effective wave functions'' of subsystems of an NYSM universe
(likewise in the de Broglie-Bohm pilot-wave theory). 

The effective wave function of a subsystem in NYSM (like in the de
Broglie-Bohm pilot-wave theory) follows from the definition of the
universal wave function (i.e., the \emph{N}-particle wave function
satisfying (28)), plus conditioning the universal wave function on
part of the total configuration \emph{$q(t)$}, plus an approximation
on the form of the universal wave function called the ``effective
product form'' \cite{Goldstein1987,Duerr1992,Duerr2009,Goldstein2013}.
More precisely, if we partition the total configuration into two subsystem
configurations, $q(t)=\left\{ X(t),Y(t)\right\} $, and condition
the universal wave function $\psi(q,t)=\psi(x,y,t)$ on $Y(t)$, we
obtain the ``conditional wave function'' of the \emph{X}-subsystem,
$\psi^{(s)}\left(x,t\right)\coloneqq\psi\left(x,Y(t),t\right)$; if
the universal wave function also takes the (effective product) form,
$\psi(x,y,t)=\varphi(x,t)\chi(y,t)+\psi^{\bot}(x,y,t)$, where the
\emph{y}-support of $\psi^{\bot}$ is macroscopically disjoint from
$Y(t)$ (which lies in the support of $\chi$), the $X$-subsystem
becomes effectively decoupled from its environment, the \emph{Y}-subsystem,
and the conditional wave function $\psi^{(s)}\left(x,t\right)$ evolves
by the \emph{N}-particle Schrödinger equation. In this case, the conditional
wave function of the subsystem is also called its ``effective wave
function''. In other words, effectively isolated subsystems of an
\emph{N}-particle NYSM universe are described in terms of effective
wave functions (correspondingly, effective osmotic potentials and
velocity potentials). (In addition, given that the configuration has
the single-time probability density $|\psi|^{2}$ (a.k.a. the ``quantum
equilibrium hypothesis'' in de Broglie-Bohm theory \cite{Goldstein2013}),
which evolves by the continuity equation (11), it can be shown that
the effective wave function randomly collapses according to the Born
rule under exactly the conditions corresponding to an ideal quantum
measurement \cite{Goldstein2013,Duerr2009,Duerr1992,Blanchard1992,Goldstein1987}.)

Since the PIP applies to effective wave functions for composite subsystems,
effective wave functions qualify as ``ontic'' PBR's sense, i.e.,
independently prepared effective wave functions of a composite subsystem
imply non-overlapping distributions for $\lambda$. Yet the fact that
effective wave functions in NYSM are ontic in PBR's sense is not logically
inconsistent with said effective wave functions not being beables
in Bell's sense (footnote 1) and having epistemic features in the
same sense as the universal wave function: The effective wave function
$\psi^{(s)}$ can be said to be epistemic in the precise sense that
it's defined in terms of $\rho^{(s)}$ and $S^{(s)}$, and these latter
two variables reflect our ignorance about ontic properties of the
actual particles (their actual positions and velocities); in other
words, $\psi^{(s)}$ ``represents our knowledge of the underlying
reality'' \cite{Leifer2014}, rather than being an element of the
underlying reality.

To gain a better understanding of the conceptual and technical interplay
between the $\psi$, $R$, and $S$ fields, it is worthwhile to study
the example of an entangled state.

Consider the case of 2 distinguishable particles, where particle 1
is associated with a wavepacket $\psi_{A}$ and particle 2 is associated
with a packet $\psi_{B}$. If, initially, the particles are classically
non-interacting and there are no correlations between them, then the
joint wave function is the product state (suppressing the $t$ variable
for simplicity) 
\begin{equation}
\psi_{f}(\mathbf{q}_{1},\mathbf{q}_{2})\coloneqq\psi_{A}(\mathbf{q}_{1})\psi_{B}(\mathbf{q}_{2}).
\end{equation}
We can also construct a non-factorizable solution of (28) by writing
\begin{equation}
\psi_{nf}(\mathbf{q}_{1},\mathbf{q}_{2})\coloneqq Norm\left[\psi_{A}(\mathbf{q}_{1})\psi_{B}(\mathbf{q}_{2})+\psi_{C}(\mathbf{q}_{1})\psi_{D}(\mathbf{q}_{2})\right].
\end{equation}
If the summands in (30) negligibly overlap by virtue of either $\psi_{A}\cap\psi_{C}\approx\varnothing$
or $\psi_{B}\cap\psi_{D}\approx\varnothing$ (\emph{Norm} = normalization
factor), then the system wave function is `effectively factorizable';
that is, the 2-particle wave function associated with the actual particles
at time $t$ is effectively either $\psi_{f}=\psi_{A}(\mathbf{q}_{1})\psi_{B}(\mathbf{q}_{2})$
or $\psi_{f}=\psi_{C}(\mathbf{q}_{1})\psi_{D}(\mathbf{q}_{2})$. On
the other hand, if we `turn on' the classical interaction $\Phi_{c}^{int}$,
evolution by (28) will make the overlap of the summands non-negligible,
and the system wave function will not be effectively factorizable
\cite{Holland1993}. Consequently, from (30), we will have a non-separable
2-particle velocity potential given by 
\begin{equation}
S_{nf}(\mathbf{q}_{1},\mathbf{q}_{2},)\coloneqq-\frac{i\hbar}{2}\ln\left(\frac{\psi_{nf}(\mathbf{q}_{1},\mathbf{q}_{2})}{\psi_{nf}^{\ast}(\mathbf{q}_{1},\mathbf{q}_{2})}\right).
\end{equation}
The probability density will also be non-factorizable since it becomes
\begin{equation}
\begin{aligned}\rho_{nf}\left(\mathbf{q}_{1},\mathbf{q}_{2}\right) & \coloneqq|\psi_{nf}(\mathbf{q}_{1},\mathbf{q}_{2})|^{2}=Norm^{2}\left\{ e^{2\left(R_{A1}+R_{B2}\right)/\hbar}+e^{2\left(R_{C1}+R_{D2}\right)/\hbar}\right.\\
 & \left.+2e^{\left(R_{A1}+R_{C1}+R_{B2}+R_{D2}\right)/\hbar}cos\left[\left(S_{A1}+S_{B2}-S_{C1}-S_{D2}\right)/\hbar\right]\right\} .
\end{aligned}
\end{equation}
And the corresponding non-separable 2-particle osmotic potential takes
the form 
\begin{equation}
R_{nf}(\mathbf{q}_{1},\mathbf{q}_{2})\coloneqq\hbar\ln\left(|\psi_{nf}(\mathbf{q}_{1},\mathbf{q}_{2})|\right),
\end{equation}
where $|\psi_{nf}(\mathbf{q}_{1},\mathbf{q}_{2})|=\sqrt{\rho_{nf}\left(\mathbf{q}_{1},\mathbf{q}_{2}\right)}$.

By the mathematical equivalence of (28) with the equation set (11)-(23)-(27),
we can see that (33) and (31) will be coupled solutions of (11) and
(23), respectively. On the other hand, when the summands of $\psi_{nf}$
have effectively disjoint support in configuration space (e.g., in
the case of particles sufficiently separated that their classical
interaction can be neglected), the system wave function becomes effectively
factorizable again. In this case, the system velocity potential is
either $S_{f}=S_{A1}+S_{B2}$ or $S_{f}=S_{C1}+S_{D2}$, the probability
density reduces to $\rho_{f}\approx N^{2}\left(e^{2\left(R_{A1}+R_{B2}\right)/\hbar}+e^{2\left(R_{C1}+R_{D2}\right)/\hbar}\right)$,
and the system osmotic potential is either $R_{f}=R_{A1}+R_{B2}$
or $R_{f}=R_{C1}+R_{D2}$.

Incidentally, this latter case most clearly illustrates how, from
the stochastic mechanics viewpoint, the wave function plays the role
of an epistemic variable while also reflecting some of the ontic properties
of the physical system: The modulus-square of the factorizable two-particle
wave function describes the position density for a statistical ensemble
of two-particle systems, while the $R$ and $S$ functions encoded
in the factorizable two-particle wave function represent the \emph{possible}
$R$ and $S$ functions that the actual particles actually `have'
at time $t$; concurrently, the possible $R$ and $S$ functions for
the two-particle system reflect objectively real properties of Nelson's
ontic ether, insofar as $R_{A1}$ ($R_{B2}$) and $R_{C1}$ ($R_{D2}$)
correspond to (effectively) disjoint regions of the ontic osmotic
potential sourced by the ether $U_{A1}$ ($U_{B2}$) and $U_{C1}$
($U_{D2}$), and insofar as $S_{A1}$ ($S_{B2}$) and $S_{C1}$ ($S_{D2}$)
reflect the irrotationality of the ether in regions \emph{A} and \emph{B}
and regions $C$ and $D$. This confirms the properties of the osmotic
potential and its relation to the velocity potential that we observed
from the solutions of the \emph{N}-particle HJM equations, for the
cases of classically interacting and non-interacting distinguishable
particles.

However, we should note that the linearity of (28) entails non-factorizable
solutions for the case of classically non-interacting identical bosons
or fermions. (To justify the symmetrization postulates, we can import
Bacciagaluppi's finding \cite{Bacciagaluppi2003} that the symmetrization
postulates are derivable from the assumption of symmetry of the Nelsonian
particle trajectories in configuration space.) For identical bosons
or fermions, we simply replace $\psi_{C}(\mathbf{q}_{1})\psi_{D}(\mathbf{q}_{2})$
in (30) with $\pm\psi_{A}(\mathbf{q}_{2})\psi_{B}(\mathbf{q}_{1})$,
and similarly for $S_{nf}$, $\rho_{nf}$, and $R_{nf}$. Then, if
particle 1 and particle 2 start out without any classical interaction,
we will initially have $\psi_{A}\cap\psi_{B}\approx\varnothing$ (approximately,
because the wavepackets never have completely disjoint support in
configuration space, even in the non-interacting case); if the packets
of these particles then move towards each other and overlap such that
$\left(<\mathbf{q}_{1}>-<\mathbf{q}_{2}>\right)^{2}\leq\sigma_{A}^{2}+\sigma_{B}^{2}$,
where $\sigma_{A}$ and $\sigma_{B}$ are the widths of the packets,
the resulting wave function of the 2-particle system will be given
by (30) with $\psi_{A}\cap\psi_{B}\neq\varnothing$ \cite{Holland1993}.
Physically, the appreciable overlap of the wavepackets implies that
the initially independent osmotic potentials possibly associated with
particle 1 ($R_{A1}$ or $R_{B1}$) and particle 2 ($R_{A2}$ or $R_{B2}$),
respectively, become non-separable by virtue of their joint support
in configuration space becoming non-negligible. So the resulting motion
of particle 1 will have a non-separable physical dependence on part
of the osmotic potentials possibly associated with particle 2 (and
vice versa), a dependence which is instantaneous between the particles
in 3-space (since the $N$-particle quantum kinetic in (23) acts instantaneously
on the two particles at time $t$). Of course, for classically non-interacting
identical particles, the 2-particle wave function will satisfy $\psi_{A}\cap\psi_{B}=\varnothing$
again once the wavepackets pass each other and their overlap becomes
negligible; but if the particles are classically interacting via $\Phi_{c}^{int}$
the non-separability will persist until the particles are sufficiently
spatially separated that $\Phi_{c}^{int}\approx0$.

Thus the linearization of the HJM equations into Schrödinger's equation,
through the use of condition (27), makes possible non-separable/non-local
correlations between (distinguishable or identical) particles not
admitted by the HJM equations alone (since the solutions of the HJM
equations don't generally satisfy the superposition principle without
(27), as we know from Wallstrom \cite{Wallstrom1989}). \footnote{To be clear, we are not claiming that the HJM equations, without the
quantization condition, do not admit solutions that make possible
EPR-type correlations between particles. It seems plausible that they
do, considering that classical Liouville statistical mechanics (with
an epistemic restriction akin to the Heisenberg uncertainty principle)
does so \cite{Bartlett2012}, and that even without the quantization
condition stochastic mechanics reproduces the uncertainty relations.
But whether solutions exist that are non-local enough to entail violations
of the continuous-variable Bell inequality \cite{Cavalcanti2007}
seems unclear. Answering this question requires a detailed mathematical
study of the analytic solutions of the HJM equations, without the
quantization condition imposed. To the best of our knowledge, this
has yet to be done.} In fact, such solutions tell us that the two-particle wave function
for identical bosons (interacting or non-interacting) must always
be given by (30), where the joint support of the summands never completely
vanishes and can increase appreciably due to (classical or non-classical)
interactions between the particles \cite{Holland1993}.

This last realization complicates the interpretation of the space
in which Nelson's ether lives versus the space in which the particles
live: we started out by postulating that the ether lives in 3-D space,
but have found that once the constraints (19) and (27) are imposed,
the $R$ and $S$ functions (which, as we've seen, reflect objectively
real properties of the ether) are in general not separable, and thus
(mathematically) always live on $3N$-dimensional configuration space.
If we take this mathematical non-factorizability of $R$ and $S$
as a literal indication about the ontic nature of the ether, then
this would seem to force us to infer that the ether must actually
live in $3N$-dimensional configuration space, and therefore regard
configuration space as an ontic space in its own right. We could then
say (to whatever extent one finds this plausible) that the ether and
osmotic potential live in configuration space, but that there are
still \emph{N} ontic particles living in an (also) ontic 3-D space,
and postulate that the two sets of beables can somehow causally interact
with each other via the set (1)-(4)-(21), despite living in independent
ontic spaces. (This situation is analogous to a common interpretation
of the de Broglie-Bohm theory, where the fundamental ontology consists
of an ontic wave function living in an ontic $3N$-dimensional configuration
space, and $N$ ontic particles living in an ontic 3-D space; one
then postulates a one-way causal relationship between the wave function
and the \emph{N} particles via the ``guiding equation'' \cite{Holland1993,Bohm1995,BellQMCosmo2004}.)

It is puzzling, however, how beables living in two independent ontic
spaces can causally interact, or rather, it is puzzling \emph{why}
there should exist any causal (or law-like) relation between beables
living in two independent ontic spaces. Perhaps this dualism is just
a brute fact of nature that we have to live with; but before resigning
ourselves to this conclusion, we might consider two possibilities
for a conception of NYSM in which there exists only one ontic space. 

First, we might consider applying David Albert's ``flat-footed''
interpretation of the de Broglie-Bohm theory \cite{Albert1996,Albert2013,Albert2015}
to NYSM. In this `Albertian' view of NYSM, the representation of \emph{N}
particles in 3-D space ``emerges'' from a (philosophical) functional
analysis of the causal relations instantiated between the 3-dimensional
coordinate components (in Albert's terminology, ``shadows'') of
a single ``world particle'' floating in $3N$-dimensional configuration
space. For this functional analysis to work, the total energy (23)
or action (18) of the world particle must be written in a preferred
time-independent coordinatization \emph{C} such that the 3-dimensional
location of the \emph{i}-th shadow of the world particle is given
by $q_{i}=\left(x_{(3i-2)},x_{(3i-1)},x_{3i}\right)_{C}$, the 3-dimensional
distance between the \emph{i}-th and \emph{j}-th shadows is given
by $\left(\left(x_{(3i-2)}-x_{(3j-2)}\right)^{2}+\left(x_{(3i-1)}-x_{(3j-1)}\right)^{2}+\left(x_{(3i)}-x_{(3j)}\right)^{2}\right)_{C}$,
and the location of the world particle in 3N-dimensional configuration
space is the 3-dimensional configuration of the shadows. On the one
hand, this approach has the virtue that it is straightforward to understand
how the world particle can causally interact with Nelson's ether,
since they both live in the same ontic space, 3N-dimensional (configuration)
space. On the other hand, insofar as the shadows of the world particle
are 3-dimensional objects that objectively exist (i.e., they're beables
in 3-D space, even if they're `components' of the world particle beable),
and insofar as the shadows instantiate 3-dimensional distance relations,
and insofar as the functional analysis relies on the causal relations
instantiated by the shadows, and insofar as the 3-dimensional causal
relations follow from the dynamics generated by (23) or (18) written
in \emph{C}, and insofar as \emph{C} is chosen precisely so as to
make possible the decomposition of the world particle's configuration
into \emph{N} 3-dimensional shadows instantiating 3-dimensional spatial
and causal relations, it seems that 3-D space (where the shadows live)
and local beables (the shadows themselves) are effectively being \emph{postulated}
as primitives along with 3N-dimensional configuration space and the
world particle therein. That is, 3-D space and local beables aren't
``emerging'' from functional analysis, but rather are playing primitive
roles in \emph{setting up} the functional analysis. (What we agree
\emph{does} emerge from the functional analysis, on a coarse-grained
level, are macroscopic objects, i.e., relatively stable 3-dimensional
arrangements of shadow components in the shapes of tables, chairs,
baseballs, etc., with table-like, chair-like, and baseball-like behaviors.)
So the Albertian view seems ultimately no different than the picture
in which there are two fundamental ontic spaces, with the beables
in each space causally interacting. Additionally, the idea that 3N-dimensional
configuration space should be regarded as more fundamental than 3-D
space seems inconsistent with the fact that, in NYSM, the non-separablility
of the \emph{R} and \emph{S} functions on 3N-dimensional configuration
space \emph{follows from} extremizing the action (18) (and imposing
condition (27)), an action that's defined from a sum of $N$ terms
and motivated from a physical \emph{postulate} of $N$ massive particles
conservatively diffusing through an ether living in a 3-D space. One
could reformulate NYSM with the physical postulate that there's only
one particle in 3N-dimensional space, conservatively diffusing through
an ether that also lives in 3N-dimensional space, but this would be
a rather post-hoc revision of NYSM.

The other possibility is that the configuration-space representation
of $R$ and $S$ is somehow an abstract encoding of a complicated
array of ontic fields in space-time that nonlocally connect the motions
of the particles. In practice, we might implement this by analogy
with Norsen's ``TELB'' approach to the de Broglie-Bohm theory \cite{Norsen2010,Norsen2014}:
Taylor-expand the $R$ and $S$ functions in configuration space into
$N$ one-particle $R$ and $S$ functions, each coupled to a countably
infinite hierarchy of ``entanglement fields'' in space-time that
implement the nonlocal connections between the motions of the particles.
The upshot of this approach is that one can maintain that Nelson's
ether lives in plain old 3-D space along with \emph{N} particles.
A drawback is the immense complexity of positing a countable infinity
of ontic fields in space-time, in order to reproduce all the information
encoded in the \emph{R} and \emph{S} functions in configuration space.
To be sure, this last possibility is more speculative than the former
two (since it would be non-trivial to actually construct such a variant
of NYSM); but we think it is ultimately the most intelligible and
fruitful one for stochastic mechanics (for reasons discussed in sections
4 and 5).

Of course, the validity of constructing the non-separable solutions
(30-33) in NYSM depends on the plausibility of imposing (27). But
such a condition is arbitrary from the point of view of (11) and (23),
insofar as we have reconstructed those equations from the Nelson-Yasue
assumptions. This, in essence, is Wallstrom's criticism applied to
the $N$-particle case. Our task then is to reformulate \emph{N}-particle
NYSM into \emph{N}-particle ZSM.

\section{Classical Model of Constrained Zitterbewegung Motion for Many Particles}

In developing $N$-particle ZSM, it will be helpful to first develop
the $N$-particle version of our classical $\emph{zbw}$ model, for
free particles, particles interacting with external fields, and particles
interacting with each other through Coulomb forces. As we will see,
even at the classical level, the $N$-particle extension turns out
to be non-trivial.

\subsection{Free \emph{zbw} particles}

Let us now suppose we have $N$ identical, non-interacting \emph{zbw}
particles in space-time, and no external fields present. In other
words, the $i$-th particle has rest mass $m_{i}$ (taking $i=1,...,N$)
and is rheonomically constrained to undergo an unspecified oscillatory
process with constant angular frequency $\omega_{ci}$ about some
fixed point in 3-space $\mathbf{q}_{0i}$ in a Lorentz frame where
$\mathbf{v}_{i}=d\mathbf{q}_{0i}/dt=0$. Then, in a fixed Lorentz
frame where $\mathbf{v}_{i}\neq0$, the $zbw$ phase for the $i$-th
free particle takes the form (using $\theta_{i}\eqqcolon-\frac{\omega_{ci}}{m_{i}c^{2}}S_{i}=-\frac{1}{\hbar}S_{i}$)
\begin{equation}
\delta S_{i}(\mathbf{q}_{i}(t),t)=\left(\mathbf{p}_{i}\cdot\delta\mathbf{q}_{i}(t)-E_{i}\delta t\right),
\end{equation}
where $E_{i}=\gamma_{i}m_{i}c^{2}$. So for each particle, we will
have 
\begin{equation}
\oint_{L}\delta S_{i}(\mathbf{q}_{i}(t),t)=\oint_{L}\left(\mathbf{p}_{i}\cdot\delta\mathbf{q}_{i}(t)-E_{i}\delta t\right)=nh,
\end{equation}
which implies 
\begin{equation}
\sum_{i=1}^{N}\oint_{L}\delta S_{i}(\mathbf{q}_{i}(t),t)=\sum_{i=1}^{N}\oint_{L}\left(\mathbf{p}_{i}\cdot\delta\mathbf{q}_{i}(t)-E_{i}\delta t\right)=nh.
\end{equation}
In the non-relativistic limit, the $i$-th $zbw$ phase is 
\begin{equation}
S_{i}(\mathbf{q}_{i}(t),t)\approx m_{i}\mathbf{v}_{i}\cdot\mathbf{q}_{i}(t)-\left(m_{i}c^{2}+\frac{m_{i}v_{i}(\mathbf{q}_{i}(t),t)^{2}}{2}\right)t+\hbar\phi_{i},
\end{equation}
and satisfies the classical HJ equation 
\begin{equation}
E_{i}(\mathbf{q}_{i}(t),t)=-\partial_{t}S_{i}(\mathbf{q}_{i},t)|_{\mathbf{q}_{j}=\mathbf{q}_{j}(t)}=\frac{\left(\nabla_{i}S_{i}(\mathbf{q}_{i},t)\right)^{2}}{2m_{i}}|_{\mathbf{q}_{j}=\mathbf{q}_{j}(t)}+m_{i}c^{2}.
\end{equation}
We can also define the total system energy as the sum of the individual
energies of each \emph{zbw} particle: 
\begin{equation}
E(q(t),t)=-\partial_{t}S(q,t)|_{\mathbf{q}_{j}=\mathbf{q}_{j}(t)}=\sum_{i=1}^{N}\frac{\left(\nabla_{i}S(q,t)\right)^{2}}{2m_{i}}|_{\mathbf{q}_{j}=\mathbf{q}_{j}(t)}+\sum_{i=1}^{N}m_{i}c^{2},
\end{equation}
where we have used $E=-\partial_{t}S=\sum_{i=1}^{N}E_{i}=-\sum_{i=1}^{N}\partial_{t}S_{i}=-\partial_{t}\sum_{i=1}^{N}S_{i}$.
Accordingly, we can define the `joint phase' of the \emph{N}-particle
system as the sum 
\begin{equation}
S(q(t),t)=\sum_{i=1}^{N}S_{i}(\mathbf{q}_{i}(t),t)\approx\sum_{i=1}^{N}m_{i}\mathbf{v}_{i}(q(t),t)\cdot\mathbf{q}_{i}(t)-\left(\sum_{i=1}^{N}m_{i}c^{2}+\sum_{i=1}^{N}\frac{m_{i}v_{i}(q(t),t)^{2}}{2}\right)t+\hbar\sum_{i=1}^{N}\phi_{i},
\end{equation}
which satisfies (39). Correspondingly, we can rewrite (36) as 
\begin{equation}
\sum_{i=1}^{N}\oint_{L}\nabla_{i}S|_{\mathbf{q}_{j}=\mathbf{q}_{j}(t)}\cdot\delta\mathbf{q}_{i}(t)=nh,
\end{equation}
for displacements along closed loops with time held fixed. We are
now ready to formulate the HJ statistical mechanics for \emph{N} free
particles.

\subsection{Classical Hamilton-Jacobi statistical mechanics for free \emph{zbw}
particles}

If the actual positions of the $zbw$ particles are unknown, then
$\mathbf{q}_{i}(t)$ gets replaced by $\mathbf{q}_{i}$, and the non-relativistic
joint \emph{zbw} phase becomes a field over the possible positions
of the actual \emph{zbw} particles, namely 
\begin{equation}
S(q,t)\approx\sum_{i=1}^{N}m_{i}\mathbf{v}_{i}(q,t)\cdot\mathbf{q}_{i}-\sum_{i=1}^{N}\left(m_{i}c^{2}+\frac{m_{i}v_{i}(q,t)^{2}}{2}\right)t+\sum_{i=1}^{N}\hbar\phi_{i},
\end{equation}
where $\mathbf{v}_{i}(q,t)=\nabla_{i}S(q,t)/m_{i}$ and satisfies
\begin{equation}
\sum_{i=1}^{N}\oint_{L}\nabla_{i}S\cdot d\mathbf{q}_{i}=nh,
\end{equation}
and 
\begin{equation}
E(q,t)=-\partial_{t}S=\sum_{i=1}^{N}\left[\frac{\left(\nabla_{i}S\right)^{2}}{2m_{i}}+m_{i}c^{2}\right].
\end{equation}
The physical independence of the particles further implies 
\begin{equation}
E_{i}=-\partial_{t}S_{i}=\frac{\left(\nabla_{i}S_{i}\right)^{2}}{2m_{i}}+m_{i}c^{2},
\end{equation}
where 
\begin{equation}
S(q,t)=\sum_{i=1}^{N}S_{i}(\mathbf{q}_{i},t),
\end{equation}
and 
\begin{equation}
\oint_{L}\nabla_{i}S_{i}\cdot d\mathbf{q}_{i}=nh.
\end{equation}

As (42) is defined from the sum of \emph{N} independent phase fields,
Eq. (46), the corresponding velocity fields, $\mathbf{v}_{i}(q,t)$,
are also physically independent of one another. Consequently, for
the trajectory fields obtained from integrating $\mathbf{v}_{i}(q,t)$,
the associated \emph{N}-particle probability density $\rho(q,t)=n(q,t)/N$
can be taken in most cases to be factorizable into a product of \emph{N}
independent probability densities (for simplicity, we ignore the special
case of classical correlations corresponding to when $\rho$ is a
mixture of factorizable densities; but see \cite{Bacciagaluppi2012}
for a discussion of classical correlations in a related context):
\begin{equation}
\rho(q,t)=\prod_{i}^{N}\rho_{i}(\mathbf{q}_{i},t),
\end{equation}
where (48) satisfies $\rho(q,t)\geq0$, the normalization condition
$\int_{\mathbb{R}^{3N}}\rho_{0}(q)d^{3N}q=1$, and evolves by the
\emph{N}-particle continuity equation 
\begin{equation}
\frac{\partial\rho}{\partial t}=-\sum_{i=1}^{N}\nabla_{i}\cdot\left[\left(\mathbf{\frac{\nabla_{\mathit{i}}\mathrm{\mathit{S}}}{\mathit{m_{i}}}}\right)\rho\right],
\end{equation}
which by (48) implies 
\begin{equation}
\frac{\partial\rho_{i}}{\partial t}=-\nabla_{i}\cdot\left[\left(\mathbf{\frac{\nabla_{\mathit{i}}\mathrm{\mathit{S_{i}}}}{\mathit{m_{i}}}}\right)\rho_{i}\right].
\end{equation}

We can then combine (44) and (49) to obtain a single-valued \emph{N}-particle
classical wave function $\psi(q,t)=\sqrt{\rho_{0}(\mathbf{q}_{1}-\mathbf{v}_{1}t,...,\mathbf{q}_{N}-\mathbf{v}_{N}t)}e^{iS(q,t)/\hbar}$
satisfying the \emph{N}-particle nonlinear Schrödinger equation 
\begin{equation}
i\hbar\frac{\partial\psi}{\partial t}=\sum_{i=1}^{N}\left[-\frac{\hbar^{2}}{2m_{i}}\nabla_{i}^{2}+\frac{\hbar^{2}}{2m_{i}}\frac{\nabla_{i}^{2}|\psi|}{|\psi|}+m_{i}c^{2}\right]\psi,
\end{equation}
which implies 
\begin{equation}
i\hbar\frac{\partial\psi_{i}}{\partial t}=\left[-\frac{\hbar^{2}}{2m_{i}}\nabla_{i}^{2}+\frac{\hbar^{2}}{2m_{i}}\frac{\nabla_{i}^{2}|\psi_{i}|}{|\psi_{i}|}+m_{i}c^{2}\right]\psi_{i},
\end{equation}
since 
\begin{equation}
\psi(q,t)=\prod_{i}^{N}\psi_{i}(\mathbf{q}_{i},t).
\end{equation}
Having completed the description of \emph{N} free particles, we now
develop the slightly less trivial case of \emph{zbw} particles interacting
with external fields.

\subsection{External fields interacting with \emph{zbw} particles}

To describe the interaction of our \emph{zbw} particles with external
fields, consider first the change in the \emph{zbw} phase of the $i$-th
particle in its rest frame: 
\begin{equation}
\delta\theta_{i}(t_{0})=\omega_{ci}\delta t_{0}=\frac{1}{\hbar}\left(m_{i}c^{2}\right)\delta t_{0}.
\end{equation}
The coupling of the particle to (say) the Earth's external gravitational
field leads to a small correction (in the now instantaneous rest frames
of the particles) as follows: 
\begin{equation}
\delta\theta_{i}(\mathbf{q}_{0i},t_{0})=\left[\omega_{ci}+\kappa_{i}(\mathbf{q}_{0i})\right]\delta t_{0}=\frac{1}{\hbar}\left[m_{i}c^{2}+m_{i}\Phi_{gi}^{ext}(\mathbf{q}_{0i})\right]\delta t_{0},
\end{equation}
where $\kappa_{i}=\omega_{ci}\Phi_{gi}^{ext}/c^{2}$. As in the single
particle case, we have approximated the coupling as point-like since
we assume $|\mathbf{q}_{i}|\gg\lambda_{ci}$. Supposing also that
the \emph{zbw} particles carry charge $e_{i}$ (so that they now become
classical charged oscillators of some identical type), their point-like
couplings to a space-time varying external electric field lead to
additional (small) phase shifts of the form 
\begin{equation}
\delta\theta_{i}(\mathbf{q}_{0i},t_{0})=\left[\omega_{ci}+\kappa_{i}(\mathbf{q}_{0i})+\varepsilon_{i}(\mathbf{q}_{0i},t_{0})\right]\delta t_{0}=\frac{1}{\hbar}\left[m_{i}c^{2}+m_{i}\Phi_{gi}^{ext}(\mathbf{q}_{0i})+e_{i}\Phi_{ei}^{ext}(\mathbf{q}_{0i},t_{0})\right]\delta t_{0},
\end{equation}
where $\varepsilon_{i}=\omega_{ci}\left(e_{i}/m_{i}c^{2}\right)\Phi_{ei}^{ext}$.

Transforming to the lab frame where the $i$-th \emph{zbw} particle
has nonzero but variable translational velocity, (56) becomes 
\begin{equation}
\begin{aligned}\delta\theta_{i}(\mathbf{q}_{i}(t),t) & =\left[\left(\omega_{dBi}+\kappa_{i}(\mathbf{q}_{i}(t))+\varepsilon_{i}(\mathbf{q}_{i}(t),t)\right)\gamma_{i}\left(\delta t-\frac{\mathbf{v}_{0i}(\mathbf{q}_{i}(t),t)\cdot\delta\mathbf{q}_{i}(t)}{c^{2}}\right)\right]\\
 & =\frac{1}{\hbar}\left[\left(\gamma_{i}m_{i}c^{2}+\gamma_{i}m_{i}\Phi_{gi}^{ext}+e_{i}\Phi_{ei}^{ext}\right)\delta t-\left(\gamma_{i}m_{i}c^{2}+\gamma_{i}m_{i}\Phi_{gi}^{ext}+e_{i}\Phi_{ei}^{ext}\right)\frac{\mathbf{v}_{0i}\cdot\delta\mathbf{q}_{i}(t)}{c^{2}}\right]\\
 & =\frac{1}{\hbar}\left(E_{i}\delta t-\mathbf{p}_{i}\cdot\delta\mathbf{q}_{i}(t)\right),
\end{aligned}
\end{equation}
where $E_{i}=\gamma_{i}m_{i}c^{2}+\gamma_{i}m_{i}\Phi_{gi}^{ext}+e_{i}\Phi_{ei}^{ext}$
and $\mathbf{p}_{i}=m_{i}\mathbf{v}_{i}=\left(\gamma_{i}m_{i}c^{2}+\gamma_{i}m_{i}\Phi_{gi}^{ext}+e_{i}\Phi_{ei}^{ext}\right)\left(\mathbf{v}_{0i}/c^{2}\right)$.
Incorporating coupling to an external vector potential, we have $\mathbf{v}_{i}\rightarrow\mathbf{v}_{i}'=\mathbf{v}_{i}+e_{i}\mathbf{A}_{i}^{ext}/\gamma_{i}m_{i}c$
(where $\gamma_{i}$ depends on the time-dependent $v_{i}$).

Now, even under the physical influence of the external fields, the
phase of the $i$-th particle's oscillation is a well-defined function
of its space-time location. Thus, if we displace the $i$-th particle
around a closed loop, the phase change is still given by 
\begin{equation}
\oint_{L}\delta\theta_{i}=\frac{1}{\hbar}\oint_{L}\left[E_{i}\delta t-\mathbf{p}_{i}'\cdot\delta\mathbf{q}_{i}(t)\right]=2\pi n,
\end{equation}
or 
\begin{equation}
\oint_{L}\delta S_{i}=\oint_{L}\left[\mathbf{p}_{i}'\cdot\delta\mathbf{q}_{i}(t)-E_{i}\delta t\right]=nh.
\end{equation}
Accordingly, we will also have 
\begin{equation}
\sum_{i=1}^{N}\oint_{L}\delta S_{i}=\sum_{i=1}^{N}\oint_{L}\left[\mathbf{p}_{i}'\cdot\delta\mathbf{q}_{i}(t)-E_{i}\delta t\right]=nh.
\end{equation}
Moreover, for the special case of a loop in which time is held fixed,
we have 
\begin{equation}
\oint_{L}\nabla_{i}S_{i}|_{\mathbf{q}_{i}=\mathbf{q}_{i}(t)}\cdot\delta\mathbf{q}_{i}(t)=\oint_{L}\mathbf{p}_{i}'\cdot\delta\mathbf{q}_{i}(t)=nh,
\end{equation}
or 
\begin{equation}
\oint_{L}m_{i}\mathbf{v}_{i}\cdot\delta\mathbf{q}_{i}(t)=nh-\frac{e_{i}}{c}\oint_{L}\mathbf{A}_{i}^{ext}\cdot\delta\mathbf{q}_{i}(t).
\end{equation}
Likewise 
\begin{equation}
\sum_{i=1}^{N}\oint_{L}\nabla_{i}S_{i}|_{\mathbf{q}_{i}=\mathbf{q}_{i}(t)}\cdot\delta\mathbf{q}_{i}(t)=\sum_{i=1}^{N}\oint_{L}\mathbf{p}_{i}'\cdot\delta\mathbf{q}_{i}(t)=nh,
\end{equation}
which is equivalent to 
\begin{equation}
\sum_{i=1}^{N}\oint_{L}m_{i}\mathbf{v}_{i}\cdot\delta\mathbf{q}_{i}(t)=nh-\sum_{i=1}^{N}\frac{e_{i}}{c}\oint_{L}\mathbf{A}_{i}^{ext}\cdot\delta\mathbf{q}_{i}(t).
\end{equation}

Integrating (57) and rewriting in terms of $S_{i}$, we obtain 
\begin{equation}
S_{i}=\int\left[\mathbf{p}_{i}'\cdot d\mathbf{q}_{i}(t)-E_{i}dt\right]-\hbar\phi_{i},
\end{equation}
and thus 
\begin{equation}
S=\sum_{i=1}^{N}S_{i}=\sum_{i=1}^{N}\int\left[\mathbf{p}_{i}'\cdot d\mathbf{q}_{i}(t)-E_{i}dt\right]-\sum_{i=1}^{N}\hbar\phi_{i}.
\end{equation}
When $v_{i}\ll c$ 
\begin{equation}
\begin{aligned}S & \approx\sum_{i=1}^{N}\int m_{i}\mathbf{v}_{i}'\cdot d\mathbf{q}_{i}(t)-\\
 & -\sum_{i=1}^{N}\int\left(m_{i}c^{2}+\frac{1}{2m_{i}}\left[\mathbf{p}_{i}-\frac{e_{i}}{c}\mathbf{A}_{i}^{ext}\right]^{2}+m_{i}\Phi_{gi}^{ext}+e_{i}\Phi_{ei}^{ext}\right)dt-\sum_{i=1}^{N}\hbar\phi_{i},
\end{aligned}
\end{equation}
and satisfies 
\begin{equation}
-\partial_{t}S|_{\mathbf{q}_{j}=\mathbf{q}_{j}(t)}=\sum_{i=1}^{N}\frac{\left(\nabla_{i}S-\frac{e_{i}}{c}\mathbf{A}_{i}^{ext}\right)^{2}}{2m_{i}}|_{\mathbf{q}_{j}=\mathbf{q}_{j}(t)}+\sum_{i=1}^{N}\left[m_{i}c^{2}+m_{i}\Phi_{gi}^{ext}+e_{i}\Phi_{ei}^{ext}\right],
\end{equation}
where the kinetic velocity, $\mathbf{v}_{i}=(1/m_{i})\nabla_{i}S|_{\mathbf{q}_{j}=\mathbf{q}_{j}(t)}-e_{i}\mathbf{A}_{i}^{ext}/m_{i}c$,
satisfies the classical Newtonian equation of motion 
\begin{equation}
\begin{aligned}m_{i}\ddot{\mathbf{q}}_{i}(t) & =\left(\frac{\partial}{\partial t}+\mathbf{v}_{i}\cdot\nabla_{i}\right)\left[\nabla_{i}S-\frac{e_{i}}{c}\mathbf{A}_{i}^{ext}\right]|_{\mathbf{q}_{j}=\mathbf{q}_{j}(t)}\\
 & =-\nabla_{i}\left[m_{i}\Phi_{gi}^{ext}+e_{i}\Phi_{ei}^{ext}\right]|_{\mathbf{q}_{j}=\mathbf{q}_{j}(t)}-\frac{e_{i}}{c}\frac{\partial\mathbf{A}_{i}^{ext}}{\partial t}|_{\mathbf{q}_{j}=\mathbf{q}_{j}(t)}+\frac{e_{i}}{c}\mathbf{v}_{i}\times\mathbf{B}_{i}^{ext}.
\end{aligned}
\end{equation}

As in the previous section, we now want to extend our model to a classical
HJ statistical mechanics for $N$-particles.

\subsection{Classical Hamilton-Jacobi statistical mechanics for \emph{zbw} particles
interacting with external fields}

If in the lab frame we do not know the actual positions of the \emph{zbw}
particles, then $\mathbf{q}_{i}(t)$ gets replaced by $\mathbf{q}_{i}$,
and the phase (67) becomes a field over the possible positions of
the \emph{zbw} particles. In the $v_{i}\ll c$ approximation 
\begin{equation}
\begin{aligned}S(q,t) & =\sum_{i=1}^{N}\int_{\mathbf{q}_{i}(t_{i})}^{\mathbf{q}_{i}(t)}m_{i}\mathbf{v}_{i}'(q(s),s)\cdot\mathbf{\mathit{d}q}_{i}(s)|_{\mathbf{q}_{j}(t)=\mathbf{q}_{j}}\\
 & -\sum_{i=1}^{N}\int_{t_{i}}^{t}\left(m_{i}c^{2}+\frac{1}{2m_{i}}\left[\mathbf{p}_{i}(q(s),s)-\frac{e_{i}}{c}\mathbf{A}_{i}^{ext}(q(s),s)\right]^{2}\right.\\
 & \left.+m_{i}\Phi_{gi}^{ext}(\mathbf{q}_{i}(s))+e_{i}\Phi_{ei}^{ext}(\mathbf{q}_{i}(s),s)\right)ds|_{\mathbf{q}_{j}(t)=\mathbf{q}_{j}}-\sum_{i=1}^{N}\hbar\phi_{i}.
\end{aligned}
\end{equation}
To obtain the equations of motion for $\ensuremath{S}$ and $\ensuremath{\mathbf{v}_{i}}$
we will now apply the classical analogue of Yasue's $N$-particle
variational principle, in anticipation of the method we will use for
constructing $N$-particle ZSM (we did not do this in the free-particles
case because there the dynamics of the particles is trivial).

First we define the ensemble-averaged $N$-particle phase/action (inputting
limits between initial and final states), 
\begin{equation}
\begin{aligned}J & =\mathrm{E}\left[\sum_{i=1}^{N}\left[\int_{\mathbf{q}_{iI}}^{\mathbf{q}_{iF}}m_{i}\mathbf{v}_{i}'\cdot\mathbf{\mathit{d}q}_{i}(t)-\int_{t_{I}}^{t_{F}}\left(m_{i}c^{2}+\frac{1}{2m_{i}}\left[\mathbf{p}_{i}-\frac{e_{i}}{c}\mathbf{A}_{i}^{ext}\right]^{2}+m_{i}\Phi_{gi}^{ext}+e_{i}\Phi_{ei}^{ext}\right)dt-\hbar\phi_{i}\right]\right]\\
 & =\mathrm{E}\left[\int_{t_{I}}^{t_{F}}\sum_{i=1}^{N}\left\{ \frac{1}{2}m\mathbf{v}_{i}^{2}+\frac{e_{i}}{c}\mathbf{A}_{i}^{ext}\cdot\mathbf{v}_{i}-m_{i}c^{2}-m_{i}\Phi_{gi}^{ext}-e_{i}\Phi_{ei}^{ext}\right\} dt-\sum_{i=1}^{N}\hbar\phi_{i}\right],
\end{aligned}
\end{equation}
where the equated expressions are related by the usual Legendre transformation.
Imposing the variational constraint 
\begin{equation}
J=extremal,
\end{equation}
a straightforward computation exactly along the lines of the Appendix
yields (69). And, upon replacing $\mathbf{q}_{i}(t)$ by $\mathbf{q}_{i}$,
we obtain the equation of motion for the acceleration field $\mathbf{a}(q,t)$:
\begin{equation}
\begin{aligned}m_{i}\mathbf{a}_{i} & =\left(\frac{\partial}{\partial t}+\mathbf{v}_{i}\cdot\nabla_{i}\right)\left[\nabla_{i}S-\frac{e_{i}}{c}\mathbf{A}_{i}^{ext}\right]\\
 & =-\nabla_{i}\left[m_{i}\Phi_{gi}^{ext}+e_{i}\Phi_{ei}^{ext}\right]-\frac{e_{i}}{c}\frac{\partial\mathbf{A}_{i}^{ext}}{\partial t}+\frac{e_{i}}{c}\mathbf{v}_{i}\times\mathbf{B}_{i}^{ext},
\end{aligned}
\end{equation}
where $\mathbf{v}_{i}=(1/m_{i})\nabla_{i}S-e_{i}\mathbf{A}_{i}^{ext}/m_{i}c$
corresponds to the kinetic velocity field associated with the \emph{i}-th
particle.

Integrating both sides of (73), summing over all $N$ terms, and setting
the integration constants equal to the rest masses, we then obtain
the classical $N$-particle Hamilton-Jacobi equation for (70) 
\begin{equation}
-\partial_{t}S=\sum_{i=1}^{N}\frac{\left(\nabla_{i}S-\frac{e_{i}}{c}\mathbf{A}_{i}^{ext}\right)^{2}}{2m_{i}}+\sum_{i=1}^{N}\left[m_{i}c^{2}+m_{i}\Phi_{gi}^{ext}+e_{i}\Phi_{ei}^{ext}\right].
\end{equation}
Correspondingly, the probability density $\rho(q,t)$ now evolves
by the modified $N$-particle continuity equation 
\begin{equation}
\frac{\partial\rho}{\partial t}=-\sum_{i=1}^{N}\nabla_{i}\cdot\left[\left(\frac{\nabla_{i}S}{m_{i}}-\frac{e_{i}}{m_{i}c}\mathbf{A}_{i}^{ext}\right)\rho\right],
\end{equation}
which preserves the normalization, $\int\rho_{0}d^{3N}q=1$. As in
the free particle case, since $S$ is a field over the possible positions
that the actual \emph{zbw} particles can occupy at a time \emph{t},
and since for each possible position the phase of each \emph{zbw}
particle satisfies the condition (63), it follows that $S$ is a single-valued
function of $q$ and $t$ (up to an additive integer multiple of $2\pi$)
and satisfies 
\begin{equation}
\sum_{i=1}^{N}\oint_{L}\nabla_{i}S\cdot d\mathbf{q}_{i}=nh.
\end{equation}
Then we can combine (74-75) into the nonlinear Schrödinger equation
\begin{equation}
i\hbar\frac{\partial\psi}{\partial t}=\sum_{i=1}^{N}\left[\frac{\left[-i\hbar\nabla_{i}-\frac{e_{i}}{c}\mathbf{A}_{i}^{ext}\right]^{2}}{2m_{i}}+\frac{\hbar^{2}}{2m_{i}}\frac{\nabla_{i}^{2}|\psi|}{|\psi|}+m_{i}\Phi_{gi}^{ext}+e_{i}\Phi_{ei}^{ext}+m_{i}c^{2}\right]\psi,
\end{equation}
with $N$-particle wave function $\psi(q,t)=\sqrt{\rho(q,t)}e^{iS(q,t)/\hbar}$,
which is single-valued because of (76). We can also obtain the single-particle
versions of (74-77) in the case that $S$, $\rho$, and $\psi$ satisfy
the factorization conditions (46), (48), and (53), respectively.

We are now ready to develop the more involved case of classically
interacting \emph{zbw} particles.

\subsection{Classically \textcolor{black}{interacting }\textcolor{black}{\emph{zbw}}\textcolor{black}{{}
particles }}

For simplicity we will consider just two \emph{zbw} particles classically
interacting through a scalar potential in the lab frame, under the
assumptions that $v_{i}\ll c$ and no external potentials are present.
(Restricting the particles to the non-relativistic regime also avoids
complications associated with potentials sourced by relativistic particles
\cite{Komar1978,Rohrlich1979}.) In particular, we suppose that the
particles interact through the Coulomb potential 
\begin{equation}
V_{c}^{int}(\mathbf{q}_{1}(t),\mathbf{q}_{2}(t))=\sum_{i=1}^{2}e_{i}\Phi_{c}^{int}(\mathbf{q}_{1}(t),\mathbf{q}_{2}(t))=\frac{e_{1}e_{2}}{|\mathbf{q}_{1}(t)-\mathbf{q}_{2}(t)|},
\end{equation}
where we recall $\Phi_{c}^{int}(\mathbf{q}_{i}(t),\mathbf{q}_{j}(t))=\frac{1}{2}\sum_{j=1}^{2(j\neq i)}\frac{e_{j}}{|\mathbf{q}_{i}(t)-\mathbf{q}_{j}(t)|}$.
Note that we make the point-like interaction assumption $|\mathbf{q}_{1}(t)-\mathbf{q}_{2}(t)|\gg\lambda_{c}$.
So the motions of the particles are not physically independent in
the lab frame, and this implies that the \emph{zbw} oscillation of
particle 1 (particle 2) in the lab frame is physically dependent on
the position of particle 2 (particle 1), through the interaction potential
(78). We can represent this physical dependence of the \emph{zbw}
oscillations by a non-separable joint phase change, which involves
contributions from both particles in the form 
\begin{equation}
\begin{aligned}\delta\theta_{joint}^{lab}(\mathbf{q}_{1}(t),\mathbf{q}_{2}(t),t) & =\left[\sum_{i=1}^{2}\omega_{ic}+\sum_{i=1}^{2}\omega_{ci}\frac{\mathbf{v}_{i}^{2}}{2c^{2}}+\sum_{i=1}^{2}\omega_{ci}\left(\frac{e_{i}\Phi_{c}^{int}}{m_{i}c^{2}}\right)\right]|_{\mathbf{q}_{j}=\mathbf{q}_{j}(t)}\\
 & \times\left(\delta t-\sum_{i=1}^{2}\frac{\mathbf{v}_{0i}}{c^{2}}\cdot\delta\mathbf{q}_{i}(t)\right)|_{\mathbf{q}_{j}=\mathbf{q}_{j}(t)}\\
 & =\sum_{i=1}^{2}\left[\omega_{ic}+\omega_{ci}\frac{\mathbf{v}_{i}^{2}}{2c^{2}}+\omega_{ci}\left(\frac{e_{i}\Phi_{c}^{int}}{m_{i}c^{2}}\right)\right]|_{\mathbf{q}_{j}=\mathbf{q}_{j}(t)}\delta t\\
 & -\sum_{i=1}^{2}\omega_{ci}\left(\frac{\mathbf{v}_{i}}{c^{2}}\right)|_{\mathbf{q}_{j}=\mathbf{q}_{j}(t)}\cdot\delta\mathbf{q}_{i}(t)\\
 & =\frac{1}{\hbar}\left[\left(\sum_{i=1}^{2}m_{i}c^{2}+\sum_{i=1}^{2}\frac{m_{i}\mathbf{v}_{i}^{2}}{2}+V_{c}^{int}\right)|_{\mathbf{q}_{j}=\mathbf{q}_{j}(t)}\delta t-\sum_{i=1}^{2}\mathbf{p}_{i}|_{\mathbf{q}_{j}=\mathbf{q}_{j}(t)}\cdot\delta\mathbf{q}_{i}(t)\right].
\end{aligned}
\end{equation}
Not surprisingly, when $|\mathbf{q}_{1}(t)-\mathbf{q}_{2}(t)|$ becomes
sufficiently great that $V_{c}^{int}$ is negligible, (79) reduces
to a sum of the physically independent phase changes associated with
particle 1 and particle 2, respectively.

Now, even though the particles don't have physically independent phases
because of $V_{c}^{int}$, it is clear that the \emph{zbw} oscillation
of particle 1 (particle 2) still has a well-defined individual phase
at all times. Moreover, we can deduce from (79) the individual (`conditional')
phase of a particle, given its physical interaction with the other
particle via (78), in much the same way that conditional wave functions
for subsystems of particles can be deduced from the universal wave
function in the de Broglie-Bohm theory \cite{Duerr1992,Norsen2010}.

To motivate this, let us first ask: in the instantaneous rest frame
(IRF) of (say) particle 1, how will the phase associated with its
\emph{zbw} oscillation change in time for a co-moving observer that's
continously monitoring the oscillation? The phase change associated
with particle 1 in its IRF can be obtained from (79) simply by subtracting
$\omega_{c2}\delta t$ and setting $\mathbf{v}_{1}=0$, giving 
\begin{equation}
\begin{aligned}\delta\theta_{1}^{rest}(\mathbf{q}_{01}(t),\mathbf{q}_{2}(t),t) & =\left[\omega_{c1}+\omega_{c2}\left(\frac{\mathbf{v}_{2}^{2}}{2c^{2}}\right)+\sum_{i=1}^{2}\omega_{ci}\left(\frac{e_{i}\Phi_{c}^{int}}{m_{i}c^{2}}\right)\right]|_{\mathbf{q}_{j}=\mathbf{q}_{j}(t)}\delta t-\omega_{c2}\left(\frac{\mathbf{v}_{2}}{c^{2}}\right)|_{\mathbf{q}_{j}=\mathbf{q}_{j}(t)}\cdot\delta\mathbf{q}_{2}(t)\\
 & =\frac{1}{\hbar}\left[\left(m_{1}c^{2}+\frac{m_{2}\mathbf{v}_{2}^{2}}{2}+V_{c}^{int}\right)|_{\mathbf{q}_{j}=\mathbf{q}_{j}(t)}\delta t-\mathbf{p}_{2}|_{\mathbf{q}_{j}=\mathbf{q}_{j}(t)}\cdot\delta\mathbf{q}_{2}(t)\right],
\end{aligned}
\end{equation}
where $\mathbf{q}_{01}(t)$ denotes the translational coordinate of
particle 1 in its IRF (which, of course, changes as a function of
time due to the Coulomb interaction). In other words, (80) tells us
how the Compton frequency of particle 1, $\omega_{c1}$, gets modulated
by the physical coupling of particle 1 to particle 2, in the IRF of
particle 1. Thus (80) represents the conditional phase change of particle
1 in its IRF. We can also confirm that when $\Phi_{c}^{int}\approx0$
the velocity of particle 2 no longer depends on the position of particle
1 at time $t$, leaving $\delta\theta_{1}^{rest}=\omega_{c1}\delta t_{0}$.
Likewise we can obtain the conditional \emph{zbw} phase of particle
2 in its IRF.

The conditional \emph{zbw} phase of particle 1 in the lab frame where
$\mathbf{v}_{1}\neq0$ is just 
\begin{equation}
\begin{aligned}\delta\theta_{1}^{lab}(\mathbf{q}_{1}(t),\mathbf{q}_{2}(t),t) & =\left[\omega_{c1}+\sum_{i=1}^{2}\omega_{ci}\left(\frac{\mathbf{v}_{i}^{2}}{2c^{2}}\right)+\sum_{i=1}^{2}\omega_{ci}\left(\frac{e_{i}\Phi_{c}^{int}}{m_{i}c^{2}}\right)\right]|_{\mathbf{q}_{j}=\mathbf{q}_{j}(t)}\delta t\\
 & -\sum_{i=1}^{2}\omega_{ci}\left(\frac{\mathbf{v}_{i}}{c^{2}}\right)|_{\mathbf{q}_{j}=\mathbf{q}_{j}(t)}\cdot\delta\mathbf{q}_{i}(t)\\
 & =\frac{1}{\hbar}\left[\left(m_{1}c^{2}+\sum_{i=1}^{2}\frac{m_{i}\mathbf{v}_{i}^{2}}{2}+V_{c}^{int}\right)|_{\mathbf{q}_{j}=\mathbf{q}_{j}(t)}\delta t-\sum_{i=1}^{2}\mathbf{p}_{i}|_{\mathbf{q}_{j}=\mathbf{q}_{j}(t)}\cdot\delta\mathbf{q}_{i}(t)\right].
\end{aligned}
\end{equation}
Equivalently, we can obtain (81) by just subtracting $\omega_{c2}\delta t$
from (79). And likewise for the conditional \emph{zbw} phase of particle
2 in the lab frame.

Recall that, by hypothesis, each \emph{zbw} particle is essentially
a harmonic oscillator. This means that when $V_{c}^{int}\approx0$
each particle has its own well-defined phase at each point along its
space-time trajectory. Consistency with this hypothesis also means
that when $V_{c}^{int}>0$ the joint phase must be a well-defined
function of the space-time trajectories of \emph{both} particles (since
we posit that both particles remain harmonic oscillators despite having
their oscillations physically coupled by $V_{c}^{int}$). Then for
a closed loop \emph{L,} along which each particle can be physically
or virtually displaced, the joint phase in the lab frame will satisfy
\begin{equation}
\sum_{i=1}^{2}\oint_{L}\delta_{i}\theta_{joint}^{lab}=2\pi n,
\end{equation}
and for a loop in which time is held fixed, 
\begin{equation}
\sum_{i=1}^{2}\oint_{L}\mathbf{p}_{i}\cdot\delta\mathbf{q}_{i}(t)=nh.
\end{equation}
It also follows from (82) and (83) that 
\begin{equation}
\oint_{L}\delta_{1}\theta_{joint}^{lab}=2\pi n,
\end{equation}
and 
\begin{equation}
\oint_{L}\mathbf{p}_{1}\cdot\delta\mathbf{q}_{1}(t)=nh,
\end{equation}
where this time the closed-loop integration involves keeping the coordinate
of particle 2 fixed while particle 1 is displaced along \emph{L}.
From (82-85), it will also be the case that 
\begin{equation}
\sum_{i=1}^{2}\oint_{L}\delta_{i}\theta_{1}^{lab}=2\pi n,
\end{equation}
and 
\begin{equation}
\oint_{L}\delta_{1}\theta_{1}^{lab}=2\pi n.
\end{equation}

Integrating (79) and multiplying through by $\hbar$ yields (using
$S_{joint}^{lab}\eqqcolon S)$ 
\begin{equation}
S=\sum_{i=1}^{2}\int_{\mathbf{q}_{i}(t_{i})}^{\mathbf{q}_{i}(t)}\mathbf{p}_{i}\cdot d\mathbf{q}_{i}(s)-\sum_{i=1}^{2}\int_{t_{i}}^{t}\left(m_{i}c^{2}+\frac{m_{i}\mathbf{v}_{i}^{2}}{2}+e_{i}\Phi_{c}^{int}\right)ds-\sum_{i=1}^{2}\hbar\phi_{i},
\end{equation}
and evolves by 
\begin{equation}
-\partial_{t}S|_{\mathbf{q}_{j}=\mathbf{q}_{j}(t)}=\sum_{i=1}^{2}m_{i}c^{2}+\sum_{i=1}^{2}\frac{\left(\nabla_{i}S\right)^{2}}{2m_{i}}|_{\mathbf{q}_{j}=\mathbf{q}_{j}(t)}+V_{c}^{int}.
\end{equation}
The conditional phase $S_{1}^{lab}=S_{1}$ and its equation of motion
only differ from (88-89) by subtracting $m_{2}c^{2}t-\hbar\phi_{2}$.
Analogous considerations apply to particle 2. Finally, the acceleration
of the \emph{i}-th particle is obtained from the equation of motion
\begin{equation}
m_{i}\ddot{\mathbf{q}}_{i}(t)=\left[\partial_{t}\mathbf{p}_{i}+\mathbf{v}_{i}\cdot\nabla_{i}\mathbf{p}_{i}\right]|_{\mathbf{q}_{j}=\mathbf{q}_{j}(t)}=-\nabla_{i}V_{c}^{int}|_{\mathbf{q}_{j}=\mathbf{q}_{j}(t)}.
\end{equation}

Another, more convenient way of modeling the case of two classically
interacting \emph{zbw} particles is by exploiting the well-known fact
that a two-particle system with an interaction potential of the form
(78) has an equivalent Hamiltonian of the form (ignoring the trivial
CM motion) 
\begin{equation}
E_{rel}=\frac{p_{rel}^{2}}{2\mu}+V_{rel}(|\mathbf{q}_{rel}(t)|)+\mu c^{2},
\end{equation}
where the reduced mass $\mu=m_{1}m_{2}/(m_{1}+m_{2})$ and $V_{rel}(|\mathbf{q}_{rel}(t)|)=V_{c}^{int}(|\mathbf{q}_{1}(t)-\mathbf{q}_{2}(t)|)$.
In other words, (91) describes a fictitious \emph{zbw} particle of
mass $\mu$ and relative coordinate $\mathbf{q}_{rel}(t)$, moving
in an ``external'' potential $V_{rel}(|\mathbf{q}_{rel}(t)|)$.
This fictitious particle then has a Compton frequency, $\omega_{c}^{red}=\mu c^{2}/\hbar$,
and an associated phase change in the lab frame of the form 
\begin{equation}
\begin{aligned}\delta\theta_{rel}(\mathbf{q}_{rel}(t)) & =\left(\omega{}_{c}^{red}+\omega{}_{c}^{red}\frac{\mathbf{v}_{rel}^{2}(\mathbf{q}_{rel}(t))}{2c^{2}}+\omega_{c}^{red}\frac{V_{rel}(|\mathbf{q}_{rel}(t)|)}{\mu c^{2}}\right)\left(\delta t-\frac{\mathbf{v}_{0rel}(\mathbf{q}_{rel}(t))\cdot\delta\mathbf{q}_{rel}(t)}{c^{2}}\right)\\
 & =\frac{1}{\hbar}\left[\left(\mu c^{2}+\frac{\mu\mathbf{v}_{rel}^{2}}{2}+V_{rel}\right)\delta t-\mathbf{p}_{rel}\cdot\delta\mathbf{q}_{rel}(t)\right].
\end{aligned}
\end{equation}
Upon integration, this of course gives 
\begin{equation}
S_{rel}\coloneqq-\hbar\theta_{rel}=\int\left[\mathbf{p}_{rel}\cdot d\mathbf{q}_{rel}(t)-E_{rel}dt\right]-\hbar\phi_{rel},
\end{equation}
which evolves in time by the HJ equation 
\begin{equation}
-\partial_{t}S_{rel}|_{\mathbf{q}_{rel}=\mathbf{q}_{rel}(t)}=\mu c^{2}+\frac{\left(\nabla_{rel}S_{rel}\right)^{2}}{2\mu}|_{\mathbf{q}_{rel}=\mathbf{q}_{rel}(t)}+V_{rel},
\end{equation}
and gives the equation of motion 
\begin{equation}
\mu\ddot{\mathbf{q}}_{rel}(t)=\left[\partial_{t}\mathbf{p}_{rel}+\mathbf{v}_{rel}\cdot\nabla_{rel}\mathbf{p}_{rel}\right]|_{\mathbf{q}_{rel}=\mathbf{q}_{rel}(t)}=-\nabla_{rel}V_{rel}|_{\mathbf{q}_{rel}=\mathbf{q}_{rel}(t)}.
\end{equation}

Since this situation is formally equivalent to the case of a single
\emph{zbw} particle moving in an external field, we can immediately
see that it follows 
\begin{equation}
\oint_{L}\delta S_{rel}=nh,
\end{equation}
and 
\begin{equation}
\oint_{L}\mathbf{p}_{rel}\cdot\delta\mathbf{q}_{rel}(t)=nh.
\end{equation}
Furthermore, the physical equivalence between this coordinatization
and the original two-particle coordinatization establishes that if
phase quantization holds in one coordinatization it must hold in the
other.

While we considered here only two \emph{zbw} particles classically
interacting through an electric scalar potential, all our considerations
straightforwardly generalize to the case of many \emph{zbw} particles
classically interacting through electric scalar potentials as well
as magnetic vector potentials (and likewise for the gravitational
analogues).

\subsection{Classical \textcolor{black}{Hamilton-Jacobi statistical mechanics
for two interacting }\textcolor{black}{\emph{zbw}}\textcolor{black}{{}
particles }}

For a statistical mechanical description of two classically interacting
\emph{zbw} particles, the trajectories $\{\mathbf{q}_{1}(t),\mathbf{q}_{2}(t)\}$
get replaced with the coordinates $\{\mathbf{q}_{1},\mathbf{q}_{2}\}$,
and the non-relativistic joint phase field in the lab frame is obtained
from (88) as 
\begin{equation}
\begin{aligned}S(\mathbf{q}_{1},\mathbf{q}_{2},t) & =\sum_{i=1}^{2}\int_{\mathbf{q}_{i}(t_{i})}^{\mathbf{q}_{i}(t)}\mathbf{p}_{i}\cdot d\mathbf{q}_{i}(s)|_{\mathbf{q}_{j}(t)=\mathbf{q}_{j}}\\
 & -\sum_{i=1}^{2}\int_{t_{i}}^{t}\left[m_{i}c^{2}+\frac{m_{i}\mathbf{v}_{i}^{2}(\mathbf{q}_{1}(s),\mathbf{q}_{2}(s),s)}{2}+e_{i}\Phi_{c}^{int}(\mathbf{q}_{1}(s),\mathbf{q}_{2}(s))\right]ds|_{\mathbf{q}_{j}(t)=\mathbf{q}_{j}}-\sum_{i=1}^{2}\hbar\phi_{i},
\end{aligned}
\end{equation}
and evolves by 
\begin{equation}
-\partial_{t}S=\sum_{i=1}^{2}m_{i}c^{2}+\sum_{i=1}^{2}\frac{\left(\nabla_{i}S\right)^{2}}{2m_{i}}+V_{c}^{int},
\end{equation}
where $\mathbf{v}_{i}(\mathbf{q}_{1},\mathbf{q}_{2})=\nabla_{i}S(\mathbf{q}_{1},\mathbf{q}_{2},t)/m_{i}$
. Since (98) is a field over the possible positions of the actual
\emph{zbw} particles, and since for each possible initial position
the phase of each \emph{zbw} particle will satisfy relation (83),
it follows that 
\begin{equation}
\sum_{i=1}^{2}\oint_{L}\nabla_{i}S\cdot d\mathbf{q}_{i}=nh,
\end{equation}
where \emph{L} is now a mathematical loop in the 2-particle configuration
space.

The two-particle probability density $\rho(\mathbf{q}_{1},\mathbf{q}_{2},t)\geq0$
evolves by the two-particle continuity equation 
\begin{equation}
\frac{\partial\rho}{\partial t}=-\sum_{i=1}^{2}\nabla_{i}\cdot\left[\left(\mathbf{\frac{\nabla_{\mathit{i}}\mathrm{\mathit{S}}}{\mathit{m_{i}}}}\right)\rho\right],
\end{equation}
and the ensemble-averaged two-particle action is defined by 
\begin{equation}
\begin{aligned}J & =\mathrm{E}\left[\sum_{i=1}^{2}\left[\int_{\mathbf{q}_{iI}}^{\mathbf{q}_{iF}}m_{i}\mathbf{v}_{i}\cdot\mathbf{\mathit{d}q}_{i}(t)-\int_{t_{I}}^{t_{F}}\left(m_{i}c^{2}+\frac{\mathbf{p}_{i}^{2}}{2m_{i}}+e_{i}\Phi_{ci}^{int}\right)dt-\hbar\phi_{i}\right]\right]\\
 & =\mathrm{E}\left[\int_{t_{I}}^{t_{F}}\sum_{i=1}^{N}\left(\frac{1}{2}m\mathbf{v}_{i}^{2}-m_{i}c^{2}-e_{i}\Phi_{ci}^{int}\right)dt-\sum_{i=1}^{N}\hbar\phi_{i}\right],
\end{aligned}
\end{equation}
where the equated expressions are related by the usual Legendre transformation.
Imposing 
\begin{equation}
J=extremal,
\end{equation}
straightforward manipulations along the lines of those in the Appendix
yield (90). And, upon replacing $\mathbf{q}_{i}(t)$ with $\mathbf{q}_{i}$,
we obtain the classical Newtonian equation for the acceleration field
$\mathbf{a}_{i}(\mathbf{q}_{1},\mathbf{q}_{2},t)$: 
\begin{equation}
m_{i}\mathbf{a}_{i}=\partial_{t}\mathbf{p}_{i}+\mathbf{v}_{i}\cdot\nabla_{i}\mathbf{p}_{i}=-\nabla_{i}V_{c}^{int}.
\end{equation}

Now, we can obtain the conditional \emph{zbw} phase field for particle
1 by evaluating the joint phase field at the actual position of particle
2 at time $t$, i.e., $S(\mathbf{q}_{1},\mathbf{q}_{2}(t),t)\eqqcolon S_{1}(\mathbf{q}_{1},t)$.
Taking the total time derivative we have 
\begin{equation}
\partial_{t}S_{1}(\mathbf{q}_{1},t)=\partial_{t}S(\mathbf{q}_{1},\mathbf{q}_{2},t)|_{\mathbf{q}_{2}=\mathbf{q}_{2}(t)}+\frac{d\mathbf{q}_{2}(t)}{dt}\cdot\nabla_{2}S(\mathbf{q}_{1},\mathbf{q}_{2},t)|_{\mathbf{q}_{2}=\mathbf{q}_{2}(t)},
\end{equation}
where the conditional velocities 
\begin{equation}
\frac{d\mathbf{q}_{1}(t)}{dt}=\mathbf{v}_{1}(\mathbf{q}_{1},t)|_{\mathbf{q}_{1}=\mathbf{q}_{1}(t)}=\frac{\nabla_{1}S_{1}(\mathbf{q}_{1},t)}{m_{1}}|_{\mathbf{q}_{1}=\mathbf{q}_{1}(t)},
\end{equation}
and 
\begin{equation}
\frac{d\mathbf{q}_{2}(t)}{dt}=\mathbf{v}_{2}(\mathbf{q}_{2},t)|_{\mathbf{q}_{2}=\mathbf{q}_{2}(t)}=\frac{\nabla_{2}S_{2}(\mathbf{q}_{2},t)}{m_{2}}|_{\mathbf{q}_{2}=\mathbf{q}_{2}(t)},
\end{equation}
the latter defined from the conditional phase field $S_{2}(\mathbf{q}_{2},t)$
for particle 2. Inserting (105) into the left hand side of (99) and
adding the corresponding term on the right hand side, we then find
that the conditional phase field for particle 1 evolves by a `conditional
HJ equation', namely 
\begin{equation}
-\partial_{t}S_{1}=m_{1}c^{2}+\frac{\left(\nabla_{1}S_{1}\right)^{2}}{2m_{1}}+\frac{\left(\nabla_{2}S\right)^{2}}{2m_{2}}|_{\mathbf{q}_{2}=\mathbf{q}_{2}(t)}-\frac{d\mathbf{q}_{2}(t)}{dt}\cdot\nabla_{2}S|_{\mathbf{q}_{2}=\mathbf{q}_{2}(t)}+V_{c}^{int}(\mathbf{q}_{1},\mathbf{q}_{2}(t)),
\end{equation}
where $V_{c}^{int}(\mathbf{q}_{1},\mathbf{q}_{2}(t))$ is the `conditional
potential' for particle 1; that is, the potential field that particle
1, at location $\mathbf{q}_{1}$, would `feel' given the actual location
of particle 2. The solution of (108) can be verified as 
\begin{equation}
S_{1}=\int\mathbf{p}_{1}\cdot d\mathbf{q}_{1}-\int\left[m_{1}c^{2}+\frac{m_{1}\mathbf{v}_{1}^{2}}{2}+\frac{m_{1}\mathbf{v}_{2}^{2}}{2}-\mathbf{p}_{2}\cdot\frac{d\mathbf{q}_{2}(t)}{dt}+V_{c}^{int}\right]dt-\hbar\phi_{1}.
\end{equation}
Notice here that the conditional phase field is a field on 3-D space.
This makes perfect sense since, after all, the conditional phase refers
to the phase associated to the \emph{zbw} oscillation of particle
1, a real physical oscillation in 3-D space. It can also be verified
that when (109) is evaluated at $\mathbf{q}_{1}=\mathbf{q}_{1}(t)$,
it is equivalent to $S_{joint}^{lab}(\mathbf{q}_{1}(t),\mathbf{q}_{2}(t),t)-m_{2}c^{2}t+\hbar\phi_{2}$.

Once again, since the conditional \emph{zbw} phase field for particle
1 is a field over the possible positions that \emph{zbw} particle
1 could actually occupy at time $t$, it will be the case that 
\begin{equation}
\oint_{L}\nabla_{1}S_{1}\cdot d\mathbf{q}_{1}=nh,
\end{equation}
where \emph{L} is a mathematical loop in 3-D space.

Likewise, we can obtain the conditional probability density for particle
1 by writing $\rho(\mathbf{q}_{1},\mathbf{q}_{2}(t),t)\eqqcolon\rho_{1}(\mathbf{q}_{1},t)$.
Taking the total time derivative gives 
\begin{equation}
\partial_{t}\rho_{1}(\mathbf{q}_{1},t)=\partial_{t}\rho(\mathbf{q}_{1},\mathbf{q}_{2},t)|_{\mathbf{q}_{2}=\mathbf{q}_{2}(t)}+\frac{d\mathbf{q}_{2}(t)}{dt}\cdot\nabla_{2}\rho(\mathbf{q}_{1},\mathbf{q}_{2},t)|_{\mathbf{q}_{2}=\mathbf{q}_{2}(t)}.
\end{equation}
Inserting this on the left hand side of (101) and adding the corresponding
term on the right hand side, we obtain the conditional continuity
equation for particle 1: 
\begin{equation}
\partial_{t}\rho_{1}=-\nabla_{1}\cdot\left[\left(\frac{\nabla_{1}S_{1}}{m_{1}}\right)\rho_{1}\right]-\nabla_{2}\cdot\left[\left(\frac{\nabla_{2}S}{m_{2}}\right)\rho\right]|_{\mathbf{q}_{2}=\mathbf{q}_{2}(t)}+\frac{d\mathbf{q}_{2}(t)}{dt}\cdot\nabla_{2}\rho|_{\mathbf{q}_{2}=\mathbf{q}_{2}(t)},
\end{equation}
which implies $\rho_{1}(\mathbf{q}_{1},t)\geq0$ and (upon suitable
redefinition of $\rho_{1}(\mathbf{q}_{1},t)$) preservation of the
normalization $\int_{\mathbb{R}^{3}}\rho_{1}(\mathbf{q}_{1},0)=1$.

The ensemble-averaged conditional action for particle 1 is defined
as 
\begin{equation}
\begin{aligned}J_{1} & =\mathrm{E}\left[\int_{\mathbf{q}_{1I}}^{\mathbf{q}_{1F}}m_{1}\mathbf{v}_{1}\cdot\mathbf{\mathit{d}q}_{1}(t)-\int_{t_{I}}^{t_{F}}\left(m_{1}c^{2}+\frac{m_{1}\mathbf{v}_{1}^{2}}{2}+\frac{m_{2}\mathbf{v}_{2}^{2}}{2}-\mathbf{p}_{2}\cdot\frac{d\mathbf{q}_{2}(t)}{dt}+V_{c}^{int}\right)dt-\hbar\phi_{1}\right]\\
 & =\mathrm{E}\left[\int_{t_{I}}^{t_{F}}\left[\frac{1}{2}m_{1}\mathbf{v}_{1}^{2}+\frac{1}{2}m_{2}\mathbf{v}_{2}^{2}-m_{1}c^{2}-V_{c}^{int}\right]dt-\hbar\phi_{1}\right],
\end{aligned}
\end{equation}
where it can be readily confirmed that the equated lines are related
by the Legendre transformation. Imposing 
\begin{equation}
J_{1}=extremal,
\end{equation}
where the subscript 1 denotes that the variation is only with respect
to $\mathbf{q}_{1}(t)$, straightforward manipulations analogous to
those in the Appendix yield, upon replacing $\mathbf{q}_{1}(t)$ with
$\mathbf{q}_{1}$, the classical equation of motion for the conditional
acceleration field of particle 1: 
\begin{equation}
m_{1}\mathbf{a}_{1}(\mathbf{q}_{1},t)=\left[\partial_{t}\mathbf{p}_{1}+\mathbf{v}_{1}\cdot\nabla_{i}\mathbf{p}_{1}\right](\mathbf{q}_{1},t)=-\nabla_{1}V_{c}^{int}(\mathbf{q}_{1},\mathbf{q}_{2}(t)).
\end{equation}
The conditional phase field, probability density, etc., for particle
2, are developed analogously.

We now turn to the formulation of our classical statistical mechanics
in terms of the reduced mass \emph{zbw} particle. Replacing $\mathbf{q}_{rel}(t)$
with $\mathbf{q}_{rel}$, the reduced mass \emph{zbw} phase field
\begin{equation}
\begin{aligned}S_{rel}(\mathbf{q}_{rel},t) & =\int_{\mathbf{q}_{rel}(t_{i})}^{\mathbf{q}_{rel}(t)}\mathbf{p}_{rel}\cdot d\mathbf{q}{}_{rel}(s)|_{\mathbf{q}{}_{rel}(t)=\mathbf{q}_{rel}}\\
 & -\int_{t_{i}}^{t}\left(\mu c^{2}+\frac{\mathbf{p}_{rel}^{2}}{2\mu}+V_{rel}\right)ds|_{\mathbf{q}{}_{rel}(t)=\mathbf{q}_{rel}}-\hbar\phi_{rel},
\end{aligned}
\end{equation}
evolves by the reduced mass HJ equation 
\begin{equation}
-\partial_{t}S_{rel}=\mu c^{2}+\frac{\left(\nabla_{rel}S_{rel}\right)^{2}}{2\mu}+V_{rel},
\end{equation}
and satisfies 
\begin{equation}
\oint_{L}\nabla_{rel}S_{rel}\cdot d\mathbf{q}_{rel}=nh,
\end{equation}
where \emph{L} is a mathematical loop in 3-D space. Introducing the
probability density for the reduced mass \emph{zbw} particle, $\rho_{rel}(\mathbf{q}_{rel},t)\geq0$,
it is straightforward to show it evolves by the continuity equation
\begin{equation}
\frac{\partial\rho_{rel}}{\partial t}=-\nabla_{rel}\cdot\left[\left(\mathbf{\frac{\nabla_{\mathit{rel}}\mathrm{\mathit{S_{rel}}}}{\mathit{m_{rel}}}}\right)\rho_{rel}\right],
\end{equation}
which preserves the normalization $\intop_{\mathbb{R}^{3}}d^{3}\mathbf{q}_{rel}\rho_{rel}(\mathbf{q}_{rel},0)=1$.
The corresponding ensemble-averaged action for the reduced mass particle
is defined by 
\begin{equation}
\begin{aligned}J_{rel} & =\mathrm{E}\left[\int_{\mathbf{q}_{relI}}^{\mathbf{q}_{relF}}\mu\mathbf{v}_{rel}\cdot\mathbf{\mathit{d}q}_{rel}(t)-\int_{t_{I}}^{t_{F}}\left(\mu c^{2}+\frac{\mathbf{p}_{rel}^{2}}{2\mu}+V_{rel}\right)dt-\hbar\phi_{rel}\right]\\
 & =\mathrm{E}\left[\int_{t_{I}}^{t_{F}}\left(\frac{1}{2}\mu\mathbf{v}_{rel}^{2}-\mu c^{2}-V_{rel}\right)dt-\hbar\phi_{rel}\right].
\end{aligned}
\end{equation}
Imposing the constraint 
\begin{equation}
J_{rel}=extremal,
\end{equation}
we obtain after manipulations (and replacing $\mathbf{q}_{rel}(t)$
by $\mathbf{q}_{rel}$) the equation of motion 
\begin{equation}
\mu\mathbf{a}_{rel}(\mathbf{q}_{rel},t)=\partial_{t}\mathbf{p}_{rel}+\mathbf{v}_{rel}\cdot\nabla_{rel}\mathbf{p}_{rel}=-\nabla_{rel}V_{rel}(|\mathbf{q}_{rel}|).
\end{equation}

Let us now recover the nonlinear Schrödinger equations for each of
the three cases we've considered.

The combination of (99)-(101) gives 
\begin{equation}
i\hbar\frac{\partial\psi}{\partial t}=\sum_{i=1}^{2}\left[-\frac{\hbar^{2}}{2m_{i}}\nabla_{i}^{2}+\frac{\hbar^{2}}{2m_{i}}\frac{\nabla_{i}^{2}|\psi|}{|\psi|}+m_{i}c^{2}\right]\psi+V_{c}^{int}\psi,
\end{equation}
where $\psi(\mathbf{q}_{1},\mathbf{q}_{2},t)=\sqrt{\rho(\mathbf{q}_{1},\mathbf{q}_{2},t)}e^{iS(\mathbf{q}_{1},\mathbf{q}_{2},t)/\hbar}$
is single-valued by (100).

Combining (108) and (112) gives the conditional nonlinear Schrödinger
equation for particle 1: 
\begin{equation}
\begin{aligned}i\hbar\frac{\partial\psi_{1}}{\partial t} & =-\frac{\hbar^{2}}{2m_{1}}\nabla_{1}^{2}\psi_{1}-\frac{\hbar^{2}}{2m_{1}}\nabla_{2}^{2}\psi|_{\mathbf{q}_{2}=\mathbf{q}_{2}(t)}+V_{c}^{int}(\mathbf{q}_{1},\mathbf{q}_{2}(t))\psi_{1}+m_{1}c^{2}\psi_{1}\\
 & +i\hbar\frac{d\mathbf{q}_{2}(t)}{dt}\cdot\nabla_{2}\psi|_{\mathbf{q}_{2}=\mathbf{q}_{2}(t)}+\left(\frac{\hbar^{2}}{2m_{1}}\frac{\nabla_{1}^{2}|\psi_{1}|}{|\psi_{1}|}\right)\psi_{1}+\left(\frac{\hbar^{2}}{2m_{2}}\frac{\nabla_{2}^{2}|\psi|}{|\psi|}\right)|_{\mathbf{q}_{2}=\mathbf{q}_{2}(t)}\psi_{1},
\end{aligned}
\end{equation}
where $\psi(\mathbf{q}_{1},\mathbf{q}_{2}(t),t)\eqqcolon\psi_{1}(\mathbf{q}_{1},t)=\sqrt{\rho_{1}(\mathbf{q}_{1},t)}e^{iS_{1}(\mathbf{q}_{1},t)/\hbar}$
is the conditional classical wave function for particle 1, and satisfies
single-valuedness as a consequence of (110). Here $d\mathbf{q}_{2}(t)/dt=(\hbar/m_{2})\mathrm{Im}\{\nabla_{2}ln(\psi_{2})\}|_{\mathbf{q}_{2}=\mathbf{q}_{2}(t)}$,
where $\psi_{2}=\psi_{2}(\mathbf{q}_{2},t)$ is the conditional wave
function for particle 2 and satisfies a conditional nonlinear Schrödinger
equation analogous to (124). Note also that (124) can be obtained
by taking the total time derivative of the conditional wave function
for particle 1 
\begin{equation}
\partial_{t}\psi_{1}(\mathbf{q}_{1},t)=\partial_{t}\psi(\mathbf{q}_{1},\mathbf{q}_{2},t)|_{\mathbf{q}_{2}=\mathbf{q}_{2}(t)}+\frac{d\mathbf{q}_{2}(t)}{dt}\cdot\nabla_{2}\psi(\mathbf{q}_{1},\mathbf{q}_{2},t)|_{\mathbf{q}_{2}=\mathbf{q}_{2}(t)},
\end{equation}
inserting this on the left hand side of (123), adding the corresponding
term on the right hand side, and subtracting $m_{2}c^{2}\psi_{1}$.

Finally, combining (117-119) gives the nonlinear Schrödinger equation
for the fictitious reduced mass particle: 
\begin{equation}
i\hbar\frac{\partial\psi_{rel}}{\partial t}=\left[-\frac{\hbar^{2}}{2\mu}\nabla_{rel}^{2}+\frac{\hbar^{2}}{2\mu}\frac{\nabla_{rel}^{2}|\psi_{rel}|}{|\psi_{rel}|}+\mu c^{2}\right]\psi_{rel}+V(|\mathbf{q}_{rel}|)\psi_{rel},
\end{equation}
where $\psi_{rel}(\mathbf{q}_{rel},t)=\sqrt{\rho_{rel}(\mathbf{q}_{rel},t)}e^{iS_{rel}(\mathbf{q}_{rel},t)/\hbar}$
is a single-valued classical wave function. As with the linear Schrödinger
equation of quantum mechanics, it is easily verified that (126) can
be obtained from (123) by transforming the two-particle Hamiltonian
operator to the center of mass and relative coordinates.

This completes the development of the classical HJ statistical mechanics
for two classically interacting \emph{zbw} particles. The generalization
to \emph{N} \emph{zbw} particles interacting through their electric
scalar and magnetic vector potentials (and the gravitational analogues
thereof) is straightforward, but will not be given here due to unnecessary
mathematical complexity.

\subsection{Remarks on close-range interactions}

Throughout we have assumed the point-like interaction case, $q_{rel}(t)=|\mathbf{q}_{1}(t)-\mathbf{q}_{2}(t)|\gg\lambda_{c}$.
But what changes when $q_{rel}(t)=|\mathbf{q}_{1}(t)-\mathbf{q}_{2}(t)|\sim\lambda_{c}$?
Not much. To show this, we adopt the approach of Zelevinsky \cite{Zelevinsky2011}
in modeling the deviation from point-like interactions with a Darwin
interaction term as follows. Consider the (hypothesized) 3-D \emph{zbw}
oscillation/fluctuation around the relative coordinate, $\mathbf{q}_{rel}(t)+\delta\mathbf{q}(t)$,
where $\delta q_{max}=|\mathbf{\delta q}_{max}(t)|=\lambda_{c}$.
Taylor expand the (Coulomb or Newtonian) interaction potential into
$V_{int}(|\mathbf{q}_{rel}(t)+\delta\mathbf{q}(t)|)\thickapprox V_{int}(|\mathbf{q}_{rel}(t)|)+\delta\mathbf{q}(t)\cdot\nabla V_{int}(|\mathbf{q}_{rel}(t)|)+\frac{1}{2}\sum_{i,j}\delta q^{i}(t)\delta q^{j}(t)\partial^{i}\partial^{j}V_{int}(|\mathbf{q}_{rel}(t)|)$.
Then, under the reasonable assumptions that the mean and variance
of the fluctuations are given by $<\delta\mathbf{q}(t)>=0$ and $<\delta q(t)^{i}\delta q(t)^{j}>=\frac{1}{3}<\delta q(t)^{2}>\delta_{ij}$,
respectively, the fluctuation-averaged potential $<V_{int}(|\mathbf{q}_{rel}(t)+\delta\mathbf{q}(t)|)>=V_{int}(|\mathbf{q}_{rel}(t)|)+\frac{1}{6}<\delta q(t)^{2}>\nabla^{2}V_{int}(|\mathbf{q}_{rel}(t)|).$
Finally, approximating $<\delta q(t)^{2}>=\frac{1}{2}\lambda_{c}^{2}$,
we find that the perturbation of the potential due to the fluctuations
is $\delta V\thickapprox\frac{1}{12}\lambda_{c}^{2}\nabla^{2}V_{int}=\frac{1}{12}\lambda_{c}^{2}4\pi K\delta(\mathbf{q})$,
if the interaction potential is of the general form, $V_{int}(q)=K\mathbf{\hat{q}}/q$,
where \emph{K} is a constant.

Note that because the \emph{zbw} oscillation is a (rheonomic) constraint
on each particle, the Coulomb interaction between them never causes
their oscillations to deviate from simple harmonic motion (even though
their oscillation frequencies can slightly shift by an amount of the
order $(\omega_{c}V_{int})/\hbar$); so phase/momentum quantization
for each particle is not altered, even when $q_{rel}(t)\sim\lambda_{c}$.
Alternatively, we could relax the \emph{zbw} constraint by assuming
that when $q_{rel}(t)\sim\lambda_{c}$, a slight deviation from simple
harmonic motion occurs because the Coulomb repulsion is sufficiently
strong to impart a nonlinear perturbation to the internal harmonic
potential of each \emph{zbw} particle; but this perturbation should
drop off rapidly as the particles move away from each other so that
simple harmonic motion is quickly restored and the momentum quantization
is stable again. Ideally, a physical model of the \emph{zbw} particle
would implement this latter possibility, but for the purposes of this
paper, it will simply be assumed throughout that the Coulomb interaction
does not alter the simple harmonic nature of the \emph{zbw} oscillations.

\section{Zitterbewegung Stochastic Mechanics}

\subsection{Free \emph{zbw} particles}

We take as our starting point the hypothesis that \emph{N} particles
of rest masses, $m_{i}$, and 3-D space positions, $\mathbf{q}_{i}(t)$,
are immersed in Nelson's hypothesized ether and undergo conservative
diffusion processes according to the stochastic differential equations
\begin{equation}
d\mathbf{q}_{i}(t)=\mathbf{b}_{i}(q(t),t)dt+d\mathbf{W}_{i}(t),
\end{equation}
and 
\begin{equation}
d\mathbf{q}_{i}(t)=\mathbf{b}_{i*}(q(t),t)dt+d\mathbf{W}_{i*}(t),
\end{equation}
where the forward Wiener processes $d\mathbf{W}_{i}(t)$ satisfy $\mathrm{E}_{t}\left[d\mathbf{W}_{i}\right]=0$
and $\mathrm{E}_{t}\left[d\mathbf{W}_{i}^{2}\right]=\left(\hbar/m_{i}\right)dt$,
and analogously for the backward Wiener processes. Note that we take
the $\mathbf{b}_{i}$ $(\mathbf{b}_{i*})$ to be functions of all
the particle positions, $q(t)=\{\mathbf{q}_{1}(t),\mathbf{q}_{2}(t),...,\mathbf{q}_{N}(t)\}$
$\in$ $\mathbb{R}^{3N}$. The reasons for this are: (i) all the particles
are continuously exchanging energy-momentum with a common background
medium (Nelson's ether) and thus are in general physically connected
in their motions through the ether via $\mathbf{b}_{i}$ $(\mathbf{b}_{i*})$,
insofar as the latter are constrained by the physical properties of
the ether; and (ii) the dynamical equations and initial conditions
for the $\mathbf{b}_{i}$ $(\mathbf{b}_{i*})$ are what will determine
the specific situations under which the latter will be effectively
separable functions of the particle positions and when they cannot
be effectively separated. Hence, at this level, it is only sensible
to write $\mathbf{b}_{i}$ $(\mathbf{b}_{i*})$ as functions of all
the particle positions at a single time.

As in the single particle case, in order to incorporate the \emph{zbw}
oscillation as a property of each particle, we must amend Nelson's
original phenomenological hypotheses about his ether and particles
with the \emph{N}-particle generalizations of the new phenomenological
hypotheses we introduced in Part I: 
\begin{enumerate}
\item Nelson's ether is not only a stochastically fluctuating medium in
space-time, but an oscillating medium with a spectrum of angular frequencies
superposed at each point in 3-space. More precisely, we imagine the
ether as a continuous (or effectively continuous) medium composed
of a countably infinite number of fluctuating, stationary, spherical
waves superposed at each point in space, with each wave having a different
fixed angular frequency, $\omega_{0}^{k}$, where $k$ denotes the
\emph{k}-th ether mode. The relative phases between the modes are
taken to be random so that each mode is effectively uncorrelated with
every other mode. 
\item The particles of rest masses $m_{i}$, located at positions $\mathbf{q}_{0i}$
in their respective instantaneous mean forward translational rest
frames (IMFTRFs), i.e., the frames in which $D\mathbf{q}_{i}(t)=\mathbf{b}_{i}(q(t),t)=0$,
are bounded to harmonic oscillator potentials with fixed natural frequencies
$\omega_{0i}=\omega_{ci}=\left(1/\hbar\right)m_{i}c^{2}$. In keeping
with the phenomenological approach of ZSM, and the approach taken
by de Broglie and Bohm with their \emph{zbw} models, we need not specify
the precise physical nature of these harmonic oscillator potentials;
this is task is left for a future physical model of the ZSM particle. 
\item Each particle's center of mass, as a result of being immersed in the
ether, undergoes approximately frictionless translational Brownian
motion (due to the homogeneous and isotropic ether fluctuations that
couple to the particles by possibly electromagnetic, gravitational,
or some other means), as modeled by Eqs. (127) and (128); and, in
their respective IMFTRFs, undergo driven oscillations about $\mathbf{q}_{0i}$
by coupling to a narrow band of ether modes that resonantly peak around
their natural frequencies. However, in order that the oscillation
of each particle doesn't become unbounded in kinetic energy, there
must be some mechanism by which the particles dissipate energy back
into the ether so that, on the average, a steady-state equilibrium
regime is reached for their oscillations. So we posit that on short
relaxation time-scales, $\tau$, which are identical for particles
of identical rest masses, the average energy absorbed from the driven
oscillation by the resonant ether modes equals the average energy
dissipated back to the ether by a given particle. The average, in
the present sense, would be over the random phases of the ether modes.
(Here we are taking inspiration from stochastic electrodynamics \cite{Boyer1975,Boyer1980},
where it has been shown that a classical charged harmonic oscillator
immersed in a classical electromagnetic zero-point field has a steady-state
condition where the phase-averaged power absorbed by the oscillator
balances the phase-averaged power radiated by the oscillator back
to the zero-point field; this yields a steady-state oscillation at
the natural frequency of the oscillator \cite{Boyer1975,Boyer1980,Puthoff1987,Puthoff2016}.
However, in keeping with our phenomenological approach, we do not
propose a specific mechanism for this energy exchange, only provisionally
assume that it occurs somehow.) Thus, in the steady-state regime,
each particle undergoes a steady-state\emph{ zbw} oscillation of angular
frequency $\omega_{ci}$ about its location $\mathbf{q}_{0i}$ in
its IMFTRF, as characterized by the `fluctuation-dissipation' relation,
$<H_{i}>_{steady-state}=\hbar\omega_{ci}=m_{i}c^{2}$, where $<H_{i}>_{steady-state}$
is the conserved random-phase-average energy due to the steady-state
oscillation of the \emph{i}-th particle. Accordingly, if, relative
to the ether, all the particles have zero mean translational motion,
then we will have $\sum_{i}^{N}<H_{i}>_{steady-state}=\sum_{i}^{N}\hbar\omega_{ci}=\sum_{i}^{N}m_{i}c^{2}=const$. 
\end{enumerate}
It follows then that, in the IMFTRF of the \emph{i}-th particle, the
mean forward steady-state \emph{zbw} phase change is given by 
\begin{equation}
\delta\bar{\theta}_{i+}\coloneqq\omega_{ci}\delta t_{0i}=\frac{m_{i}c^{2}}{\hbar}\delta t_{0i},
\end{equation}
and the corresponding cumulative mean forward steady-state phase at
proper time $t_{0i}$ is 
\begin{equation}
\bar{\theta}_{i+}=\omega_{ci}t_{0i}+\phi_{i}=\frac{m_{i}c^{2}}{\hbar}t_{0i}+\phi_{i+}.
\end{equation}
Then the joint cumulative mean forward steady-state phase for all
the particles will just be 
\begin{equation}
\bar{\theta}_{+}=\sum_{i=1}^{N}\bar{\theta}_{i+}=\sum_{i=1}^{N}\left(\omega_{ci}t_{0i}+\phi_{i+}\right)=\sum_{i=1}^{N}\left(\frac{m_{i}c^{2}}{\hbar}t_{0i}+\phi_{i+}\right).
\end{equation}

The reason for starting our analysis with the IMFTRFs goes back to
the fact that, before constraining the diffusion process to simultaneous
solutions of the forward and backward Fokker-Planck equations associated
to (127-128), neither the forward nor the backward stochastic differential
equations (127-128) have well-defined time reversals. So the forward
and backward stochastic differential equations describe independent,
time-asymmetric diffusion processes in opposite time directions, and
we must start by considering the steady-state \emph{zbw} phases in
each time direction separately. So we chose to start with the more
intuitive forward time direction.

For the \emph{i}-th \emph{zbw} particle in its instantaneous mean
backward translational rest frame (IMBTRF), i.e., the frame defined
by $D_{*}\mathbf{q}_{i}(t)=\mathbf{b}_{i*}(q(t),t)=0$, its mean backward
steady-state \emph{zbw} phase change is given by 
\begin{equation}
\delta\bar{\theta}_{i-}\coloneqq-\omega_{ci}\delta t_{0i}=-\frac{m_{i}c^{2}}{\hbar}\delta t_{0i},
\end{equation}
and 
\begin{equation}
\bar{\theta}_{i-}=\left(-\omega_{ci}t_{0i}\right)+\phi_{i-}=\left(-\frac{m_{i}c^{2}}{\hbar}t_{0i}\right)+\phi_{i-}.
\end{equation}
Then the cumulative joint mean backward steady-state phase for all
the particles will just be 
\begin{equation}
\bar{\theta}_{-}=\sum_{i=1}^{N}\bar{\theta}_{i-}=\sum_{i=1}^{N}\left(\omega_{ci}t_{0i}+\phi_{i-}\right)=\sum_{i=1}^{N}\left(\frac{m_{i}c^{2}}{\hbar}t_{0i}+\phi_{i-}\right).
\end{equation}

As in the single particle case, we note that both the diffusion coefficient
$\nu_{i}=\hbar/2m_{i}$ and the (reduced) \emph{zbw} period $T_{ci}=1/\omega_{ci}=\hbar/m_{i}c^{2}$
are scaled by $\hbar$. This is consistent with our hypothesis that
the ether is the common physical cause of both the frictionless diffusion
processes and the steady-state \emph{zbw} oscillations of the particles.
Had we not proposed Nelson's ether as the physical cause of the \emph{zbw}
oscillations as well as the frictionless diffusions, the occurrence
of $\hbar$ in both of these properties of the particles would be
inexplicable and compromising for the plausibility of our proposed
modification of NYSM.

As also in the single particle case, we cannot talk of the \emph{zbw}
phases in rest frames other than the IMFTRFs or IMBTRFs of the particles,
because we cannot transform to a frame in which $d\mathbf{q}_{i}(t)/dt=0$,
as this expression is undefined for the Wiener process.

Now suppose we Lorentz transform back to the lab frame. For the forward
time direction, this corresponds to a boost of (129) by $-\mathbf{b}_{i}(q(t),t)$.
Approximating the transformation for non-relativistic velocities so
that $\gamma=1/\sqrt{\left(1-\mathbf{b}_{i}^{2}/c^{2}\right)}\approx1+\mathbf{b}_{i}^{2}/2c^{2},$
the mean forward steady-state joint phase change becomes 
\begin{equation}
\begin{aligned}\delta\bar{\theta}_{+}(q(t),t) & =\sum_{i=1}^{N}\frac{\omega_{ci}}{m_{i}c^{2}}\mathrm{E}_{t}\left[E_{i+}(D\mathbf{q}_{i}(t))\delta t-m_{i}D\mathbf{q}_{i}(t)\cdot\left(D\mathbf{q}_{i}(t)\right)\delta t\right]\\
 & =\frac{1}{\hbar}\mathrm{E}_{t}\left[\sum_{i=1}^{N}E_{i+}(D\mathbf{q}_{i}(t))\delta t-\sum_{i=1}^{N}m_{i}\mathbf{b}_{i}(q(t),t)\cdot\delta\mathbf{q}_{i+}(t)\right],
\end{aligned}
\end{equation}
where 
\begin{equation}
E_{i+}(D\mathbf{q}_{i}(t))=m_{i}c^{2}+\frac{1}{2}m_{i}\left(D\mathbf{q}_{i}(t)\right)^{2}=m_{i}c^{2}+\frac{1}{2}m_{i}\mathbf{b}_{i}^{2},
\end{equation}
neglecting the momentum terms proportional to $\mathbf{b}_{i}^{3}/c^{2}$.
We emphasize that the $\delta\mathbf{q}_{i+}(t)$ in (135) corresponds
to the physical, translational, mean forward displacement of the \emph{i}-th
\emph{zbw} particle, defined by 
\begin{equation}
\delta\mathbf{q}_{i+}(t)=\left(D\mathbf{q}_{i}(t)\right)\delta t=\mathbf{b}_{i}(q(t),t)\delta t.
\end{equation}
This will be important later.

For the backward time direction, the Lorentz transformation to the
lab frame corresponds to a boost of (132) by $-\mathbf{b}_{i*}(q(t),t)$.
Then the mean backward steady-state joint phase change becomes 
\begin{equation}
\begin{aligned}\delta\bar{\theta}_{-}(q(t),t) & =\sum_{i=1}^{N}\frac{\omega_{ci}}{m_{i}c^{2}}\mathrm{E}_{t}\left[-E_{i-}(D_{*}\mathbf{q}_{i}(t))\delta t+m_{i}D_{*}\mathbf{q}_{i}(t)\cdot\left(D_{*}\mathbf{q}_{i}(t)\right)\delta t\right]\\
 & =\frac{1}{\hbar}\mathrm{E}_{t}\left[-\sum_{i=1}^{N}E_{i-}(D_{*}\mathbf{q}_{i}(t))\delta t+\sum_{i=1}^{N}m_{i}\mathbf{b}_{i*}(q(t),t)\cdot\delta\mathbf{q}_{i-}(t)\right],
\end{aligned}
\end{equation}
where 
\begin{equation}
E_{i-}(D_{*}\mathbf{q}_{i}(t))=m_{i}c^{2}+\frac{1}{2}m_{i}\left(D_{*}\mathbf{q}_{i}(t)\right)^{2}=m_{i}c^{2}+\frac{1}{2}m_{i}\mathbf{b}_{i*}^{2}.
\end{equation}
The $\delta\mathbf{q}_{i-}(t)$ in (138) corresponds to the physical,
translational, mean backward displacement of the \emph{i}-th \emph{zbw}
particle, as defined by 
\begin{equation}
\delta\mathbf{q}_{i-}(t)=\left(D_{*}\mathbf{q}_{i}(t)\right)\delta t=\mathbf{b}_{i*}(q(t),t)\delta t.
\end{equation}
(Notice that $\delta\mathbf{q}_{i+}(t)$ and $\delta\mathbf{q}_{i-}(t)$
are not equal in general since $\delta\mathbf{q}_{i+}(t)-\delta\mathbf{q}_{i-}(t)=(\mathbf{b}_{i}-\mathbf{b}_{i*})\delta t\neq0$
in general.) Now since each \emph{zbw} particle is essentially a harmonic
oscillator, each particle has its own, effectively independent, well-defined
forward steady-state phase at each point along its forward space-time
trajectory, when $\mathbf{b}_{i}(q,t)\approx\sum_{i}^{N}\mathbf{b}_{i}(\mathbf{q}_{i},t)$.
Consistency with this hypothesis also means that when $\mathbf{b}_{i}(q,t)\neq\sum_{i}^{N}\mathbf{b}_{i}(\mathbf{q}_{i},t)$,
the forward steady-state joint phase must be a well-defined function
of the space-time trajectories of\emph{ all} \emph{the particles}
(since we posit that all particles remain harmonic oscillators despite
having their oscillations physically coupled through the common ether
medium they interact with). Furthermore, since, at this stage, the
forward and backward steady-state joint \emph{zbw} phase changes,
(135) and (138), are independent of one another, each must equal $2\pi n$
when integrated along a closed loop $L$ in which both time and position
change. Otherwise, we will contradict our hypothesis that the system
of \emph{zbw} particles has a well-defined steady-state joint phase
in each time direction.

In the lab frame, the forward and backward stochastic differential
equations for the translational motion are again given by (127) and
(128), and the corresponding forward and backward Fokker-Planck equations
take the form 
\begin{equation}
\frac{\partial\rho(q,t)}{\partial t}=-\sum_{i=1}^{N}\nabla_{i}\cdot\left[\mathbf{b}_{i}(q,t)\rho(q,t)\right]+\sum_{i=1}^{N}\frac{\hbar}{2m_{i}}\nabla_{i}^{2}\rho(q,t),
\end{equation}
and 
\begin{equation}
\frac{\partial\rho(q,t)}{\partial t}=-\sum_{i=1}^{N}\nabla_{i}\cdot\left[\mathbf{b}_{i*}(q,t)\rho(q,t)\right]-\sum_{i=1}^{N}\frac{\hbar}{2m_{i}}\nabla_{i}^{2}\rho(q,t).
\end{equation}
Restricting to simultaneous solutions of (137) and (138) entails the
current velocity field 
\begin{equation}
\mathbf{v}_{i}(q,t)\coloneqq\frac{1}{2}\left[\mathbf{b}_{i}(q,t)+\mathbf{b}_{i*}(q,t)\right]=\frac{\nabla_{i}S(q,t)}{m_{i}},
\end{equation}
and the osmotic velocity field 
\begin{equation}
\mathbf{u}_{i}(q,t)\coloneqq\frac{1}{2}\left[\mathbf{b}_{i}(q,t)-\mathbf{b}_{i*}(q,t)\right]=\frac{\hbar}{2m_{i}}\frac{\nabla_{i}\rho(q,t)}{\rho(q,t)}.
\end{equation}
Then (141) and (142) reduce to the continuity equation 
\begin{equation}
\frac{\partial\rho({\normalcolor q},t)}{\partial t}=-\sum_{i=1}^{N}\nabla_{i}\cdot\left[\frac{\nabla_{i}S(q,t)}{m_{i}}\rho(q,t)\right],
\end{equation}
with $\mathbf{b}_{i}=\mathbf{v}_{i}+\mathbf{u}_{i}$ and $\mathbf{b}_{i*}=\mathbf{v}_{i}-\mathbf{u}_{i}$.

As we did for \emph{N}-particle NYSM, we now postulate here the presence
of an external (to the particle) osmotic potential, $U(q,t)$, which
couples to the $i$-th particle as $R(q(t),t)=\mu U(q(t),t)$ (assuming
that the coupling constant $\mu$ is identical for particles of the
same species), and imparts to the $i$-th particle a momentum, $\nabla_{i}R(q,t)|_{\mathbf{q}_{j}=\mathbf{q}_{j}(t)}$.
This momentum then gets counter-balanced by the ether fluid's osmotic
impulse pressure, $\left(\hbar/2m_{i}\right)\nabla_{i}\ln[n(q,t)]|_{\mathbf{q}_{j}=\mathbf{q}_{j}(t)}$,
leading to the equilibrium condition $\nabla_{i}R/m_{i}=\left(\hbar/2m_{i}\right)\nabla_{i}\rho/\rho$
(using $\rho=n/N$), which implies $\rho=e^{2R/\hbar}$ for all times.
As discussed in section 2, it is expected that $R$ generally depends
on the coordinates of all the other particles. The reasons, to remind
the reader, are that: (i) we argued, for reasons of consistency, that
$U$ should be sourced by the ether, and (ii) since the particles
continuously exchange energy-momentum with the ether, the functional
dependence of $U$ will be determined by the dynamical coupling of
the ether to the particles as well as the magnitude of the inter-particle
physical interactions (whether through a classical inter-particle
potential or, in the free particle case, just through the ether).
To make this last point more explicit, suppose two classically non-interacting
\emph{zbw} particles of identical mass, each initially driven in their
oscillations and translational motions by effectively independent
regions of oscillating ether, each region sourcing the osmotic potentials
$U_{1}(\mathbf{q}_{1},t)$ and $U_{2}(\mathbf{q}_{2},t)$, move along
trajectories that cause the spatial support of their dynamically relevant
regions of oscillating ether to significantly overlap; then the particles
will be exchanging energy-momentum with a common region of oscillating
ether modes, leading to an osmotic potential sourced by this common
region of oscillating ether that depends on the motions (hence positions)
of both particles, i.e., $U(\mathbf{q}_{1},\mathbf{q}_{2},t)$. Indeed,
this common region of oscillating ether will drive the subsequent
steady-state \emph{zbw} oscillations and translational Brownian motions
of both particles, leading (after the constraint of conservative diffusions
is imposed, as we will see) to a time-symmetrized steady-state joint
phase $S(\mathbf{q}_{1},\mathbf{q}_{2},t)$ whose gradient with respect
to the \emph{i}-th particle coordinate gives rise to the current velocity
of the \emph{i}-th particle, and to an osmotic counter-balancing of
$\nabla_{i}U(\mathbf{q}_{1},\mathbf{q}_{2},t)$, which gives rise
to the osmotic velocity of the \emph{i}-th particle (as we've already
seen). Mathematically, the non-linear coupling between the osmotic
potential and the evolution of the (conservative-diffusions-constrained)
time-symmetrized joint phase of the \emph{zbw} particles can be seen
by writing the solution to (145), which from section 2 is 
\begin{equation}
\rho(q,t)=\rho_{0}(q_{0})exp[-\int_{0}^{t}\left(\sum_{i}^{N}\nabla_{i}\cdot\mathbf{v}_{i}\right)dt'=\rho_{0}(q_{0})exp[-\int_{0}^{t}\left(\sum_{i}^{N}\frac{\nabla_{i}^{2}S}{m_{i}}\right)dt',
\end{equation}
giving 
\begin{equation}
R(q,t)=R_{0}(q_{0})-(\hbar/2)\int_{0}^{t}\left(\sum_{i}^{N}\frac{\nabla_{i}^{2}S}{m_{i}}\right)dt',
\end{equation}
Then we can infer from (146) that if a narrow bandwidth of common
ether modes is driving the \emph{zbw} oscillations of both particles
(as described in hypothesis 3 above), the evolution of the osmotic
potential (sourced by the common ether modes) will develop functional
dependence on the positions of both particles. The precise form of
this functional dependence and how it evolves in time will depend
on the evolution equation for $S$, which we of course need to specify
(but already know will end up being the $N$-particle quantum HJ equation).

To obtain the 2nd-order time-symmetric dynamics for the mean translational
motions of the \emph{N} particles, we will define the ensemble-averaged
action Eq. (18) in terms of a symmetric combination of the forward
and backward steady-state joint \emph{zbw} phase changes (135) and
(138). This is natural to do since (135) and (138) correspond to the
same frame (the lab frame), and since (135) and (138) are no longer
independent of one another as a result of the constraints (143-144).

First, we take the difference between (135) and (138) to get (replacing
$\delta t\rightarrow dt$ and $\delta\mathbf{q}_{i+,-}(t)\rightarrow d\mathbf{q}_{i+,-}(t)$)
\begin{equation}
\begin{aligned}d\bar{\theta}(q(t),t) & \coloneqq\frac{1}{2}\left[d\bar{\theta}_{+}(q(t),t)-d\bar{\theta}_{-}(q(t),t)\right]\\
 & =\sum_{i=1}^{N}\frac{\omega_{ci}}{m_{i}c^{2}}\mathrm{E}_{t}\left[E_{i}(D\mathbf{q}_{i}(t),D_{*}\mathbf{q}_{i}(t))dt-\frac{m_{i}}{2}\left(\mathbf{b}_{i}(q(t),t)\cdot d\mathbf{q}_{i+}(t)+\mathbf{b}_{i*}(q(t),t)\cdot d\mathbf{q}_{i-}(t)\right)\right]+\phi\\
 & =\frac{1}{\hbar}\mathrm{E}_{t}\left[\sum_{i=1}^{N}E_{i}dt-\sum_{i=1}^{N}\frac{m_{i}}{2}\left(\mathbf{b}_{i}\cdot\frac{d\mathbf{q}_{i+}(t)}{dt}+\mathbf{b}_{i*}\cdot\frac{d\mathbf{q}_{i-}(t)}{dt}\right)dt\right]+\phi\\
 & =\frac{1}{\hbar}\mathrm{E}_{t}\left[\left(\sum_{i=1}^{N}E_{i}-\sum_{i=1}^{N}\frac{m_{i}}{2}\left(\mathbf{b}_{i}\cdot\frac{d\mathbf{q}_{i+}(t)}{dt}+\mathbf{b}_{i*}\cdot\frac{d\mathbf{q}_{i-}(t)}{dt}\right)\right)dt\right]+\phi\\
 & =\frac{1}{\hbar}\mathrm{E}_{t}\left[\left(\sum_{i=1}^{N}E_{i}-\sum_{i=1}^{N}\frac{m_{i}}{2}\left(\mathbf{b}_{i}^{2}+\mathbf{b}_{i*}^{2}\right)\right)dt\right]+\phi\\
 & =\frac{1}{\hbar}\mathrm{E}_{t}\left[\left(\sum_{i=1}^{N}E_{i}-\sum_{i=1}^{N}\left(m_{i}\mathbf{v}_{i}\cdot\mathbf{v}_{i}+m_{i}\mathbf{u}_{i}\cdot\mathbf{u}_{i}\right)\right)dt\right]+\phi\\
 & =\frac{1}{\hbar}\mathrm{E}_{t}\left[\sum_{i=1}^{N}\left(m_{i}c^{2}-\frac{1}{2}m_{i}\mathbf{v}_{i}^{2}-\frac{1}{2}m_{i}\mathbf{u}_{i}^{2}\right)dt\right]+\phi,
\end{aligned}
\end{equation}
where $\phi=\sum_{i=1}^{N}\left(\phi_{i+}-\phi_{i-}\right)$, and
from (136) and (139), we have 
\begin{equation}
E_{i}(D\mathbf{q}_{i}(t),D_{*}\mathbf{q}_{i}(t))\coloneqq m_{i}c^{2}+\frac{1}{2}\left[\frac{1}{2}m_{i}\mathbf{b}_{i}^{2}+\frac{1}{2}m_{i}\mathbf{b}_{i*}^{2}\right]=m_{i}c^{2}+\frac{1}{2}m_{i}\mathbf{v}_{i}^{2}+\frac{1}{2}m_{i}\mathbf{u}_{i}^{2}.
\end{equation}
Equation (148) is the time-symmetrized steady-state joint phase change
of the \emph{zbw} particles in the lab frame, before the constraint
of conservative diffusions is imposed. Note that because $\bar{\theta}_{+}$
and $\bar{\theta}_{-}$ are no longer independent of one another,
it is no longer consistent to have that $\oint_{L}\delta\bar{\theta}_{+}$
and $\oint_{L}\delta\bar{\theta}_{-}$ both equal $2\pi n$. However,
the consistency of our theory does require that $\oint_{L}\delta\bar{\theta}=2\pi n$,
otherwise we will contradict our hypothesis that the system of \emph{N}
\emph{zbw} particles, after imposing (143-144) has a well-defined
and unique steady-state joint phase that functionally depends on the
3-space trajectories of the \emph{zbw} particles.

Now, defining the steady-state joint phase-principal function 
\begin{equation}
I(q(t),t)=-\hbar\bar{\theta}(q(t),t)=\mathrm{E}\left[\int_{t_{I}}^{t}\sum_{i=1}^{N}\left(\frac{1}{2}m_{i}\mathbf{v}_{i}^{2}+\frac{1}{2}m_{i}\mathbf{u}_{i}^{2}-m_{i}c^{2}\right)dt'\left|\mathbf{q}_{j}(t)\right.\right]-\hbar\sum_{i=1}^{N}\left(\phi_{i+}-\phi_{i-}\right),
\end{equation}
we can use (150) to define the steady-state joint phase-action 
\begin{equation}
\begin{aligned}J & =I_{IF}=\mathrm{E}\left[\int_{t_{I}}^{t_{F}}\sum_{i=1}^{N}\left(\frac{1}{2}m_{i}\mathbf{v}_{i}^{2}+\frac{1}{2}m_{i}\mathbf{u}_{i}^{2}-m_{i}c^{2}\right)dt'-\hbar\phi\right].\end{aligned}
\end{equation}
It is straightforward to see that (151) is just Eq. (18) in section
2, with the potentials set equal to zero, and modulo the rest-energy
terms and the time-symmetrized initial joint phase constant $\phi$.

Note, also, that from the second to last line of (148), we can write
the cumulative, time-symmetric, steady-state joint phase at time \emph{t}
as 
\begin{equation}
\begin{aligned}\bar{\theta}(q(t),t) & =\frac{1}{\hbar}\mathrm{E}\left[\int_{t_{I}}^{t}\left(\sum_{i=1}^{N}E_{i}-\sum_{i=1}^{N}\left(m_{i}\mathbf{v}_{i}\cdot\mathbf{v}_{i}+m_{i}\mathbf{u}_{i}\cdot\mathbf{u}_{i}\right)\right)dt'\left|\mathbf{q}_{j}(t)\right.\right]+\phi\\
 & =\frac{1}{\hbar}\mathrm{E}\left[\int_{t_{I}}^{t}\left(\sum_{i=1}^{N}\left(E_{i}-m_{i}\mathbf{u}_{i}\cdot\mathbf{u}_{i}\right)-\sum_{i=1}^{N}m_{i}\mathbf{v}_{i}\cdot\mathbf{v}_{i}\right)dt'\left|\mathbf{q}_{j}(t)\right.\right]+\phi\\
 & =\frac{1}{\hbar}\mathrm{E}\left[\int_{t_{I}}^{t}\left(H-\sum_{i=1}^{N}m_{i}\mathbf{v}_{i}\cdot\mathbf{v}_{i}\right)dt'\left|\mathbf{q}_{j}(t)\right.\right]+\phi\\
 & =\frac{1}{\hbar}\mathrm{E}\left[\int_{t_{I}}^{t}\left(H-\sum_{i=1}^{N}\frac{m_{i}}{4}\left(D\mathbf{q}_{i}(t')+D_{*}\mathbf{q}_{i}(t')\right)\cdot\left(D+D_{*}\right)\mathbf{q}_{i}(t')\right)dt'\left|\mathbf{q}_{j}(t)\right.\right]+\phi\\
 & =\frac{1}{\hbar}\mathrm{E}\left[\int_{t_{I}}^{t}Hdt'-\sum_{i=1}^{N}\frac{m_{i}}{2}\int_{\mathbf{q}_{i}(t_{I})}^{\mathbf{q}_{i}(t)}\left(D\mathbf{q}_{i}(t')+D_{*}\mathbf{q}_{i}(t')\right)\cdot\mathrm{D}\mathbf{q}_{i}(t')\left|\mathbf{q}_{j}(t)\right.\right]+\phi,
\end{aligned}
\end{equation}
where 
\begin{equation}
H\coloneqq\sum_{i=1}^{N}\left(E_{i}-m_{i}\mathbf{u}_{i}\cdot\mathbf{u}_{i}\right)=\sum_{i=1}^{N}\left(m_{i}c^{2}+\frac{1}{2}m_{i}\mathbf{v}_{i}^{2}-\frac{1}{2}m_{i}\mathbf{u}_{i}^{2}\right),
\end{equation}
and where we have used the fact that $0.5\left(D+D_{*}\right)\mathbf{q}_{i}(t)=\left(\partial_{t}+\sum_{j}\mathbf{v}_{j}(q(t),t)\cdot\nabla_{j}\right)\mathbf{q}_{i}(t)$,
and $\mathbf{v}_{i}(q(t),t)=\left(\partial_{t}+\sum_{j}\mathbf{v}_{j}\cdot\nabla_{j}\right)\mathbf{q}_{i}(t)\eqqcolon\mathrm{D}\mathbf{q}_{i}(t)/\mathrm{D}t$,
and $\mathrm{D}\mathbf{q}_{i}(t)=\left(\mathrm{D}\mathbf{q}_{i}(t)/\mathrm{D}t\right)dt$.
Now, consider an integral curve $\mathbf{Q}_{i}(t)$ of the \emph{i}-th
current velocity/momentum field, i.e., a solution of 
\begin{equation}
m_{i}\frac{d\mathbf{Q}_{i}(t)}{dt}=m_{i}\mathbf{v}_{i}(Q(t),t)=\mathbf{p}_{i}(Q(t),t)=\nabla_{i}S(q,t)|_{\mathbf{q}_{j}=\mathbf{Q}_{j}(t)}.
\end{equation}
Then we can replace the functional dependence of (152) on $q(t)$
by $Q(t)$, obtaining 
\begin{equation}
\begin{aligned}\bar{\theta}(Q(t),t) & =\frac{1}{\hbar}\int_{t_{I}}^{t}\left[H-\sum_{i=1}^{N}m_{i}\mathbf{v}_{i}(Q(t'),t')\cdot\frac{d\mathbf{Q}_{i}(t')}{dt'}\right]dt'+\phi\\
 & =\frac{1}{\hbar}\left[\int_{t_{I}}^{t}Hdt'-\sum_{i=1}^{N}\int_{\mathbf{Q}_{i}(t_{I})}^{\mathbf{Q}_{i}(t)}\mathbf{p}_{i}\cdot d\mathbf{Q}_{i}(t')\right]+\phi,
\end{aligned}
\end{equation}
where it should be noticed that we've dropped the conditional expectation.
So (155) denotes the cumulative, time-symmetric, steady-state joint
phase of the \emph{zbw} particles, evaluated along the time-symmetric
mean trajectories of the \emph{zbw} particles, i.e., the integral
curves of (154). That the time-symmetric mean trajectories of the
\emph{zbw} particles should correspond to the integral curves of (154)
can be seen from the fact that the single-time joint probability density
$\rho(q,t)$, after imposing the time-symmetric constraints (143-144),
is a solution of the continuity equation (145), from which it follows
that the possible mean trajectories of the \emph{zbw} particles are
the flow lines of the probability current $\rho\mathbf{v}_{i}$, i.e.,
the solutions of (154) for all possible initial conditions $\mathbf{Q}_{i}(0)$.)

Now, taking the total differential of the left hand side of (155)
gives 
\begin{equation}
d\bar{\theta}=\sum_{i=1}^{N}\nabla_{i}\bar{\theta}|_{\mathbf{q}_{j}=\mathbf{Q}_{j}(t)}d\mathbf{Q}_{i}(t)+\partial_{t}\bar{\theta}|_{\mathbf{q}_{j}=\mathbf{Q}_{j}(t)}dt.
\end{equation}
This allows us to identify 
\begin{equation}
\mathbf{p}_{i}(Q(t),t)=-\hbar\nabla_{i}\bar{\theta}|_{\mathbf{q}_{j}=\mathbf{Q}_{j}(t)}=\nabla_{i}S|_{\mathbf{q}_{j}=\mathbf{Q}_{j}(t)},
\end{equation}
using (156) along with (155) and (143). Thus the \emph{i}-th current
velocity in the lab frame corresponds the gradient of the time-symmetrized
steady-state joint phase of the \emph{zbw} particles at the location
of the \emph{i}-th \emph{zbw} particle, and $S$ can be identified
with the cumulative, time-symmetric, steady-state joint phase function
of the \emph{zbw} particles in the lab frame. In addition, we have
\begin{equation}
H(Q(t),t)=\hbar\partial_{t}\bar{\theta}|_{\mathbf{q}_{j}=\mathbf{Q}_{j}(t)}=-\partial_{t}S|_{\mathbf{q}_{j}=\mathbf{Q}_{j}(t)}.
\end{equation}
From (158), (157), and (155), it follows that 
\begin{equation}
\begin{aligned}S(Q(t),t) & =\sum_{i=1}^{N}\int_{\mathbf{Q}_{i}(t_{I})}^{\mathbf{Q}_{i}(t)}\mathbf{p}_{i}\cdot d\mathbf{Q}_{i}(t')-\int_{t_{I}}^{t}Hdt'-\hbar\phi\\
 & =\int_{t_{I}}^{t}\sum_{i=1}^{N}\left[\frac{1}{2}m_{i}\mathbf{v}_{i}^{2}+\frac{1}{2}m_{i}\mathbf{u}_{i}^{2}-m_{i}c^{2}\right]dt'-\hbar\phi=I(Q(t),t),
\end{aligned}
\end{equation}
and 
\begin{equation}
\oint_{L}\delta S(Q(t),t)=\sum_{i=1}^{N}\oint_{L}\left[\mathbf{p}_{i}(Q(t),t)\cdot\delta\mathbf{Q}_{i}(t)-E_{i}(Q(t),t)\delta t\right]=nh.
\end{equation}
We shall use these last two expressions for later comparisons.

Recall that after restricting the forward and backward diffusions
to simultaneous solutions of (141-142), we have $\mathbf{b}_{i}=\mathbf{v}_{i}+\mathbf{u}_{i}$
and $\mathbf{b}_{i*}=\mathbf{v}_{i}-\mathbf{u}_{i}$. So the IMFTRF
and the IMBTRF will not coincide since, for $\mathbf{b}_{i}=\mathbf{v}_{i}+\mathbf{u}_{i}=0$,
it will generally not be the case that $\mathbf{b}_{i*}=\mathbf{v}_{i}-\mathbf{u}_{i}=0$.
Nevertheless, we can define an instantaneous mean (time-)symmetric
rest frame (IMSTRF) as the frame in which $\mathbf{b}_{i}+\mathbf{b}_{i*}=2\mathbf{v}_{i}=0$.
And the lab frame remains the lab frame.

Applying the conservative diffusion constraint through the extremality
of (151), we obtain the mean acceleration equation 
\begin{equation}
\sum_{i=1}^{N}\frac{m_{i}}{2}\left[D_{*}D+DD_{*}\right]\mathbf{q}_{i}(t)=0.
\end{equation}
Moreover, since the $\delta\mathbf{q}_{i}(t)$ are independent (as
shown in the Appendix), it follows from (161) that we have the individual
equations of motion 
\begin{equation}
m_{i}\mathbf{a}_{i}(q(t),t)=\frac{m_{i}}{2}\left[D_{*}D+DD_{*}\right]\mathbf{q}_{i}(t)=0.
\end{equation}
By applying the mean derivatives in (161), and using that $\mathbf{b}_{i}=\mathbf{v}_{i}+\mathbf{u}_{i}$
and $\mathbf{b}_{i*}=\mathbf{v}_{i}-\mathbf{u}_{i}$, straightforward
manipulations give 
\begin{equation}
\sum_{i=1}^{N}m_{i}\left[\partial_{t}\mathbf{v}_{i}+\mathbf{v}_{i}\cdot\nabla_{i}\mathbf{v}_{i}-\mathbf{u}_{i}\cdot\nabla_{i}\mathbf{u}_{i}-\frac{\hbar}{2m_{i}}\nabla_{i}^{2}\mathbf{u}_{i}\right]|_{\mathbf{q}_{j}=\mathbf{q}_{j}(t)}=0.
\end{equation}
Using (143-144), (163) yields 
\begin{equation}
\begin{aligned}\sum_{i=1}^{N}m_{i}\mathbf{a}_{i}(q(t),t) & =\sum_{i=1}^{N}m_{i}\left[\frac{\partial\mathbf{v}_{i}(q,t)}{\partial t}+\mathbf{v}_{i}(q,t)\cdot\nabla_{i}\mathbf{v}_{i}(q,t)\right.\\
 & \left.-\mathbf{u}_{i}(q,t)\cdot\nabla_{i}\mathbf{u}_{i}(q,t)-\frac{\hbar}{2m_{i}}\nabla_{i}^{2}\mathbf{u}_{i}(q,t)\right]|_{\mathbf{q}_{j}=\mathbf{q}_{j}(t)}\\
 & =\sum_{i=1}^{N}\nabla_{i}\left[\frac{\partial S(q,t)}{\partial t}+\frac{\left(\nabla_{i}S(q,t)\right)^{2}}{2m_{i}}-\frac{\hbar^{2}}{2m_{i}}\frac{\nabla_{i}^{2}\sqrt{\rho(q,t)}}{\sqrt{\rho(q,t)}}\right]|_{\mathbf{q}_{j}=\mathbf{q}_{j}(t)}=0.
\end{aligned}
\end{equation}
Integrating both sides of (164) and setting the arbitrary integration
constants equal to the rest energies, we then have the \emph{N}-particle
quantum Hamilton-Jacobi equation 
\begin{equation}
\begin{aligned}\tilde{E}(q(t),t) & \coloneqq\sum_{i=1}^{N}\tilde{E}_{i}(q(t),t)\\
 & \coloneqq-\partial_{t}S(q(t),t)\\
 & =\sum_{i=1}^{N}m_{i}c^{2}+\sum_{i=1}^{N}\frac{\left(\nabla_{i}S(q,t)\right)^{2}}{2m_{i}}|_{\mathbf{q}_{j}=\mathbf{q}_{j}(t)}-\sum_{i=1}^{N}\frac{\hbar^{2}}{2m_{i}}\frac{\nabla_{i}^{2}\sqrt{\rho(q,t)}}{\sqrt{\rho(q,t)}}|_{\mathbf{q}_{j}=\mathbf{q}_{j}(t)},
\end{aligned}
\end{equation}
describing the total energy of the actual particles along their stochastic
trajectories $q(t)$. Alternatively, given the integral curves $\mathbf{Q}_{i}(t)$
of the reformulated mean acceleration equation 
\begin{equation}
m_{i}\frac{d^{2}\mathbf{Q}_{i}(t)}{dt^{2}}=m_{i}\left(\partial_{t}\mathbf{v}_{i}+\mathbf{v}_{i}\cdot\nabla_{i}\mathbf{v}_{i}\right)|_{\mathbf{q}_{j}=\mathbf{Q}_{j}(t)}=-\nabla_{i}\left(-\frac{\hbar^{2}}{2m_{i}}\frac{\nabla_{i}^{2}\sqrt{\rho(q,t)}}{\sqrt{\rho(q,t)}}\right)|_{\mathbf{q}_{j}=\mathbf{Q}_{j}(t)},
\end{equation}
for $i=1,...,N$, we can replace $q(t)$ by $Q(t)$ and thereby obtain
the total energy $\tilde{E}(Q(t),t)$ of the actual \emph{zbw} particles
along their time-symmetric mean trajectories, the latter now given
by solutions of (166). The corresponding general solution of (165)
is then given by 
\begin{equation}
\begin{aligned}S(Q(t),t) & =\sum_{i=1}^{N}\int_{\mathbf{Q}_{i}(t_{I})}^{\mathbf{Q}_{i}(t)}\mathbf{p}_{i}(Q(t'),t')\cdot d\mathbf{Q}_{i}(t')-\sum_{i=1}^{N}\int_{t_{I}}^{t}\tilde{E}_{i}(Q(t'),t')dt'-\sum_{i=1}^{N}\hbar\phi_{i}\\
 & =\int_{t_{I}}^{t}\sum_{i=1}^{N}\left[\frac{1}{2}m_{i}\mathbf{v}_{i}^{2}-\left(-\frac{\hbar^{2}}{2m_{i}}\frac{\nabla_{i}^{2}\sqrt{\rho}}{\sqrt{\rho}}\right)-m_{i}c^{2}\right]dt'-\sum_{i=1}^{N}\hbar\phi_{i}\\
 & =\int_{t_{I}}^{t}\sum_{i=1}^{N}\left[\frac{1}{2}m_{i}\mathbf{v}_{i}^{2}+\frac{1}{2}m_{i}\mathbf{u}_{i}^{2}+\frac{\hbar}{2}\nabla_{i}\cdot\mathbf{u}_{i}-m_{i}c^{2}\right]dt'-\sum_{i=1}^{N}\hbar\phi_{i}.
\end{aligned}
\end{equation}
We identify (167) as the conservative-diffusion-constrained, time-symmetric,
steady-state joint phase associated with the \emph{zbw} particles
in the lab frame. Notice that the last line of (167) differs from
the last line of (159) only by addition of the terms involving $\nabla_{i}\cdot\mathbf{u}_{i}$.

Notice also that the dynamics for (167) clearly differs from the dynamics
of the joint phase of the free classical \emph{zbw} particles by the
presence of the quantum kinetic in (165-166). As in the single-particle
case, the two phases are formally connected by the `classical limit'
$(\hbar/2m_{i})\rightarrow0$, but this is only formal since such
a limit corresponds to deleting the presence of the ether, thereby
also deleting the physical mechanism that causes the \emph{zbw} particles
to oscillate at their Compton frequencies. The physically realistic
`classical limit' for the phase (167) corresponds to situations where
the quantum kinetic and its gradient are negligible, which will occur
(as in the dBB theory) whenever the center of mass of a system of
interacting particles is sufficiently large and environmental decoherence
is appreciable \cite{Allori2001,Bowm2005,Oriols2016,Derakhshani2017b}.

Since each \emph{zbw} particle is posited to essentially be a harmonic
oscillator of (unspecified) identical type, each particle has its
own, effectively independent, well-defined phase at each point along
its time-symmetric mean space-time trajectory, when $\mathbf{v}_{i}(q,t)\approx\sum_{i}^{N}\mathbf{v}_{i}(\mathbf{q}_{i},t)$.
Consistency with this means that when $\mathbf{v}_{i}(q,t)\neq\sum_{i}^{N}\mathbf{v}_{i}(\mathbf{q}_{i},t)$,
the time-symmetric steady-state joint phase must be a well-defined
function of the time-symmetric mean trajectories of\emph{ all} the
particles (since we posit that all the particles remain harmonic oscillators,
despite having their oscillations physically coupled through the common
ether medium they interact with). Then, for a closed loop \emph{L}
along which each particle can be physically or virtually displaced,
it follows that 
\begin{equation}
\oint_{L}\delta S(Q(t),t)=\sum_{i=1}^{N}\oint_{L}\left[\mathbf{p}_{i}(Q(t),t)\cdot\delta\mathbf{Q}_{i}(t)-\tilde{E_{i}}(Q(t),t)\delta t\right]=nh.
\end{equation}
And for a closed loop $L$ with $\delta t=0$, we have 
\begin{equation}
\sum_{i=1}^{N}\oint_{L}\mathbf{p}_{i}\cdot\delta\mathbf{Q}_{i}(t)=\sum_{i=1}^{N}\oint_{L}\mathbf{\nabla}_{i}S(q,t)|_{\mathbf{q}_{j}=\mathbf{Q}_{j}(t)}\cdot\delta\mathbf{Q}_{i}(t)=nh.
\end{equation}
If we also consider the joint phase field $S(q,t)$, a field over
the possible positions of the \emph{zbw} particles, then, as a result
of the same physical reasoning applied to the \emph{i}-th particle
at any possible initial position it can occupy, we will have 
\begin{equation}
\oint_{L}dS\left(q,t\right)=\sum_{i=1}^{N}\oint_{L}\mathbf{p}_{i}\cdot d\mathbf{q}_{i}=\sum_{i=1}^{N}\oint_{L}\nabla_{i}S(q,t)\cdot d\mathbf{q}_{i}=nh.
\end{equation}
Notice that (170) constrains the osmotic potential as well, due to
the coupling of $S$ to $R$ (hence $U$) via (147). This makes physical
sense since, as we observed earlier, the oscillating ether drives
the \emph{zbw} oscillations of the particles while also sourcing the
osmotic potential that imparts the osmotic velocities to the particles.

Combining (170), (165), and (145), we can construct the \emph{N}-particle
Schrödinger equation 
\begin{equation}
i\hbar\frac{\partial\psi(q,t)}{\partial t}=\sum_{i=1}^{N}\left[-\frac{\hbar^{2}}{2m_{i}}\nabla_{i}^{2}+m_{i}c^{2}\right]\psi(q,t),
\end{equation}
where the \emph{N}-particle wave function $\psi(q,t)=\sqrt{\rho(q,t)}e^{iS(q,t)/\hbar}$
is single-valued by (170).

How does the interpretation of the ZSM wave function differ from that
of the NYSM wave function? The only difference comes from $S(q,t)$
being the conservative-diffusion-constrained, time-symmetrized, steady-state,
joint phase of the \emph{zbw} particles in ZSM, as opposed to being
an \emph{N}-particle velocity potential satisfying a law-like quantization
constraint of the form (170) in NYSM. This difference means that,
in ZSM, $S(q,t)$ reflects not only ontic aspects such as the irrotationality
of the ZSM version of the ether, and the influence of classical fields
on the \emph{zbw} particles, it also reflects the steady-state oscillations
of \emph{zbw} particles immersed in the ether, as well as the (hypothesized)
oscillations of the ether at each point in 3-D space. And it is a
consequence of these last two ontic aspects of an \emph{N}-particle
ZSM system that the quantization condition (170) follows; in other
words, the quantization condition is no longer a law-like constraint
on $S(q,t)$, but a consequence of certain ontic properties of an
\emph{N}-particle ZSM system.

Note that since the solution space of the combination of (170), (165),
and (145) is equivalent to the solution space of (171), any non-factorizable
wave functions that can be constructed as solutions of (171) will
also be solutions (in $\rho$ and $S$ variables) of the combination
of (170), (165), and (145). As an example, let us consider two identical,
classically non-interacting bosons or fermions with initial wave function
\footnote{The Nelsonian derivation of the symmetry postulates given by Bacciagaluppi
in \cite{Bacciagaluppi2003}, which allows us to write down a wave
function like (172) (or its anti-symmetric counterpart), is consistent
with the assumptions of ZSM and carries over without any change.} 
\begin{equation}
\psi_{nf}(\mathbf{q}_{1},\mathbf{q}_{2})\coloneqq Norm_{\pm}\left[\psi_{A}(\mathbf{q}_{1})\psi_{B}(\mathbf{q}_{2})\pm\psi_{A}(\mathbf{q}_{2})\psi_{B}(\mathbf{q}_{1})\right],
\end{equation}
where particle 1 is associated with wavepacket $\psi_{A}$ and particle
2 is associated with packet $\psi_{B}$, and the wavepackets satisfy
$\psi_{A}\cap\psi_{B}\approx\varnothing$. Then, if the packets of
these particles move towards each other and overlap such that $\left(<\mathbf{q}_{1}>-<\mathbf{q}_{2}>\right)^{2}\leq\sigma_{A}^{2}+\sigma_{B}^{2}$,
the subsequent wave function of the 2-particle system will be (172)
but with $\psi_{A}\cap\psi_{B}\neq\varnothing$. Moreover, in terms
of $\rho$ and $S$ variables, we have 
\begin{equation}
\begin{aligned}\rho_{nf}\left(\mathbf{q}_{1},\mathbf{q}_{2}\right) & \coloneqq|\psi_{nf}(\mathbf{q}_{1},\mathbf{q}_{2})|^{2}=Norm_{\pm}^{2}\left\{ e^{2\left(R_{A1}+R_{B2}\right)/\hbar}+e^{2\left(R_{A2}+R_{B1}\right)/\hbar}\right.\\
 & \pm e^{\left[\left(R_{A1}+R_{B2}+R_{A2}+R_{B1}\right)+i\left(S_{A2}+S_{B1}-S_{A1}-S_{B2}\right)\right]/\hbar}\\
 & \left.\pm e^{\left[\left(R_{A1}+R_{B2}+R_{A2}+R_{B1}\right)+i\left(S_{A1}+S_{B2}-S_{A2}-S_{B1}\right)\right]/\hbar}\right\} ,
\end{aligned}
\end{equation}
and 
\begin{equation}
S_{nf}(\mathbf{q}_{1},\mathbf{q}_{2},)\coloneqq-\frac{i\hbar}{2}\ln\left(\frac{\psi_{nf}(\mathbf{q}_{1},\mathbf{q}_{2})}{\psi_{nf}^{\ast}(\mathbf{q}_{1},\mathbf{q}_{2})}\right),
\end{equation}
where (173) satisfies (145), and (174) is a solution of (165) and
satisfies (170). That is, the two particles will be entangled in their
joint phase (174) and their joint osmotic potential obtained from
(172) or (173): 
\begin{equation}
R_{nf}(\mathbf{q}_{1},\mathbf{q}_{2})\coloneqq\hbar\ln\left(|\psi_{nf}(\mathbf{q}_{1},\mathbf{q}_{2})|\right).
\end{equation}
This scenario of entanglement formation between two identical bosons
or fermions is essentially equivalent to the scenario we considered
earlier for our justification of why the osmotic potential should
have functional dependence on the positions of both particles: Eq.
(174) is the conservative-diffusion-constrained, time-symmetric, steady-state
joint phase that develops between the two particles from having their
\emph{zbw} oscillations driven by a common region of oscillating ether
that forms when $\left(<\mathbf{q}_{1}>-<\mathbf{q}_{2}>\right)^{2}\leq\sigma_{A}^{2}+\sigma_{B}^{2}$.
Likewise, (175) is the joint osmotic potential that arises from this
common region of oscillating ether sourcing the osmotic potential.

Additionally, Eqs. (170), (165), and (145) tell us how the non-local
functional dependence of (175) on the positions of the two particles
changes in time: for classically non-interacting particles, the non-local
correlations become negligible when the 3-D spatial separation between
the particles becomes sufficiently large, i.e., when the overlap of
the wavepackets in the summands of (172) becomes negligible. Of course,
the correlations never completely vanish because the overlap of the
wavepackets in the summands of (172) never completely vanishes, implying
that the common region of oscillating ether that physically connects
the steady-state \emph{zbw} oscillations and translational Brownian
motions of the particles must, in some sense, extend over macroscopic
distances in 3-D space. \footnote{More precisely, we have in mind that the regions of oscillating ether
immediately surrounding each particle will directly drive their respective
\emph{zbw} oscillations, while the ether in between the two particles
will nonlocally encode physical correlations between the immediate
regions of ether surrounding each particle, in a way consistent with
the conservative diffusion constraint $J=extremal$, even if the two
particles are macroscopically separated in 3-D space. Of course, the
exact details of how Nelson's ether (under the amendments 1-3) would
accomplish this await the construction of a physical model for it.} That is, if we view the ether as a medium in 3-D space and not in
3N-dimensional configuration space, even though (174-175) are non-separable
fields on configuration space. This last (TELB) view is indeed the
one we take, since, as we stated earlier, we think it's the most conceptually
plausible one among the present options.

To be sure, the interpretive issues we discussed in section 2 for
NYSM apply just as well to ZSM. To review the options, one might view
the mathematical non-factorizability of (174-175) as indicating that
the oscillating ether medium lives in 3N-dimensional configuration
space instead of 3-D space. Or one might view the configuration space
representation (174-175) as a mathematically convenient encoding of
a much more complicated 3-D space representation of the joint phase
field and joint osmotic potential of the particles, making it conceptually
unproblematic to imagine the oscillating ether as a medium in 3-D
space. 

In the former case, we then seem to have the options of: (i) taking
the \emph{zbw} particles to live in 3-D space, and positing a law-like
dynamical relationship between the particles in 3-D space and the
oscillating ether in 3N-dimensional configuration space; and (ii)
taking the particles in 3-D space to be `functionally emergent' from
a single real \emph{zbw} particle living at a point in 3N-dimensional
configuration space and interacting with an oscillating ether that
also lives in 3N-space. 

In the latter case, since both the particles and the oscillating ether
would live in 3-D space (the TELB view), their physical interactions
would occur there as well.

As with NYSM, the drawback of option (i) in the former case is that
it seems puzzling why two sets of beables, living in completely independent
ontic spaces, should have a law-like dynamical relationship between
them (i.e., why should oscillations of an ether medium in a 3N-dimensional
configuration space `drive' the steady-state \emph{zbw} oscillations
of particles at definite positions in a 3-D space?). The drawback
of option (ii) is that while it's conceptually more transparent how
oscillations of the ether could drive the steady-state oscillations
of a \emph{zbw} particle, if both the ether and the \emph{zbw} particle
live in the same ontic space, examination of the details of so-called
`functional emergence' reveals this option to be fundamentally no
different from option (i); not to mention the inconsistency of the
claim that 3N-dimensional configuration space is more fundamental
than 3-D space, with the fact that we \emph{derived} the \emph{N}-particle
QHJ equation from an ensemble-averaged action defined from a sum of
\emph{N} contributions, under the starting hypothesis that there really
are \emph{N} particles conservatively diffusing through an ether in
a 3-D space. 

Of course, the main shortcoming of the TELB view is that it remains
speculative at the moment, since no such formulation of NYSM or ZSM
exists at present; but it is not implausible that such a formulation
can be constructed, and we have already sketched in section 2 one
way it could be done. Thus we assume, provisionally, that a TELB formulation
of ZSM exists and awaits discovery (unless shown otherwise), and base
our interpretation of the beables of ZSM on this provisional assumption.

It is interesting to observe that the existence of entangled solutions
such as (174-175) is a consequence of four physical constraints we've
used in our construction of ZSM: (i) time-reversal invariance of the
probability density via (145); (ii) the conservative diffusion constraint
on the ensemble-averaged \emph{N}-particle action (151); (iii) single-valuedness
of the conservative-diffusion-constrained, time-symmetrized, joint
phase field (up to an integer multiple of $2\pi$) via (170); and
(iv) the requirement that the particles, under the time-evolution
constraints (143-170), satisfy a natural notion of identicality under
exchange of their coordinates, thereby yielding the symmetrization
postulates associated with bosons and fermions \cite{Bacciagaluppi2003}
(though let us be clear that for classically interacting non-identical
particles, entangled solutions can also arise by virtue of the previous
three physical constraints). So ZSM offers a novel way to understand
the emergence of continuous-variable entanglement nonlocality from
deeper, `subquantum' physical constraints. One could then study how
relaxing these physical constraints might lead to experimentally testable
differences from the entangled solutions of the \emph{N}-particle
Schrödinger equation, in experimental tests of Bell inequalities for
continuous-variable correlations \cite{Cavalcanti2007}.

Now, since we wish to view the particles as living at definite points
in 3-D space, and their \emph{zbw} oscillations as occuring in 3-D
space, we should find a way of constructing the phase field associated
with the \emph{i}-th particle's \emph{zbw} oscillation in 3-D space.
To do this, we can construct the conditional phase field and conditional
osmotic potential field for the \emph{i}-th particle from the solutions
of (165) and (145) using (170). For generality and to avoid redundancy,
we will give these constructions for the case of classically-interacting
\emph{zbw} particles in the next section.

\subsection{Classical fields interacting with \emph{zbw} particles}

For completeness, we will describe \emph{zbw} particles interacting
with each other through a scalar (Coulomb) potential and with external
vector and scalar potentials. For simplicity, we will restrict our
attention to only two \emph{zbw} particles.

We begin by supposing again that each particle undergoes a steady-state
\emph{zbw} oscillation in its IMFTRF, and that each \emph{zbw} particle
carries charge, $e_{i}$, making them classical charged harmonic oscillators
of some identical type. \footnote{Which we subject again to the hypothetical constraint of no electromagnetic
radiation emitted when there is no translational motion; or the constraint
that the oscillation of the charge is radially symmetric so that there
is no net energy radiated; or, if the ether turns out to be electromagnetic
in nature as Nelson suggested \cite{Nelson1985}, then that the steady-state
\emph{zbw} oscillations of the particles are due to a balancing between
the random-phase-averaged electromagnetic energy absorbed via the
driven oscillations of the particle charges, and the random-phase-averaged
electromagnetic energy radiated back to the ether by the particles.} So the classical interaction between the particles is described by
the interaction potential $\Phi_{c}^{int}(\mathbf{q}_{i}(t),\mathbf{q}_{j}(t))=\frac{1}{2}\sum_{j=1}^{2(j\neq i)}\frac{e_{j}}{|\mathbf{q}_{i}(t)-\mathbf{q}_{j}(t)|}$,
under the point-like interaction assumption, $|\mathbf{q}_{1}(t)-\mathbf{q}_{2}(t)|\gg\lambda_{c}$.
In addition, we allow coupling to an external electric potential $\Phi_{i}^{ext}(\mathbf{q}_{i}(t),t)$
(again making the point-like approximation $|\mathbf{q}_{i}|\gg\lambda_{c}$).
Then the mean forward, steady-state, joint phase change of the particles
in the lab frame is given by 
\begin{equation}
\begin{aligned}\delta\bar{\theta}_{+}(\mathbf{q}_{1}(t),\mathbf{q}_{2}(t),t) & =\mathrm{E_{t}}\left[\sum_{i=1}^{2}\left(\omega_{ic}+\omega_{ci}\frac{\mathbf{b}_{i}^{2}}{2c^{2}}+\omega_{ci}\left(\frac{e_{i}\Phi_{i}^{ext}}{m_{i}c^{2}}+\frac{e_{i}\Phi_{c}^{int}}{m_{i}c^{2}}\right)\right)\left(\delta t-\sum_{i=1}^{2}\frac{\mathbf{b}_{0i}}{c^{2}}\cdot\delta\mathbf{q}_{i+}(t)\right)\right]\\
 & =\mathrm{E}_{t}\left[\sum_{i=1}^{2}\left(\omega_{ic}+\omega_{ci}\frac{\mathbf{b}_{i}^{2}}{2c^{2}}+\omega_{ci}\left(\frac{e_{i}\Phi_{i}^{ext}}{m_{i}c^{2}}+\frac{e_{i}\Phi_{c}^{int}}{m_{i}c^{2}}\right)\right)\delta t-\sum_{i=1}^{2}\omega_{ci}\left(\frac{\mathbf{b}_{i}}{c^{2}}\right)\cdot\delta\mathbf{q}_{i+}(t)\right]\\
 & =\frac{1}{\hbar}\mathrm{E}_{t}\left[\left(\sum_{i=1}^{2}m_{i}c^{2}+\sum_{i=1}^{2}\frac{m_{i}\mathbf{b}_{i}^{2}}{2}+\sum_{i=1}^{2}V_{i}^{ext}+V_{c}^{int}\right)\delta t-\sum_{i=1}^{2}m_{i}\mathbf{b}_{i}\cdot\delta\mathbf{q}_{i+}(t)\right].
\end{aligned}
\end{equation}
The mean backward joint phase change $\delta\bar{\theta}_{-}$ differs
by $\mathbf{b}_{i}\rightarrow-\mathbf{b}_{i*}$, $\delta t\rightarrow-\delta t$,
and $\delta\mathbf{q}_{i+}(t)\rightarrow\delta\mathbf{q}_{i-}(t)$.
Incorporating coupling to an external vector potential, we then have
$\mathbf{b}_{i}=\mathbf{b}'_{i}-(e_{i}/m_{i}c)\mathbf{A}_{i}^{ext}$
and $\mathbf{b}_{i*}=\mathbf{b}'_{i*}-(e_{i}/m_{i}c)\mathbf{A}_{i}^{ext}$.
When $|\mathbf{q}_{1}(t)-\mathbf{q}_{2}(t)|$ becomes sufficiently
great that $V_{c}^{int}$ is negligible, (176) reduces to an effectively
separable sum of the forward steady-state phase changes associated
with particle 1 and particle 2, respectively. (Effectively, because
the ether will of course still physically connect the phase changes
of the particles, even if negligibly.) We can then write 
\begin{equation}
\delta\bar{\theta}_{+}(q(t),t)=\frac{1}{\hbar}\mathrm{E}_{t}\left[E_{joint+}(q(t),Dq(t),t)\delta t-\sum_{i=1}^{N}m_{i}\mathbf{b}'_{i}(q(t),t)\cdot\delta\mathbf{q}_{i+}(t)\right],
\end{equation}
where 
\begin{equation}
E_{joint+}=\sum_{i=1}^{2}m_{i}c^{2}+\sum_{i=1}^{2}\frac{m_{i}\mathbf{b}_{i}^{2}}{2}+\sum_{i=1}^{2}V_{i}^{ext}+V_{c}^{int}.
\end{equation}
Correspondingly, 
\begin{equation}
\delta\bar{\theta}_{-}(q(t),t)=\frac{1}{\hbar}\mathrm{E}_{t}\left[-E_{joint-}(q(t),D_{*}q(t),t)\delta t+\sum_{i=1}^{2}m_{i}\mathbf{b}'_{i*}(q(t),t)\cdot\delta\mathbf{q}_{i-}(t)\right],
\end{equation}
where 
\begin{equation}
E_{joint-}=\sum_{i=1}^{2}m_{i}c^{2}+\sum_{i=1}^{2}\frac{m_{i}\mathbf{b}_{i*}^{2}}{2}+\sum_{i=1}^{2}V_{i}^{ext}+V_{c}^{int}.
\end{equation}

As in the classical case, we can readily construct from (177) or (179)
the corresponding mean forward or backward conditional phase change
for particle 1 (particle 2), in the lab frame or IMFTRF/IMBTRF of
particle 1 (particle 2). Likewise for the backward conditional phase
change for particle 1 (particle 2).

Because each \emph{zbw} particle is essentially a harmonic oscillator,
when $V_{c}^{int}\approx0$, each particle has its own well-defined
forward/backward steady-state phase at each point along its mean forward/backward
space-time trajectory. Consistency with this fact entails that for
$V_{c}^{int}>0$, the forward/backward steady-state joint phase must
be a well-defined function of the mean forward/backward space-time
trajectories of\emph{ }both particles (since we again posit that both
particles remain harmonic oscillators even when physically coupled
by $V_{c}^{int}$). Furthermore, we note that at this stage (177)
and (179) are independent of one another. Accordingly, for a closed
loop \emph{L} along which each particle can be physically or virtually
displaced, the forward steady-state joint phase in the lab frame will
satisfy 
\begin{equation}
\oint_{L}\delta\bar{\theta}_{+}=2\pi n,
\end{equation}
and likewise for the steady-state backward joint phase. It also follows
from (181) that 
\begin{equation}
\oint_{L}\delta_{1}\bar{\theta}_{+}=2\pi n,
\end{equation}
where the closed-loop integral here keeps the coordinate of particle
2 fixed while particle 1 is displaced along \emph{L}.

In the lab frame, the forward and backward stochastic differential
equations for the translational motion are then given by 
\begin{equation}
d\mathbf{q}_{i}(t)=\left(\mathbf{b}_{i}'(q(t),t)-\frac{e_{i}}{m_{i}c}\mathbf{A}_{i}^{ext}(q(t),t)\right)dt+d\mathbf{W}_{i}(t),
\end{equation}
and 
\begin{equation}
d\mathbf{q}_{i}(t)=\left(\mathbf{b}'_{i*}(q(t),t)-\frac{e_{i}}{m_{i}c}\mathbf{A}_{i}^{ext}(q(t),t)\right)dt+d\mathbf{W}_{i*}(t),
\end{equation}
with corresponding Fokker-Planck equations 
\begin{equation}
\frac{\partial\rho(q,t)}{\partial t}=-\sum_{i=1}^{2}\nabla_{i}\cdot\left[\left(\mathbf{b}_{i}'(q,t)-\frac{e_{i}}{m_{i}c}\mathbf{A}_{i}^{ext}(q,t)\right)\rho(q,t)\right]+\sum_{i=1}^{2}\frac{\hbar}{2m_{i}}\nabla_{i}^{2}\rho(q,t),
\end{equation}
and 
\begin{equation}
\frac{\partial\rho(q,t)}{\partial t}=-\sum_{i=1}^{2}\nabla_{i}\cdot\left[\left(\mathbf{b}'_{i*}(q,t)-\frac{e_{i}}{m_{i}c}\mathbf{A}_{i}^{ext}(q,t)\right)\rho(q,t)\right]-\sum_{i=1}^{2}\frac{\hbar}{2m_{i}}\nabla_{i}^{2}\rho(q,t).
\end{equation}
Restricting to simultaneous solutions of (185-186) leads us to the
modified current velocity 
\begin{equation}
\mathbf{v}_{i}\coloneqq\frac{1}{2}\left[\mathbf{b}_{i}+\mathbf{b}_{i*}\right]=\frac{\nabla_{i}S}{m_{i}}-\frac{e_{i}}{m_{i}c}\mathbf{A}_{i}^{ext},
\end{equation}
and the usual osmotic velocity 
\begin{equation}
\mathbf{u}_{i}\coloneqq\frac{1}{2}\left[\mathbf{b}_{i}-\mathbf{b}_{i*}\right]=\frac{\hbar}{2m_{i}}\frac{\nabla_{i}\rho}{\rho}.
\end{equation}
Then (185) and (186) reduce to 
\begin{equation}
\frac{\partial\rho}{\partial t}=-\sum_{i=1}^{2}\nabla_{i}\cdot\left[\left(\frac{\nabla_{i}S}{m_{i}}-\frac{e_{i}}{m_{i}c}\mathbf{A}_{i}^{ext}\right)\rho\right],
\end{equation}
where $\mathbf{b}_{i}'=\mathbf{v}_{i}'+\mathbf{u}_{i}$ and $\mathbf{b}'_{i*}=\mathbf{v}_{i}'-\mathbf{u}_{i}$
since $\mathbf{v}_{i}'=\mathbf{v}_{i}+(e_{i}/m_{i}c)\mathbf{A}_{i}^{ext}$,
and $\mathbf{b}_{i}=\mathbf{b}_{i}'-(e_{i}/m_{i}c)\mathbf{A}_{i}^{ext}$,
and $\mathbf{b}_{i*}=\mathbf{b}'_{i*}-(e_{i}/m_{i}c)\mathbf{A}_{i}^{ext}$.
The solution of (189) is just 
\begin{equation}
\rho(q,t)=\rho_{0}(q_{0})exp[-\int_{0}^{t}\left[\sum_{i}^{2}\nabla_{i}\cdot\mathbf{v}_{i}\right]dt'=\rho_{0}(q_{0})exp[-\int_{0}^{t}\left[\sum_{i}^{2}\left(\frac{\nabla_{i}^{2}S}{m_{i}}-\frac{e_{i}}{m_{i}c}\nabla_{i}\cdot\mathbf{A}_{i}^{ext}\right)\right]dt'.
\end{equation}
Here again we postulate an osmotic potential to which each particle
couples via $R(q(t),t)=\mu U(q(t),t)$, which imparts momentum $\nabla_{i}R(q,t)|_{\mathbf{q}_{j}=\mathbf{q}_{j}(t)}$
that is counter-balanced by the osmotic impulse $\left(\hbar/2m_{i}\right)\nabla_{i}\ln[n(q,t)]|_{\mathbf{q}_{j}=\mathbf{q}_{j}(t)}$,
giving the equilibrium condition $\nabla_{i}R/m_{i}=\left(\hbar/2m_{i}\right)\nabla_{i}\rho/\rho$.
Thus $\rho=e^{2R/\hbar}$ for all times and 
\begin{equation}
R(q,t)=R_{0}(q_{0})-(\hbar/2)\int_{0}^{t}\left[\sum_{i}^{2}\left(\frac{\nabla_{i}^{2}S}{m_{i}}-\frac{e_{i}}{m_{i}c}\nabla_{i}\cdot\mathbf{A}_{i}^{ext}\right)\right]dt',
\end{equation}
where $S$ will end up playing the role of the conservative-diffusion-constrained,
time-symmetrized, steady-state joint phase of the \emph{zbw} particles.

As in the free particle case, we will obtain the 2nd-order time-symmetric
mean dynamics for the \emph{zbw} particles from Yasue's variational
principle.

Since (177) and (179) correspond to the same (lab) frame and are no
longer independent because of (187-188), it is natural to define the
time-symmetrized steady-state joint \emph{zbw} particle phase in the
lab frame by taking the difference between (177) and (179) (under
the replacements $\mathbf{b}_{i}\rightarrow\mathbf{b}'_{i}$ and $\mathbf{b}_{i*}\rightarrow\mathbf{b}'_{i*}$
in the mean forward and mean backward momentum contributions to the
phases): 
\begin{equation}
\begin{aligned}d\bar{\theta}(q(t),t) & \coloneqq\frac{1}{2}\left[d\bar{\theta}_{+}(q(t),t)-d\bar{\theta}_{-}(q(t),t)\right]\\
 & =\frac{1}{\hbar}\mathrm{E}_{t}\left[\sum_{i=1}^{2}\left(E_{i}(q(t),D\mathbf{q}_{i}(t),D_{*}\mathbf{q}_{i}(t),t)dt-\frac{m_{i}}{2}\left(\mathbf{b}'_{i}\cdot d\mathbf{q}_{i+}(t)+\mathbf{b}'_{i*}\cdot d\mathbf{q}_{i-}(t)\right)\right)+\phi\right]\\
 & =\frac{1}{\hbar}\mathrm{E}_{t}\left[\sum_{i=1}^{2}E_{i}dt-\sum_{i=1}^{2}\frac{m_{i}}{2}\left(\mathbf{b}'_{i}\cdot\frac{d\mathbf{q}_{i+}(t)}{dt}+\mathbf{b}'_{i*}\cdot\frac{d\mathbf{q}_{i-}(t)}{dt}\right)dt\right]+\phi\\
 & =\frac{1}{\hbar}\mathrm{E}_{t}\left[\left(E_{joint}-\sum_{i=1}^{2}\frac{m_{i}}{2}\left(\mathbf{b}'_{i}\cdot\frac{d\mathbf{q}_{i+}(t)}{dt}+\mathbf{b}'_{i*}\cdot\frac{d\mathbf{q}_{i-}(t)}{dt}\right)\right)dt\right]+\phi\\
 & =\frac{1}{\hbar}\mathrm{E}_{t}\left[\left(E_{joint}-\sum_{i=1}^{2}\frac{m_{i}}{2}\left(\mathbf{b}'_{i}\cdot\mathbf{b}_{i}+\mathbf{b}'_{i*}\cdot\mathbf{b}_{i*}\right)\right)dt\right]+\phi\\
 & =\frac{1}{\hbar}\mathrm{E}_{t}\left[\left(E_{joint}-\sum_{i=1}^{2}\frac{m_{i}}{2}\left(\mathbf{b}{}_{i}^{2}+\frac{e_{i}}{m_{i}c}\mathbf{b}_{i}\cdot\mathbf{A}_{i}^{ext}+\mathbf{b}_{i*}^{2}+\frac{e_{i}}{m_{i}c}\mathbf{b}_{i*}\cdot\mathbf{A}_{i}^{ext}\right)\right)dt\right]+\phi\\
 & =\frac{1}{\hbar}\mathrm{E}_{t}\left[\left(E_{joint}-\sum_{i=1}^{2}\frac{m_{i}}{2}\left(\mathbf{b}_{i}{}^{2}+\mathbf{b}_{i*}^{2}\right)-\sum_{i=1}^{2}\frac{e_{i}}{c}\left(\frac{\mathbf{b}_{i}+\mathbf{b}_{i*}}{2}\right)\cdot\mathbf{A}_{i}^{ext}\right)dt\right]+\phi\\
 & =\frac{1}{\hbar}\mathrm{E}_{t}\left[\left(E_{joint}-\sum_{i=1}^{2}\left(m_{i}\mathbf{v}_{i}\cdot\mathbf{v}_{i}+m_{i}\mathbf{u}_{i}\cdot\mathbf{u}_{i}\right)-\sum_{i=1}^{2}\frac{e_{i}}{c}\mathbf{v}_{i}\cdot\mathbf{A}_{i}^{ext}\right)dt\right]+\phi\\
 & =\frac{1}{\hbar}\mathrm{E}_{t}\left[\left(\sum_{i=1}^{2}\left(m_{i}c^{2}-\frac{1}{2}m_{i}\mathbf{v}_{i}^{2}-\frac{1}{2}m_{i}\mathbf{u}_{i}^{2}-\frac{e_{i}}{c}\mathbf{v}_{i}\cdot\mathbf{A}_{i}^{ext}\right)+\sum_{i=1}^{2}V_{i}^{ext}+V_{c}^{int}\right)dt\right]+\phi.
\end{aligned}
\end{equation}
where $\phi=\sum_{i=1}^{2}\left(\phi_{i+}-\phi_{i-}\right)$ and,
using (178) and (180), along with the constraints (187) and (188),
we have 
\begin{equation}
\begin{aligned}E_{joint} & \coloneqq\sum_{i=1}^{2}E_{i}\\
 & \coloneqq\frac{1}{2}\left[E_{joint+}+E_{joint-}\right]\\
 & =\sum_{i=1}^{2}m_{i}c^{2}+\sum_{i=1}^{2}\frac{1}{2}\left[\frac{1}{2}m_{i}\mathbf{b}_{i}{}^{2}+\frac{1}{2}m\mathbf{b}_{i*}{}^{2}\right]+\sum_{i=1}^{2}V_{i}^{ext}+V_{c}^{int}\\
 & =\sum_{i=1}^{2}m_{i}c^{2}+\sum_{i=1}^{2}\left[\frac{1}{2}m_{i}\mathbf{v}{}_{i}{}^{2}+\frac{1}{2}m_{i}\mathbf{u}_{i}^{2}\right]+\sum_{i=1}^{2}V_{i}^{ext}+V_{c}^{int}.
\end{aligned}
\end{equation}
As in the free particle case, the consistency of our theory requires
that (192) satisfies 
\begin{equation}
\oint_{L}\delta\bar{\theta}=2\pi n.
\end{equation}
Otherwise we would contradict our hypothesis that, after imposing
(187-188), the \emph{zbw} particles have a well-defined, unique, steady-state
joint phase at the 3-space locations that they can occupy at a time
\emph{t}.

Defining the steady-state joint phase-principal function 
\begin{equation}
I=-\hbar\bar{\theta}=\mathrm{E}\left[\int_{t_{I}}^{t}\left(\sum_{i=1}^{2}\left[\frac{1}{2}m_{i}\mathbf{v}_{i}^{2}+\frac{1}{2}m_{i}\mathbf{u}_{i}^{2}+\frac{e_{i}}{c}\mathbf{v}_{i}\cdot\mathbf{A}_{i}^{ext}-m_{i}c^{2}-V_{i}^{ext}\right]-V_{c}^{int}\right)dt'\left|\mathbf{q}_{j}(t)\right.\right]-\hbar\phi,
\end{equation}
allows us to define the joint phase-action 
\begin{equation}
\begin{aligned}J & =I_{IF}=\mathrm{E}\left[\int_{t_{I}}^{t_{F}}\left(\sum_{i=1}^{2}\left[\frac{1}{2}m_{i}\mathbf{v}_{i}^{2}+\frac{1}{2}m_{i}\mathbf{u}_{i}^{2}+\frac{e_{i}}{c}\mathbf{v}_{i}\cdot\mathbf{A}_{i}^{ext}-m_{i}c^{2}-V_{i}^{ext}\right]-V_{c}^{int}\right)dt'\right]-\hbar\phi.\end{aligned}
\end{equation}
Equation (196) is just Eq. (18) in section 2, with the addition of
the rest-energy terms and the time-symmetrized initial joint phase
constant $\phi$.

From the second to last line of (192), we can use Fubini's theorem
in stochastic calculus to write the cumulative, time-symmetric, steady-state
joint phase at time \emph{t} as 
\begin{equation}
\begin{aligned}\bar{\theta}(q(t),t) & =\frac{1}{\hbar}\mathrm{E}\left[\int_{t_{I}}^{t}\left(E_{joint}-\sum_{i=1}^{2}\left(m_{i}\mathbf{v}_{i}\cdot\mathbf{v}_{i}+m_{i}\mathbf{u}_{i}\cdot\mathbf{u}_{i}\right)-\sum_{i=1}^{2}\frac{e_{i}}{c}\mathbf{v}_{i}\cdot\mathbf{A}_{i}^{ext}\right)dt'\left|\mathbf{q}_{j}(t)\right.\right]+\phi\\
 & =\frac{1}{\hbar}\mathrm{E}\left[\int_{t_{I}}^{t}\left(\left(E_{joint}-\sum_{i=1}^{2}m_{i}\mathbf{u}_{i}\cdot\mathbf{u}_{i}\right)-\sum_{i=1}^{2}m_{i}\mathbf{v}_{i}\cdot\mathbf{v}_{i}-\sum_{i=1}^{2}\frac{e_{i}}{c}\mathbf{v}_{i}\cdot\mathbf{A}_{i}^{ext}\right)dt'\left|\mathbf{q}_{j}(t)\right.\right]+\phi\\
 & =\frac{1}{\hbar}\mathrm{E}\left[\int_{t_{I}}^{t}\left(H-\sum_{i=1}^{2}m_{i}\mathbf{v}_{i}\cdot\mathbf{v}_{i}-\sum_{i=1}^{2}\frac{e_{i}}{c}\mathbf{v}_{i}\cdot\mathbf{A}_{i}^{ext}\right)dt'\left|\mathbf{q}_{j}(t)\right.\right]+\phi\\
 & =\frac{1}{\hbar}\mathrm{E}\left[\int_{t_{I}}^{t}\left(H-\sum_{i=1}^{2}\left(\mathbf{p}_{i}+\frac{e_{i}}{c}\mathbf{A}_{i}^{ext}\right)\cdot\mathbf{v}_{i}\right)dt'\left|\mathbf{q}_{j}(t)\right.\right]+\phi\\
 & =\frac{1}{\hbar}\mathrm{E}\left[\int_{t_{I}}^{t}Hdt'-\sum_{i=1}^{2}\int_{\mathbf{q}_{i}(t_{I})}^{\mathbf{q}_{i}(t)}\left(\mathbf{p}_{i}+\frac{e_{i}}{c}\mathbf{A}_{i}^{ext}\right)\cdot\mathrm{D}\mathbf{q}_{i}(t')\left|\mathbf{q}_{j}(t)\right.\right]+\phi,
\end{aligned}
\end{equation}
where 
\begin{equation}
H\coloneqq E_{joint}-\sum_{i=1}^{2}m_{i}\mathbf{u}_{i}\cdot\mathbf{u}_{i}=\sum_{i=1}^{2}m_{i}c^{2}+\sum_{i=1}^{2}\left[\frac{1}{2}m_{i}\mathbf{v}{}_{i}{}^{2}-\frac{1}{2}m_{i}\mathbf{u}_{i}^{2}\right]+\sum_{i=1}^{2}V_{i}^{ext}+V_{c}^{int}.
\end{equation}
Now, consider an integral curve $\mathbf{Q}_{i}(t)$ obtained from
\begin{equation}
m_{i}\frac{d\mathbf{Q}_{i}(t)}{dt}=m_{i}\mathbf{v}_{i}(Q(t),t)=\mathbf{p}_{i}(Q(t),t)=\nabla_{i}S(q,t)|_{\mathbf{q}_{j}=\mathbf{Q}_{j}(t)}.
\end{equation}
Then we can replace the functional dependence of (197) on $q(t)$
by $Q(t)$, obtaining 
\begin{equation}
\begin{aligned}\bar{\theta}(Q(t),t) & =\frac{1}{\hbar}\int_{t_{I}}^{t}\left[H-\sum_{i=1}^{2}\left(m_{i}\mathbf{v}_{i}+\frac{e_{i}}{c}\mathbf{A}_{i}^{ext}\right)\cdot\frac{d\mathbf{Q}_{i}(t')}{dt'}\right]dt'+\phi\\
 & =\frac{1}{\hbar}\left[\int_{t_{I}}^{t}Hdt'-\sum_{i=1}^{2}\int_{\mathbf{Q}_{i}(t_{I})}^{\mathbf{Q}_{i}(t)}\left(\mathbf{p}_{i}+\frac{e_{i}}{c}\mathbf{A}_{i}^{ext}\right)\cdot d\mathbf{Q}_{i}(t')\right]+\phi,
\end{aligned}
\end{equation}
where we've dropped the conditional expectation.

The total differential of the left hand side of (200) gives 
\begin{equation}
d\bar{\theta}=\sum_{i=1}^{2}\nabla_{i}\bar{\theta}|_{\mathbf{q}_{j}=\mathbf{Q}_{j}(t)}d\mathbf{Q}_{i}(t)+\partial_{t}\bar{\theta}|_{\mathbf{q}_{j}=\mathbf{Q}_{j}(t)}dt,
\end{equation}
hence, 
\begin{equation}
\mathbf{p}_{i}(Q(t),t)+\frac{e_{i}}{c}\mathbf{A}_{i}^{ext}(\mathbf{Q}_{i}(t),t)=-\hbar\nabla_{i}\bar{\theta}|_{\mathbf{q}_{j}=\mathbf{Q}_{j}(t)}=\nabla_{i}S|_{\mathbf{q}_{j}=\mathbf{Q}_{j}(t)}.
\end{equation}
Thus the \emph{i}-th current velocity in the lab frame, plus the correction
due to the \emph{i}-th external vector potential, corresponds the
gradient of the time-symmetrized steady-state joint phase at the location
of the \emph{i}-th \emph{zbw} particle, and $S$ can again be identified
with the cumulative, time-symmetric, steady-state joint phase function
of the \emph{zbw} particles in the lab frame. Along with 
\begin{equation}
H(Q(t),t)=\hbar\partial_{t}\bar{\theta}|_{\mathbf{q}_{j}=\mathbf{Q}_{j}(t)}=-\partial_{t}S|_{\mathbf{q}_{j}=\mathbf{Q}_{j}(t)},
\end{equation}
it follows that 
\begin{equation}
\begin{aligned}S(Q(t),t) & =\sum_{i=1}^{2}\int_{\mathbf{Q}_{i}(t_{I})}^{\mathbf{Q}_{i}(t)}\left(\mathbf{p}_{i}+\frac{e_{i}}{c}\mathbf{A}_{i}^{ext}\right)\cdot d\mathbf{Q}_{i}(t')-\int_{t_{I}}^{t}Hdt'-\hbar\phi\\
 & =\int_{t_{I}}^{t}\left\{ \sum_{i=1}^{2}\left[\frac{1}{2}m_{i}\mathbf{v}_{i}^{2}+\frac{1}{2}m_{i}\mathbf{u}_{i}^{2}+\frac{e_{i}}{c}\mathbf{v}_{i}\cdot\mathbf{A}_{i}^{ext}-m_{i}c^{2}-V_{i}^{ext}\right]-V_{c}^{int}\right\} dt'-\hbar\phi=I(Q(t),t).
\end{aligned}
\end{equation}

The restriction to simultaneous solutions of (185-186) means that
the IMFTRF and the IMBTRF of the \emph{i}-th zbw particle will not
coincide since $\mathbf{b}_{i}=\mathbf{v}_{i}+\mathbf{u}_{i}=0$ will
generally not entail $\mathbf{b}_{i*}=\mathbf{v}_{i}-\mathbf{u}_{i}=0$.
So we define an instantaneous mean (time-)symmetric rest frame (IMSTRF)
as the frame in which $\mathbf{b}_{i}+\mathbf{b}_{i*}=2\mathbf{v}_{i}=0$,
and the lab frame remains unchanged.

Applying $J=extremal$, we have 
\begin{equation}
\sum_{i=1}^{2}\frac{m_{i}}{2}\left[D_{*}D+DD_{*}\right]\mathbf{q}_{i}(t)=\sum_{i=1}^{2}e_{i}\left[-\frac{1}{c}\partial_{t}\mathbf{A}_{i}^{ext}-\nabla_{i}\left(\Phi_{i}^{ext}+\Phi_{c}^{int}\right)+\frac{\mathbf{v}_{i}}{c}\times\left(\nabla_{i}\times\mathbf{A}_{i}^{ext}\right)\right]|_{\mathbf{q}_{j}=\mathbf{q}_{j}(t)},
\end{equation}
and from the independent $\delta\mathbf{q}_{i}(t)$, the individual
equations of motion 
\begin{equation}
\begin{aligned}m_{i}\mathbf{a}_{i}(q(t),t) & =\frac{m_{i}}{2}\left[D_{*}D+DD_{*}\right]\mathbf{q}_{i}(t)\\
 & =\left[-\frac{e_{i}}{c}\partial_{t}\mathbf{A}_{i}^{ext}-e_{i}\nabla_{i}\left(\Phi_{i}^{ext}+\Phi_{c}^{int}\right)+\frac{e_{i}}{c}\mathbf{v}_{i}\times\left(\nabla_{i}\times\mathbf{A}_{i}^{ext}\right)\right]|_{\mathbf{q}_{j}=\mathbf{q}_{j}(t)}.
\end{aligned}
\end{equation}
Applying the mean derivatives and using that $\mathbf{b}_{i}=\mathbf{v}_{i}+\mathbf{u}_{i}$
and $\mathbf{b}_{i*}=\mathbf{v}_{i}-\mathbf{u}_{i}$, (206) becomes
\begin{equation}
\begin{aligned} & \sum_{i=1}^{2}m_{i}\left[\partial_{t}\mathbf{v}_{i}+\mathbf{v}_{i}\cdot\nabla_{i}\mathbf{v}_{i}-\mathbf{u}_{i}\cdot\nabla_{i}\mathbf{u}_{i}-\frac{\hbar}{2m_{i}}\nabla_{i}^{2}\mathbf{u}_{i}\right]|_{\mathbf{q}_{j}=\mathbf{q}_{j}(t)}\\
 & =\sum_{i=1}^{2}\left[-\frac{e_{i}}{c}\partial_{t}\mathbf{A}_{i}^{ext}-e_{i}\nabla_{i}\left(\Phi_{i}^{ext}+\Phi_{c}^{int}\right)+\frac{e_{i}}{c}\mathbf{v}_{i}\times\left(\nabla_{i}\times\mathbf{A}_{i}^{ext}\right)\right]|_{\mathbf{q}_{j}=\mathbf{q}_{j}(t)}.
\end{aligned}
\end{equation}
Integrating and setting the integration constants equal to the particle
rest energies, we then get 
\begin{equation}
\begin{aligned}\tilde{E}(q(t),t) & =\sum_{i=1}^{2}\tilde{E}_{i}(q(t),t)\\
 & =-\partial_{t}S(q(t),t)\\
 & =\sum_{i=1}^{2}m_{i}c^{2}+\sum_{i=1}^{2}\frac{\left[\nabla_{i}S(q,t)-\frac{e_{i}}{c}\mathbf{A}_{i}^{ext}(\mathbf{q}_{i},t)\right]^{2}}{2m_{i}}|_{\mathbf{q}_{j}=\mathbf{q}_{j}(t)}\\
 & +\sum_{i=1}^{2}e_{i}\left[\Phi_{i}^{ext}(\mathbf{q}_{i}(t),t)+\Phi_{c}^{int}(\mathbf{q}_{i}(t),\mathbf{q}_{j}(t))\right]-\sum_{i=1}^{2}\frac{\hbar^{2}}{2m_{i}}\frac{\nabla_{i}^{2}\sqrt{\rho(q,t)}}{\sqrt{\rho(q,t)}}|_{\mathbf{q}_{j}=\mathbf{q}_{j}(t)},
\end{aligned}
\end{equation}
where the $q(t)$ in (208) corresponds to the solution set of the
stochastic differential equations (183-184). Alternatively, given
the integral curves $\mathbf{Q}_{i}(t)$ of the reformulated mean
acceleration equation (206), 
\begin{equation}
\begin{aligned}m_{i}\frac{d^{2}\mathbf{Q}_{i}(t)}{dt^{2}} & =m_{i}\left(\partial_{t}\mathbf{v}_{i}+\mathbf{v}_{i}\cdot\nabla_{i}\mathbf{v}_{i}\right)|_{\mathbf{q}_{j}=\mathbf{Q}_{j}(t)}=-\nabla_{i}\left(-\frac{\hbar^{2}}{2m_{i}}\frac{\nabla_{i}^{2}\sqrt{\rho(q,t)}}{\sqrt{\rho(q,t)}}\right)|_{\mathbf{q}_{j}=\mathbf{Q}_{j}(t)}\\
 & +e_{i}\left[-\frac{1}{c}\partial_{t}\mathbf{A}_{i}^{ext}-\nabla_{i}\left(\Phi_{i}^{ext}+\Phi_{c}^{int}\right)+\frac{\mathbf{v}_{i}}{c}\times\mathbf{B}_{ext}^{i}\right]|_{\mathbf{q}_{j}=\mathbf{Q}_{j}(t)},
\end{aligned}
\end{equation}
we can also obtain $\tilde{E}(Q(t),t)$. The general solution of (208),
written in terms of $Q(t)$, is given by 
\begin{equation}
\begin{aligned}S(Q(t),t) & =\sum_{i=1}^{2}\int_{\mathbf{Q}_{i}(t_{I})}^{\mathbf{Q}_{i}(t)}\mathbf{p}_{i}'\cdot d\mathbf{Q}_{i}(s)-\sum_{i=1}^{2}\int_{t_{I}}^{t}\tilde{E}_{i}ds-\hbar\phi\\
 & =\int_{t_{I}}^{t}\left\{ \sum_{i=1}^{2}\left[\frac{1}{2}m_{i}\mathbf{v}_{i}^{2}-\left(-\frac{\hbar^{2}}{2m_{i}}\frac{\nabla_{i}^{2}\sqrt{\rho}}{\sqrt{\rho}}\right)+\frac{e_{i}}{c}\mathbf{v}_{i}\cdot\mathbf{A}_{i}^{ext}-m_{i}c^{2}-V_{i}^{ext}\right]-V_{c}^{int}\right\} ds-\hbar\phi\\
 & =\int_{t_{I}}^{t}\left\{ \sum_{i=1}^{2}\left[\frac{1}{2}m_{i}\mathbf{v}_{i}^{2}+\frac{1}{2}m_{i}\mathbf{u}_{i}^{2}+\frac{\hbar}{2}\nabla_{i}\cdot\mathbf{u}_{i}+\frac{e_{i}}{c}\mathbf{v}_{i}\cdot\mathbf{A}_{i}^{ext}-m_{i}c^{2}-V_{i}^{ext}\right]-V_{c}^{int}\right\} ds-\hbar\phi,
\end{aligned}
\end{equation}
and corresponds to the conservative-diffusion-constrained, time-symmetric,
steady-state joint phase for the \emph{zbw} particles in the lab frame
(hereafter, just the steady-state joint phase), evaluated along the
time-symmetric mean trajectory of the \emph{zbw} particles, i.e.,
solutions of (209) for initial conditions $\mathbf{Q}_{i}(0)$, and
for $i=1,..,N$. Replacing $Q(t)$ with $q$ on both sides of (210)
yields the steady-state joint phase field over the possible positions
of the \emph{zbw} particles. Note the difference between the last
lines of (210) and (204) via the terms involving $\nabla_{i}\cdot\mathbf{u}_{i}$.

As in the classical model, we make the natural assumption that the
presence of classical external potentials doesn't alter the harmonic
nature of the steady-state \emph{zbw} oscillations. Moreover, since
each \emph{zbw} particle is a harmonic oscillator, each particle has
its own well-defined steady-state phase at each point along its time-symmetric
mean trajectory. Accordingly, when $V_{c}^{int}$ is not negligible,
the steady-state joint phase must be a well-defined function of the
mean trajectories of\emph{ }both particles (since we posit that all
particles remain harmonic oscillators despite having their oscillations
physically coupled through $\Phi_{c}^{int}$ and through the common
ether medium they interact with). So for a closed loop \emph{L} along
which each particle can be physically or virtually displaced, it follows
that 
\begin{equation}
\oint_{L}\delta S=\sum_{i=1}^{2}\oint_{L}\left[\mathbf{p}_{i}'\cdot\delta\mathbf{Q}_{i}(t)-\tilde{E}_{i}\delta t\right]=nh,
\end{equation}
and 
\begin{equation}
\sum_{i=1}^{2}\oint_{L}\mathbf{p}_{i}'\cdot\delta\mathbf{Q}_{i}(t)=\sum_{i=1}^{2}\oint_{L}\mathbf{\nabla}_{i}S|_{\mathbf{q}_{j}=\mathbf{Q}_{j}(t)}\cdot\delta\mathbf{Q}_{i}(t)=nh,
\end{equation}
for a closed loop $L$ with $\delta t=0$. For the steady-state joint
phase field $S(q,t)$, we can apply the same physical reasoning above
to each \emph{zbw} particle for each possible 3-space position that
can be occupied at time \emph{t}, thereby implying 
\begin{equation}
\oint_{L}dS\left(q,t\right)=\sum_{i=1}^{2}\oint_{L}\mathbf{p}_{i}'\cdot d\mathbf{q}_{i}=\sum_{i=1}^{2}\oint_{L}\mathbf{\nabla}_{i}S\cdot d\mathbf{q}_{i}=nh.
\end{equation}
Clearly (212-213) implies `phase quantization' for each individual
\emph{zbw} particle, upon keeping all but the \emph{i}-th coordinate
fixed and performing the closed-loop integration. Combining (213),
(208), and (189), we can construct the 2-particle Schrödinger equation
for classically interacting \emph{zbw} particles in the presence of
external fields 
\begin{equation}
i\hbar\frac{\partial\psi(\mathbf{q}_{1},\mathbf{q}_{2},t)}{\partial t}=\sum_{i=1}^{2}\left[\frac{\left[-i\hbar\nabla_{i}-\frac{e_{i}}{c}\mathbf{A}_{i}^{ext}(\mathbf{q}_{i},t)\right]^{2}}{2m_{i}}+m_{i}c^{2}+e_{i}\left(\Phi_{i}^{ext}(\mathbf{q}_{i},t)+\Phi_{c}^{int}(\mathbf{q}_{i},\mathbf{q}_{j})\right)\right]\psi(\mathbf{q}_{1},\mathbf{q}_{2},t),
\end{equation}
where $\psi(\mathbf{q}_{1},\mathbf{q}_{2},t)=\sqrt{\rho(\mathbf{q}_{1},\mathbf{q}_{2},t)}e^{iS(\mathbf{q}_{1},\mathbf{q}_{2},t)/\hbar}$
is single-valued via (213).

We would now like to specify the evolution of the conditional steady-state
phase field and conditional probability density associated to each\emph{
zbw} particle. For simplicity, we first set $\mathbf{A}_{i}^{ext}=\Phi_{i}^{ext}=0$.
We then obtain the conditional steady-state phase field for particle
1 by writing $S(\mathbf{q}_{1},\mathbf{Q}_{2}(t),t)\eqqcolon S_{1}(\mathbf{q}_{1},t)$.
Taking the total time derivative gives 
\begin{equation}
\partial_{t}S_{1}(\mathbf{q}_{1},t)=\partial_{t}S(\mathbf{q}_{1},\mathbf{q}_{2},t)|_{\mathbf{q}_{2}=\mathbf{Q}_{2}(t)}+\frac{d\mathbf{Q}_{2}(t)}{dt}\cdot\nabla_{2}S(\mathbf{q}_{1},\mathbf{q}_{2},t)|_{\mathbf{q}_{2}=\mathbf{Q}_{2}(t)},
\end{equation}
where the conditional velocities 
\begin{equation}
\mathbf{v}_{1}(\mathbf{q}_{1},t)|_{\mathbf{q}_{1}=\mathbf{Q}_{1}(t)}\coloneqq\frac{\nabla_{1}S_{1}(\mathbf{q}_{1},t)}{m_{1}}|_{\mathbf{q}_{1}=\mathbf{Q}_{1}(t)}=\frac{d\mathbf{Q}_{1}(t)}{dt},
\end{equation}
and 
\begin{equation}
\mathbf{v}_{2}(\mathbf{q}_{2},t)|_{\mathbf{q}_{2}=\mathbf{Q}_{2}(t)}\coloneqq\frac{\nabla_{2}S_{2}(\mathbf{q}_{2},t)}{m_{2}}|_{\mathbf{q}_{2}=\mathbf{Q}_{2}(t)}=\frac{d\mathbf{Q}_{2}(t)}{dt},
\end{equation}
the latter defined from the conditional steady-state phase field,
$S_{2}(\mathbf{q}_{2},t)$, for particle 2. Likewise, for the conditional
density for particle 1, $\rho(\mathbf{q}_{1},\mathbf{Q}_{2}(t),t)\eqqcolon\rho_{1}(\mathbf{q}_{1},t)$
and 
\begin{equation}
\partial_{t}\rho_{1}(\mathbf{q}_{1},t)=\partial_{t}\rho(\mathbf{q}_{1},\mathbf{q}_{2},t)|_{\mathbf{q}_{2}=\mathbf{Q}_{2}(t)}+\frac{d\mathbf{Q}_{2}(t)}{dt}\cdot\nabla_{2}\rho(\mathbf{q}_{1},\mathbf{q}_{2},t)|_{\mathbf{q}_{2}=\mathbf{Q}_{2}(t)}.
\end{equation}
Inserting (218) on the left hand side of (189) and adding the corresponding
term on the right hand side, we obtain the conditional continuity
equation for particle 1: 
\begin{equation}
\partial_{t}\rho_{1}=-\nabla_{1}\cdot\left[\left(\frac{\nabla_{1}S_{1}}{m_{1}}\right)\rho_{1}\right]-\nabla_{2}\cdot\left[\left(\frac{\nabla_{2}S}{m_{2}}\right)\rho\right]|_{\mathbf{q}_{2}=\mathbf{Q}_{2}(t)}+\frac{d\mathbf{Q}_{2}(t)}{dt}\cdot\nabla_{2}\rho|_{\mathbf{q}_{2}=\mathbf{Q}_{2}(t)},
\end{equation}
which implies $\rho_{1}(\mathbf{q}_{1},t)\geq0$ and (upon suitable
redefinition of $\rho_{1}(\mathbf{q}_{1},t)$) preservation of the
normalization $\int_{\mathbb{R}^{3}}\rho_{1}(\mathbf{q}_{1},0)=1$.
Similarly, inserting (215) into the left hand side of (208) and adding
the corresponding term on the right hand side, we find that the conditional
steady-state phase field for particle 1 evolves by the conditional
quantum Hamilton-Jacobi equation 
\begin{equation}
\begin{aligned}-\partial_{t}S_{1} & =m_{1}c^{2}+\frac{\left(\nabla_{1}S_{1}\right)^{2}}{2m_{1}}+\frac{\left(\nabla_{2}S\right)^{2}}{2m_{2}}|_{\mathbf{q}_{2}=\mathbf{Q}_{2}(t)}-\frac{d\mathbf{Q}_{2}(t)}{dt}\cdot\nabla_{2}S|_{\mathbf{q}_{2}=\mathbf{Q}_{2}(t)}\\
 & +V_{c}^{int}(\mathbf{q}_{1},t)-\frac{\hbar^{2}}{2m_{1}}\frac{\nabla_{1}^{2}\sqrt{\rho_{1}}}{\sqrt{\rho_{1}}}-\frac{\hbar^{2}}{2m_{2}}\frac{\nabla_{2}^{2}\sqrt{\rho}}{\sqrt{\rho}}|_{\mathbf{q}_{2}=\mathbf{Q}_{2}(t)}
\end{aligned}
\end{equation}
where $V_{c}^{int}(\mathbf{q}_{1},t)$ is the `conditional interaction
potential' for particle 1. The solution of (219) can be verified as
\begin{equation}
\rho_{1}=\rho_{01}exp[-\int_{0}^{t}\left[\nabla_{1}\cdot\mathbf{v}_{1}(\mathbf{q}_{1},t)+\nabla_{2}\cdot\mathbf{v}_{2}(\mathbf{q}_{1},\mathbf{q}_{2},t)|_{\mathbf{q}_{2}=\mathbf{Q}_{2}(t)}\right]dt',
\end{equation}
from which we extract the conditional osmotic potential 
\begin{equation}
R_{1}=R_{01}-(\hbar/2)\int_{0}^{t}\left[\nabla_{1}\cdot\mathbf{v}_{1}(\mathbf{q}_{1},t)+\nabla_{2}\cdot\mathbf{v}_{2}(\mathbf{q}_{1},\mathbf{q}_{2},t)|_{\mathbf{q}_{2}=\mathbf{Q}_{2}(t)}\right]dt',
\end{equation}
while the solution of (220) is 
\begin{equation}
\begin{aligned}S_{1} & =\int_{\mathbf{Q}_{1}(t_{I})}^{\mathbf{Q}_{1}(t)}\mathbf{p}_{1}\cdot d\mathbf{Q}_{1}(t')|_{\mathbf{Q}_{1}(t)=\mathbf{q}_{1}}\\
 & -\int_{0}^{t}\left[m_{1}c^{2}+\frac{m_{1}\mathbf{v}_{1}^{2}}{2}+\frac{m_{1}\mathbf{v}_{2}^{2}}{2}-\mathbf{p}_{2}\cdot\frac{d\mathbf{Q}_{2}(t)}{dt}+V_{c}^{int}\right.\\
 & \left.+\frac{\hbar^{2}}{2m_{1}}\frac{\nabla_{1}^{2}\sqrt{\rho_{1}}}{\sqrt{\rho_{1}}}+\frac{\hbar^{2}}{2m_{2}}\frac{\nabla_{2}^{2}\sqrt{\rho}}{\sqrt{\rho}}|_{\mathbf{q}_{2}=\mathbf{Q}_{2}(t)}\right]dt'|_{\mathbf{Q}_{1}(t)=\mathbf{q}_{1}}-\hbar\phi_{1}.
\end{aligned}
\end{equation}
Hence (222) allows us to consistently ascribe a region of oscillating
ether in 3-D space that sources a local (i.e., in 3-D space) osmotic
potential that imparts the osmotic momentum to particle 1. Likewise,
(223) lets us ascribe a region of oscillating ether in 3-D space that
directly drives the steady-state \emph{zbw} oscillation of particle
1 in 3-D space. Note that when (223) is evaluated at $\mathbf{q}_{1}=\mathbf{Q}_{1}(t)$,
it is equivalent to $S(\mathbf{q}_{1}(t),\mathbf{Q}_{2}(t),t)-m_{2}c^{2}t+\hbar\phi_{2}$.
As in the classical model, since the conditional steady-state phase
field for particle 1 is a field over the possible positions of the
\emph{zbw} particles, it follows that 
\begin{equation}
\oint_{L}\nabla_{1}S_{1}\cdot d\mathbf{q}_{1}=nh,
\end{equation}
where \emph{L} is a mathematical loop in 3-D space.

With these results in hand, the conditional forward and backward stochastic
differential equations for particle 1 can be straightforwardly obtained
by writing $\mathbf{b}_{1}=\mathbf{v}_{1}+\mathbf{u}_{1}$, $\mathbf{b}_{1*}=\mathbf{v}_{1}-\mathbf{u}_{1}$,
and inserting these expressions into (183) and (184), respectively.

Also like in the classical model, we can define the steady-state conditional
phase-action 
\begin{equation}
\begin{aligned}J_{1} & =I_{1}^{IF}=\mathrm{E}\left[\int_{t_{I}}^{t_{F}}\left[m_{1}c^{2}+\frac{1}{2}m_{1}\mathbf{v}_{1}^{2}+\frac{1}{2}m_{2}\mathbf{v}_{2}^{2}+\frac{1}{2}m_{1}\mathbf{u}_{1}^{2}+\frac{1}{2}m_{2}\mathbf{u}_{2}^{2}-V_{c}^{int}\right]dt-\hbar\phi_{1}\right],\end{aligned}
\end{equation}
and then impose 
\begin{equation}
J_{1}=extremal,
\end{equation}
we get the conditional mean acceleration for particle 1: 
\begin{equation}
m_{1}\mathbf{a}_{1}(\mathbf{q}_{1}(t),t)=\frac{m_{1}}{2}\left[D_{*}D+DD_{*}\right]\mathbf{q}_{1}(t)=-\nabla_{1}V_{c}^{int}(\mathbf{q}_{1},\mathbf{q}_{2}(t))|_{\mathbf{q}_{1}=\mathbf{q}_{1}(t)},
\end{equation}
thus 
\begin{equation}
\begin{aligned}m_{1}\frac{\mathrm{D}\mathbf{v}_{1}(\mathbf{Q}_{1}(t),t)}{\mathrm{D}t} & =\left[\partial_{t}\mathbf{p}_{1}+\mathbf{v}_{1}\cdot\nabla_{i}\mathbf{p}_{1}\right](\mathbf{q}_{1},t)|_{\mathbf{q}_{1}=\mathbf{Q}_{1}(t)}\\
 & =-\nabla_{1}\left[V_{c}^{int}(\mathbf{q}_{1},\mathbf{Q}_{2}(t))-\frac{\hbar^{2}}{2m_{1}}\frac{\nabla_{1}^{2}\sqrt{\rho_{1}(\mathbf{q}_{1},t)}}{\sqrt{\rho_{1}(\mathbf{q}_{1},t)}}\right]|_{\mathbf{q}_{1}=\mathbf{Q}_{1}(t)},
\end{aligned}
\end{equation}
and likewise for particle 2. Equation (228) is what we would obtain
from computing the derivatives in (227) for $i=1$ (modulo the external
potentials) and subtracting the $\mathbf{u}_{i}$ dependent terms
on both sides. Of course, it should be said that we cannot obtain
(210) simply by integrating (220) and the analogous expression for
particle 2, and then summing up the terms. This is because we obtained
(220) directly from the full configuration space fields $S$ and $\rho$,
themselves obtained from extremizing (196).

For particle 2, the conditional steady-state phase field, probability
density, etc., are defined analogously.

Finally, combining (224), (220), and (219) gives us the conditional
Schrödinger equation for particle 1: 
\begin{equation}
\begin{aligned}i\hbar\frac{\partial\psi_{1}}{\partial t} & =-\frac{\hbar^{2}}{2m_{1}}\nabla_{1}^{2}\psi_{1}-\frac{\hbar^{2}}{2m_{1}}\nabla_{2}^{2}\psi|_{\mathbf{q}_{2}=\mathbf{Q}_{2}(t)}+V_{c}^{int}(\mathbf{q}_{1},\mathbf{Q}_{2}(t))\psi_{1}\\
 & +m_{1}c^{2}\psi_{1}+i\hbar\frac{d\mathbf{Q}_{2}(t)}{dt}\cdot\nabla_{2}\psi|_{\mathbf{q}_{2}=\mathbf{Q}_{2}(t)},
\end{aligned}
\end{equation}
where $\psi_{1}(\mathbf{q}_{1},t)=\sqrt{\rho_{1}(\mathbf{q}_{1},t)}e^{iS_{1}(\mathbf{q}_{1},t)/\hbar}$
is the single-valued conditional wave function for particle 1, and
$d\mathbf{Q}_{2}(t)/dt=(\hbar/m_{2})\mathrm{Im}\{\nabla_{2}\ln(\psi_{2})\}|_{\mathbf{q}_{2}=\mathbf{Q}_{2}(t)}$,
where $\psi_{2}=\psi_{2}(\mathbf{q}_{2},t)$ is the conditional wave
function for particle 2, satisfying the analogous conditional Schrödinger
equation. Like in the classical case, (229) can also be obtained from
writing 
\begin{equation}
\partial_{t}\psi_{1}(\mathbf{q}_{1},t)=\partial_{t}\psi(\mathbf{q}_{1},\mathbf{q}_{2},t)|_{\mathbf{q}_{2}=\mathbf{Q}_{2}(t)}+\frac{d\mathbf{Q}_{2}(t)}{dt}\cdot\nabla_{2}\psi(\mathbf{q}_{1},\mathbf{q}_{2},t)|_{\mathbf{q}_{2}=\mathbf{Q}_{2}(t)},
\end{equation}
inserting this on the left hand side of (214), adding the corresponding
term on the right hand side, and subtracting $m_{2}c^{2}\psi_{1}$
(again, modulo the external potentials).

The development of ZSM in relative coordinates is formally identical
to the case of a single \emph{zbw} particle in an external potential,
and need not be explicitly given here.

This completes the formulation of ZSM for \emph{N}-particles interacting
with classical fields.

\subsection{Remark on on close-range interactions}

Since the quantum kinetic doesn't depend on the inter-particle separation,
its presence in the equation of motion (209) doesn't introduce any
fundamentally new complications for the description of two-particle
scattering in ZSM. So the account we gave of two-particle scattering
in section 4.7 carries over to classically interacting particles in
ZSM.

\section{Plausibility of the Zitterbewegung Hypothesis}

Ultimately, the plausibility of our suggested answer to Wallstrom
hinges (in no particular order) on the plausibility of the \emph{zbw}
hypothesis, its incorporation into NYSM, and the generalizability
of ZSM. So we should ask if: 1) ZSM can be consistently generalized
to relativistic flat and curved spacetimes; 2) the \emph{zbw} hypothesis
can be generalized to incorporate electron spin; 3) ZSM has a conceivable
field-theoretic extension; 4) a self-consistent physical model of
the \emph{zbw} particle, Nelson's ether (suitably amended for ZSM),
and dynamical interaction between the two, can be constructed; and
5) ZSM suggests testable new predictions and/or offers novel solutions
to open problems in the foundations of quantum mechanics that justify
its mathematical and conceptual complexity (relative to other hidden
variable approaches to solving the measurement problem, such as the
dBB theory).

Can ZSM be consistently generalized to relativistic flat and curved
spacetimes? We have implicitly assumed throughout our paper that this
is possible, based on our repeated use of the next-to-leading order
approximation of the Lorentz transformation. But there is also good
reason to expect that relativistic generalizations of ZSM to flat
and curved spacetimes do exist. Stochastic mechanics based on the
Guerra-Morato variational principle has already been given a consistent
generalization to the case of relativistic spacetimes (flat and curved)
by Dohrn and Guerra \cite{Dohrn1978,Dohrn1979,Dohrn1985} as well
as Serva \cite{Serva1988}. An attempt was made by Zastawniak to give
a relativistic flat-spacetime generalization of Yasue's variational
principle \cite{Zastawniak1990}, but it seems problematic since it
doesn't address the problem of not having a normalizable spacetime
probability density when the metric is not positive-definite. Fortunately,
this problem can be resolved in the approaches of Dohrn-Guerra and
Serva, and there seems to be no obstacle in adapting Dohrn and Guerra's
methods or Serva's method to extend Yasue's variational principle
to flat and curved spacetimes (currently in progress by us). Once
done, we see no fundamental reason why a corresponding generalization
of ZSM cannot be given.

Can the \emph{zbw} hypothesis be generalized to incorporate electron
spin? It seems plausible to us that it can. As is well-known, in standard
relativistic quantum mechanics for spin-1/2 particles, the Dirac spinor
satisfying the Dirac equation implies \emph{zbw} of the corresponding
velocity operator \cite{Greiner2000}. What's more, realist versions
of relativistic quantum mechanics for spin-1/2 particles - the Bohm-Dirac
theory \cite{Holland1992,Holland1993}, the ``zig-zag'' model of
de Broglie-Bohm theory by Colin \& Wiseman \cite{Colin2011} and Struyve
\cite{Struyve2012}, and the stochastic mechanical models of the Dirac
electron by de Angelis et al. \cite{Angelis1986} and Garbaczewski
\cite{Garbaczewski1992} - all predict \emph{zbw} as a real, continuous
oscillation of the particle beable. In the de Broglie-Bohm theories,
the \emph{zbw} arises from imposing Lorentz invariance and the Dirac
spinor algebra on the dynamics of the wave function (described by
Dirac spinors in the Bohm-Dirac theory, or Weyl spinors in the zig-zag
model), and then using this wave function in the definition of the
guiding equation for the de Broglie-Bohm particle. Likewise, in the
stochastic mechanical theories, the \emph{zbw} beable arises from
constructing Nelsonian diffusion processes from the Dirac wave function.
The description of a physically real spin-based \emph{zbw} can also
be implemented in classical physics, namely in the Barut-Zanghì model
of a classical Dirac electron \cite{Barut1984,Barut1987,Barut1989,Barut1990},
which turns into the usual flat-space and curved-space versions of
the Dirac equation (in the proper-time formulation) upon first-quantization
by the standard methods \cite{BarutnPavsic1987,BarutnDuru1989}. Here
it is the imposition of relativistic covariance and the Dirac spinor
algebra that leads to classical equations of motion for a massless
(non-radiating) point charge circularly orbiting a center of mass,
the former moving with speed $c$ and the latter moving translationally
with sub-luminal relativistic speeds. So it is plausible to imagine
a relativistic generalization of ZSM in which the Barut-Zanghì model
of a \emph{zbw} particle is implemented into a relativistic version
of the Nelson-Yasue diffusion process (e.g., along the lines of Dohrn
and Guerra), under the hypothesis that Nelson's ether has vorticity
that imparts to the massless point charge a mean rotational motion
of speed $c$ and angular momentum $\hbar/2$, and derive from this
spin-based \emph{zbw} a relativistic generalization of the quantization
condition, along with the Dirac equation for a double-valued Dirac
spinor wave function. (The approaches of de Angelis et al. and Garbaczweski
don't seem adequate for this task because they don't actually derive
the zitterbewegung and Dirac equation from Nelson-Yasue diffusions;
rather, they start from the Dirac equation and Dirac spinor wave function,
and show that Nelsonian diffusions can be associated to them.) The
non-relativistic limit of this ZSM theory should presumably then recover
non-relativistic ZSM for a spinning \emph{zbw} particle with angular
momentum magnitude $\hbar/2$, along with a vorticity term added to
the current velocity (as is known to arise from the non-relativistic
limit of the relativistic guiding equation under Gordon decomposition
in the Bohm-Dirac theory \cite{HollandPhilipp2003,Bacciagaluppi1999}).
Alternatively, we might try deducing a non-relativistic ZSM theory
directly from Takabayasi's non-relativistic generalization of the
Madelung fluid to spin-1/2 motion \cite{Takabayasi1983}. These tasks
remain for a future paper.

Does ZSM have a field-theoretic generalization that recovers the predictions
of relativistic quantum field theory for fermions and bosons? A generalization
of ZSM to massive scalar or spinor fields seems in-principle unproblematic,
but a generalization to massless fields (such as to describe the photon
or gluon, which have no measured rest mass) would seem, at first sight,
difficult (though not necessarily impossible \footnote{For example, we might consider introducing small rest masses for the
photon and gluon consistent with experimental bounds, which for the
photon is $<10^{-14}eV/c^{2}$ \cite{Adelberger2007} and for the
gluon $<0.0002eV/c^{2}$ \cite{Yndurain1995}, if both masses are
to be produced by the Higgs mechanism. This would, of course, change
the gauge symmetries of QED and QCD, but not in a way that can be
experimentally discerned at energy scales above these lower-bounds
\cite{GoldhaberNieto2010}.}). Another possibility is to note that one can reproduce nearly all
\footnote{The single different prediction appears to be that this Dirac sea
pilot-wave model predicts fermion number conservation, whereas the
Standard Model predicts a violation of fermion number for sufficiently
high energies (so-called anomalies of the Standard Model). To the
best of our knowledge, no evidence has been found for fermion number
violation thus far \cite{Durieux2013}. But as Colin and Struyve point
out \cite{Colin2007}, even if fermion number violation is eventually
observed, it may still be possible to model it in a Dirac sea picture.} the predictions of the Standard Model (SM) with a pilot-wave model
for point-like fermions in which the Dirac sea is taken seriously
(i.e., taken as ontological) \cite{Colin2007}. In this model, no
beables are introduced for the massless bosons, yet it recovers nearly
all the predictions of the SM. So we might try constructing a version
of relativistic ZSM for spin-1/2 particles in which the Dirac sea
for fermions is taken seriously, and check if it can recover nearly
all the predictions of the SM as well. If one insists on adding beables
for the bosons, perhaps one could adapt the approach of Nielsen et
al. \cite{Nielsen98,Habara2008}, who show how to introduce a Dirac
sea for bosons in second-quantized field theory based on massive hypermultiplets.
Finally, it seems plausible that one could make a ZSM generalization
of bosonic string theory by constructing a Nelson-Yasue version of
the model of Santos and Escobar \cite{Santos99}, who use the Guerra-Morato
variational principle to construct a stochastic mechanics of the open
bosonic string (the idea being that the open bosonic string's instantaneous-rest-frame
oscillations would play the role of the \emph{zbw}, and would be hypothesized
to be dynamically driven by resonant coupling to the ZSM version of
Nelson's ether). All this remains for future work.

Can a self-consistent dynamical model of the \emph{zbw} particle,
Nelson's ether, and the physical interaction between the two, be constructed?
We see no principled obstacle to this possibility. Furthermore, physical
models of a real classical \emph{zbw} particle have been constructed
in the context of stochastic electrodynamics (SED), by Rueda \& Cavelleri
\cite{Rueda1983}, Rueda \cite{Rueda1993,Rueda1993a}, de la Peña
\& Cetto \cite{Pena1996}, and Haisch \& Rueda \cite{Haisch2000}.
These models involve treating the electron as a structured object
composed of a point charge with negligible (or zero) mass, harmonically
bound to some non-charged center of mass, and driven to oscillate
at near or equal to the speed of light (i.e., Compton frequency) by
resonant modes of a classically fluctuating electromagnetic zero-point
field. Additionally, in Rueda's model \cite{Rueda1993,Rueda1993a},
not only does the classical zero-point field drive the \emph{zbw}
oscillations, but the frequency cut-off generated by the \emph{zbw}
results in a non-dissipative, (effectively) Markovian diffusion process
with diffusion coefficient $\hbar/2m$. Of course, these SED-based
approaches should be cautioned; SED is know to have difficulties as
a viable theory of quantum electrodynamical phenomena \cite{Pope2000,Genovese2007},
and it is not clear that these difficulties can be resolved (but see
\cite{Valdes-Hernandez2011,Pena2012a,Pena2012,Cetto2012,Cetto2014}
for recent counter-arguments). Furthermore, we expect that any realistic
physical model of the \emph{zbw} particle should consistently incorporate
the Higgs mechanism (or some subquantum generalization thereof) \cite{Penrose2005}
as the process by which the self-stable \emph{zbw} harmonic potential
of rest-mass $m$ is formed in the first place. Nevertheless, these
SED-based models can at least be viewed as proofs of principle that
the \emph{zbw} hypothesis can be implemented in a concrete model;
and, in a future paper, we will show how one of these SED-based models
can in fact recover the quantization condition as an effective condition.
But the task of constructing a physical model of the \emph{zbw} particle,
the ZSM version of Nelson's ether, and the physical/dynamical interaction
between the two, which also incorporates spin and can be used to recover
the Dirac/Pauli/Schrödinger equation, remains for future work.

Lastly, does ZSM suggest testable new predictions and/or novel solutions
to open foundational problems in quantum mechanics? We claim it does.
Since the equilibrium density $\rho=|\psi|^{2}$, ZSM's statistical
predictions in equilibrium will agree with all the statistical predictions
of non-relativistic quantum mechanics. But if $\rho\neq|\psi|^{2}$,
we should expect differences, such as position and momentum measurements
with more precision than allowed by Heisenberg's uncertainty principle
\cite{Pearle2006}. \footnote{Everything we have said here is of course also true of the dBB theory
\cite{Pearle2006}. However, in our view, a proper understanding of
the origin of randomness in the dBB theory (the `typicality' approach
of Dürr-Goldstein-Zanghì \cite{Duerr1992}) entails that the existence
of quantum nonequilibrium subsystems in the observable universe is
extremely improbable, even in the context of early universe cosmology
(for a different view, see \cite{Valentini2010}). By contrast, we
will suggest here that this limitation of the dBB theory does not
necessarily apply to ZSM.} Accordingly, it would be possible, in principle, to experimentally
detect the stochasticity of the particle trajectories, hence deviations
from the mean trajectories satisfying the quantization condition.
Under what physical conditions might we see nonequilibrium fluctuations?
The most obvious possibility is by measuring the position or momentum
of a Nelsonian particle on time-scales comparable to or shorter than
the correlation time of the ether fluctuations. For ZSM, insofar as
it's based on Nelson's white-noise diffusion process, the correlation
timescale of the fluctuations is infinitesimal because of the assumption
that the noise is white. Nelson stressed, however, that his white-noise
(Markovian) assumption was only a simplifying one \cite{Nelson1985};
so one could instead imagine a colored-noise (non-Markovian) description
of conservative diffusions, to which Nelson's white-noise description
is a long-time approximation (as is the case with all other known
statistical fluctuation phenomena in nature \cite{Hanggi1995a}).
\footnote{Of course, this idea could also be explored in NYSM with the quantization
condition imposed ad-hoc. The advantage of ZSM, though, is that it
makes the idea worth taking seriously as a possibility since ZSM gives
an independent justification for the (more basic) quantization condition,
without which the stochastic mechanics approach would be neither empirically
viable nor plausible.} Then the true fluctuation timescale would be finite and one could
work out the expected experimental signatures of the nonequilibrium
dynamics on timescales comparable to some hypothetical finite correlation
time $\tau_{noise}$ (work on this is currently underway). In connection
to this, Montina's theorem \cite{Montin2008} says that any ontic
theory compatible with the predictions of a quantum system with finite
Hilbert space dimensionality $k$ must contain at least $2k-2$ continuous
real variables, assuming that the theory has deterministic or stochastic
Markovian dynamics (i.e., a dynamics that is local in time). Correspondingly,
$2k-2$ turns out to be the minimum number of real-valued parameters
required to describe a pure quantum state. On the other hand, Montina's
theorem implies that an ontic theory with non-Markovian dynamics (i.e.,
dynamics which is nonlocal in time) could have fewer continuous real
variables than $2k-2$. Montina has demonstrated this in a toy model
of a single ontic variable with stochastic evolution driven by time-correlated
(colored) noise that exactly reproduces any unitary evolution of a
qubit ($\psi$ for a qubit has two degrees of freedom) \cite{Montin2011,Montin2012}.
Extrapolating the implications of Montina's theorem to stochastic
mechanics, we could expect that a non-Markovian extension of stochastic
mechanics would recover an $N$-particle `wave function' that can
be described by fewer than $2k-2$ real-valued parameters, where $k$
would be the dimensionality of the Hilbert space associated to the
$N$-particle wave function of Markovian stochastic mechanics (Markovian
stochastic mechanics would be the $\tau_{noise}\rightarrow0$ limit
of non-Markovian stochastic mechanics). And insofar as the $N$-particle
wave function can be polar decomposed into $N$-particle $R$ and
$S$ fields, the $N$-particle $R$ and $S$ fields of non-Markovian
ZSM would presumably also require fewer real-valued parameters to
describe than the $N$-particle $R$ and $S$ fields of Markovian
ZSM. Moreover, since $R$ and $S$ directly reflect ontological elements
of ZSM (see sections 3 and 5.1), the reduced complexity of the $R$
and $S$ fields of non-Markovian ZSM would (presumably) imply that
the ontological complexity of non-Markovian ZSM will be less than
that of Markovian ZSM. It seems conceivable, then, that if we make
a TELB \cite{Norsen2010,Norsen2014} version of non-Markovian ZSM
by decomposing the $N$-particle $R$ and $S$ fields into $N$ single-particle
$R$ and $S$ fields (a pair for each particle), we may only require
a finite number of (or perhaps zero) supplementary continuous ontic
variables on 3-space, in order to encode non-local correlations arising
between \emph{zbw} particles that are classically interacting and
coupling to the common oscillating ether. If so, we would (arguably)
then have a reasonably ontologically parsimonious TELB version of
ZSM. This TELB version of ZSM would considerably strengthen the justification
for viewing the joint \emph{zbw} phase $S$ for an $N$-particle system
as the joint phase of real physical oscillations about the actual
3-space locations of the \emph{zbw} particles, while supporting the
hypothesis that the ether is a medium that fundamentally lives in
3-space instead of 3N-dimensional configuration space.

\section{Comparison to Other Answers }

Several other answers to Wallstrom's criticism have been offered in
the context of stochastic mechanics \cite{Carlen1989,Wallstrom1994,Smolin06,Fritsche2009,Schmelzer(2011),Groessing2011}.
Here we briefly review and assess each approach, and compare them
to ZSM.

Smolin proposed \cite{Smolin06} that Wallstrom's criticism could
be answered by allowing discontinuities in the wave function - that
is, for a given multi-valued wave function, one could introduce discontinuities
at the multi-valued points to make it single-valued. The example he
used is stochastic mechanics on $\textrm{S}^{1}$, where he argued
that although the resultant wave function is not single-valued and
smooth, it is well-known that almost every wave function in the Hilbert
space $\mathcal{L}^{2}(\mathrm{S}^{1})$ is discontinuous at one or
many points, and yet each wave function is normalizable and gives
well-defined (i.e., single-valued) current velocities. Smolin's proposal
seems incomplete, however. Even if his proposal works for the multiply
connected configuration space of the unit circle, how will it work
in the more general cases of simply connected configuration spaces
of dimensionality 3N? Wallstrom emphasizes, after all, that the inequivalence
between the HJM equations and Schrödinger 's equation applies to simply
connected configuration spaces of two dimensions or greater \cite{Wallstrom1994}.
(See also \cite{Valentini2010} for a critique of Smolin's approach.)
To compare with ZSM, these concerns don't arise - the derived wave
functions are single-valued and smooth, and ZSM works for the general
case of simply connected 3N-dimensional configuration space.

Carlen \& Loffredo \cite{Carlen1989} considered stochastic mechanics
on $\textrm{S}^{1}$ and suggested to introduce a stochastic analogue
of the quantization condition, which they argue is related in a natural
way to the topological properties of $\textrm{S}^{1}$. They then
showed that this stochastic analogue of the quantization condition
establishes mathematical equivalence between stochastic mechanics
and quantum mechanics on $\textrm{S}^{1}$. However, the difficulty
with taking their proposal as a general answer is that it seems to
only work in the special case of $\textrm{S}^{1}$, whereas Wallstrom's
criticism applies to simply connected configuration spaces of two
dimensions or greater, as mentioned earlier.

Fritsche \& Haugk \cite{Fritsche2009} attempted to answer Wallstrom
by motivating the quantization condition from the physical requirement
that the probability density, $|\psi|^{2}$, should always be normalizable.
To accomplish this, they first required that the velocity potential,
$S$, be single-valued on a closed loop (in analogy with the definition
of a single-valued magnetic scalar potential) via jump discontinuities.
Constructing the wave function from this $S$ function through an
approach equivalent to Nelson's Newtonian formulation of stochastic
mechanics, they then argued that the only way $|\psi|^{2}$ can remain
normalizable for a superposition of two eigenstates is if the phase
difference between the eigenstates satisfies the quantization condition.
The main problem with their approach lies in the their non-trivial
assumption that $S$ can have jump discontinuities. As pointed out
by Wallstrom \cite{Wallstrom1989,Wallstrom1994}, allowing jump discontinuities
in $S$ implies that $\nabla\psi=\left(\frac{1}{\hbar}\right)\left(\nabla R+i\nabla S\right)\psi$
develops a singularity, which is physically inadmissible. Accordingly,
the same technical concerns we raised towards Smolin's proposal apply
here as well. We note, by contrast, that in ZSM, $\nabla S$ is always
continuous even though $S$ is in general discontinuous (e.g., at
nodal points of the probability density).

Wallstrom made the observation \cite{Wallstrom1994} that if one takes
the quantization condition as an initial condition on the current
velocity, then the time-evolution of the HJM equations will ensure
that it is valid for all future times, in analogy with Kelvin's circulation
theorem from classical fluid mechanics. So one might think to use
this as a justification for the quantization condition in the context
of the HJM equations. As he pointed out, however, this seems to require
an extreme form of fine-tuning (why should the initial condition on
the current velocity correspond exactly to the quantization condition?),
and it is not clear that this initial condition would be stable for
interacting particles. By contrast, we saw in ZSM that the \emph{zbw}
hypothesis combined with the Lorentz transformation implies the quantization
condition so that it is not the result of fine-tuning (other than
the assumption that the steady-state oscillation frequency in the
IMFTRF/IMBTRF/IMSTRF is of fixed Compton magnitude). Moreover, we
showed that in the case of classically interacting \emph{zbw} particles,
it can be plausibly argued that the quantization condition remains
stable.

Bacciagaluppi \cite{Bacciagaluppi2005} suggested that when the external
potential $V$ has time-dependence, the complement of the nodal set
of $\rho$ may become simply connected in a neighborhood of a given
time $t$. In other words, the time-dependence of $V$ may make it
possible to eliminate the nodes of $\rho$ around which a multi-valued
$S$ accumulates values other than $nh$ (because $S$ would have
to be single-valued in that neighborhood of $t$). While Bacciagaluppi's
suggestion was intended as an abstract, mathematical argument, it
is interesting to note that his proposal seems relevant to measurement
situations when the interaction of a system with a pointer apparatus
entails a time-dependent $V$; in other words, Bacciagaluppi's suggestion
might be used to argue that energy-momentum quantization arises as
a dynamical effect of measurement interactions, as opposed to a measurement-independent
property of particles in bound states (as in ZSM). We find this an
intriguing possibility, but the technical details need to be developed
for it to become a serious proposal.

Grössing et al. \cite{Groessing2011} constructed a model of a classical
``walking bouncer'' particle (essentially a harmonic oscillator
of natural frequency $\omega_{0}$) coupled to a dissipative thermal
environment which imparts a stochastic, periodic, driving force. They
then showed that in the large friction limit the mean stochastic dynamics
of the bouncer satisfies what amounts to the quantization condition.
They claim ``this condition resolves the problem discussed by Wallstrom
{[}20{]} about the single-valuedness of the quantum mechanical wave
functions and eliminates possible contradictions arising from Nelson-type
approaches to model quantum mechanics.'' It is unclear to us that
their model involves physically consistent assumptions; \footnote{They assume that their dissipative thermal environment corresponds
to a classical ``zero-point field'' of Ornstein-Uhlenbeck statistical
type, unknown positive temperature, and that imparts to the bouncer
a total energy of $\hbar\omega_{0}/2$. But the zero-point fields
of QED and SED are, by construction, frequency-cubed-dependent in
their spectral density, non-dissipative in that they produce no Einstein-Hopf
drag force, and \emph{non-thermal} in that the zero-point motion they
induce on charged particles persists at zero temperature \cite{Boyer1980,Milonni1994}. } but setting aside this concern, the main difficulty we see with their
claim is that they don't show how to derive the HJM equations from
their model (although they do show that their model yields the energy
spectrum of the quantum harmonic oscillator), which is the context
in which Wallstrom's critique applies. In addition, it is unclear
how their model is consistent with NYSM since Nelson's diffusion process
is a conservative one while their model assumes a dissipative diffusion
process in a thermal environment. No such (apparent) inconsistency
exists for ZSM, since we implemented the \emph{zbw} hypothesis into
NYSM in a manner consistent with Nelson's (suitably generalized) ether
hypothesis. Nevertheless, in our view, Grössinget al.'s model (if
it can be shown physically consistent) has value as a proof-of-principle
that one can construct a physical model of a classical, harmonically
oscillating particle coupled to some fluctuating, oscillating, ether-like
background medium, and dynamically obtain the quantization condition.

Schmelzer \cite{Schmelzer(2011)} argued that in order to obtain empirical
equivalence with quantum mechanics, it is sufficient for stochastic
mechanics to only recover wave functions with simple zeros. He then
showed that if one invokes the postulate, $0<\Delta\rho(\mathbf{x})<\infty$
almost everywhere when $\rho(\mathbf{x})=0$, one obtains the quantization
condition for simple zeros, i.e., where $n=\pm1$. He also showed
that this postulate corresponds to an ``energy balance'' constraint,
namely, that the total energy density of the Nelsonian particle remains
finite. Schmelzer suggested that it remains for subquantum theories
to somehow dynamically justify the energy balance constraint. In our
view, Schmelzer does not adequately justify his claim that simple
zeros are sufficient to recover empirical equivalence with quantum
mechanics (e.g., how can this account for energy level shifts in the
hydrogen atom described by the Rydberg formula?); but if this can
be shown, then we would concur that his proposal seems to be a non-circular,
non-ad\emph{-}hoc, empirically adequate justification for a limited
version of the quantization condition. In ZSM, by contrast, the full
quantization condition is obtained from the phase of the hypothesized
\emph{zbw} particle(s), with the proviso that it should be understood
as a phenomenological stepping-stone to a physical theory of Nelson's
(suitable modified) ether, the \emph{zbw} particle, and the dynamical
interaction between the two.

Caticha and his collaborators \cite{Catich2011,BartolomeoCaticha2015}
have offered two routes to answering Wallstrom within the context
of his ``entropic dynamics'' (ED) framework (essentially, a Bayesian
inference version of stochastic mechanics). In the first route, Caticha
appeals to Pauli \cite{Pauli1980}, who suggested that the criterion
for admissibility for wave functions is that they must form a basis
for a representation of the transformation group for a given eigenvalue
problem. He then suggests that this criterion is ``extremely natural''
from the perspective of a theory of inference since ``in any physical
situation symmetries constitute the most common and most obviously
relevant pieces of information'' \cite{Catich2011}. However, it
should be noted that Pauli's criterion, more precisely, is that ``repeated
actions of the operators corresponding to physical quantities should
not lead outside the domain of square-integrable eigenfunctions''
\cite{Pauli1980}. In other words, Pauli's criterion just requires
that wave functions continue to satisfy the linearity of Schrödinger
's equation (i.e., the superposition principle), even after being
acted upon by operators for physical quantities. But insofar as ED
attempts to recover the Schrödinger equation from the HJM equations,
such a criterion cannot be invoked in entropic dynamics without begging
the question. In the second route, Bartolomeo and Caticha \cite{BartolomeoCaticha2015}
take inspiration from Takabayasi's generalization of the HJM equations
to a spinning fluid \cite{Takabayasi1983}; they propose to interpret
their postulated ``drift potential'', $\phi(\mathbf{x},t)$, as
an angle describing particle spin, and thereby argue that the change
of $\phi$ along a closed loop in space must equal $2\pi n$. In fact,
this argument is conceptually equivalent to the ones given by de Broglie
\cite{Broglie1925,Darrigol1994} and Bohm \cite{Bohm1957,BohmHiley1982,Bohm2002},
and which we've used in ZSM. On the other hand, it should be noted
that Bartolomeo and Caticha don't actually model spin in ED, nor do
they suggest to connect spin to the dynamical influence of an ether
or background field (in contrast to ZSM). Indeed, Bartolomeo and Caticha
admit that ``ED is a purely epistemic theory. It does not attempt
to describe the world.... In fact ED is silent on the issue of what
causative power is responsible for the peculiar motion of the particles''
\cite{BartolomeoCaticha2015}. From our point of view, this makes
their argument for the quantization condition less compelling than
the one offered by ZSM, and ED less compelling as a satisfactory theory
of quantum phenomena compared to the (programmatic) ontological approach
offered by ZSM. Nevertheless, to whatever extent one views the Bayesian
inference approach to physics as valuable and interesting, it appears
that one can give a somewhat non-ad-hoc justification for the quantization
condition via ED.

\section{Conclusion}

We have extended our classical \emph{zbw} model and ZSM to the cases
of free particles, particles in external fields, and classically interacting
particles. Along the way, we have made explicit the beables of ZSM
and suggested three possible approaches for parsing the beables into
local vs. nonlocal types. In addition, we have given arguments for
the plausibility of the \emph{zbw} hypothesis and suggested new lines
of research that could be pursued from the foundation provided here.
We have also reviewed and compared several other proposals for answering
the Wallstrom criticism, arguing that ZSM is the most general and
viable approach of all of them presently.

We wish to emphasize once more that ZSM should not be viewed as a
proposal for a fundamental physical theory of non-relativistic quantum
phenomena; rather, it should be viewed as a provisional, phenomenological
theory that provides the conceptual and mathematical scaffolding for
an eventual physical theory of Nelson's ether (amended for ZSM), the
\emph{zbw} particle, and the dynamical coupling between the two.

In his 1994 paper \cite{Wallstrom1994}, Wallstrom wrote: ``There
seems to be nothing within the particle-oriented world of stochastic
mechanics which can lead to what is effectively a condition on the
`wave function'''. Given the example of ZSM, we would suggest that
Wallstrom's claim can no longer be sustained for all formulations
of stochastic mechanics.

\section{Acknowledgments}

It is a pleasure to thank Guido Bacciagaluppi, Dieter Hartmann, and
Herman Batelaan for helpful discussions and encouragement throughout
this work. Special thanks to Guido for a careful reading of the paper
and several helpful suggestions for improvements, and to Harvey Brown,
Dennis Dieks, Bei-Lok Hu, Ward Struyve, and Nino Zanghì, for helpful
feedback.

\appendix

\section{Proof of the \emph{N}-particle Stochastic Variational Principle }

Let $\mathbf{q}_{i}'(t)=\mathbf{q}_{i}(t)+\delta\mathbf{q}_{i}(t)$
be variations of the sample paths $\mathbf{q}_{i}(t)$, with end-point
constraints $\delta\mathbf{q}_{i}(t_{I})=\delta\mathbf{q}_{i}(t_{F})=0$.
Then, using $\mathbf{b}_{i}=D\mathbf{q}_{i}(t)$ and $\mathbf{b}_{i*}=D_{*}\mathbf{q}_{i}(t)$,
the condition 
\begin{equation}
\begin{aligned}J & =\mathrm{E}\left[\int_{t_{I}}^{t_{F}}\sum_{i=1}^{N}\left\{ \frac{1}{2}\left[\frac{1}{2}m_{i}\left(D\mathbf{q}_{i}(t)\right)^{2}+\frac{1}{2}m_{i}\left(D_{*}\mathbf{q}_{i}(t)\right)^{2}\right]\right.\right.\\
 & \left.\left.+\frac{e_{i}}{c}\mathbf{A}_{i}^{ext}\cdot\frac{1}{2}\left(D+D_{*}\right)\mathbf{q}_{i}(t)-e_{i}\left(\Phi_{i}^{ext}+\Phi_{c}^{int}\right)\right\} dt\right]\\
 & =\mathrm{E}\left[\int_{t_{I}}^{t_{F}}\sum_{i=1}^{N}\left\{ \frac{1}{2}m_{i}v_{i}^{2}+\frac{1}{2}m_{i}u_{i}^{2}+\frac{e_{i}}{c}\mathbf{A}_{i}^{ext}\cdot\mathbf{v}_{i}-e_{i}\left[\Phi_{i}^{ext}+\Phi_{c}^{int}\right]\right\} dt\right]=extremal,
\end{aligned}
\end{equation}
is equivalent to the variation, 
\begin{equation}
\delta J(q)=J(q')-J(q),
\end{equation}
up to first order in $||\delta\mathbf{q}_{i}(t)||$. So (232) gives
\begin{equation}
\begin{aligned}\delta J & =\mathrm{E}\left[\int_{t_{I}}^{t_{F}}\sum_{i=1}^{N}\left\{ \left[\frac{1}{2}m_{i}\left(D\mathbf{q}_{i}(t)\cdot D\delta\mathbf{q}_{i}(t)+D_{*}\mathbf{q}_{i}(t)\cdot D_{*}\delta\mathbf{q}_{i}(t)\right)\right]\right.\right.\\
 & \left.\left.+\frac{e_{i}}{c}\mathbf{A}_{i}^{ext}\cdot\frac{1}{2}\left(D\delta\mathbf{q}_{i}(t)+D_{*}\delta\mathbf{q}_{i}(t)\right)+\frac{e_{i}}{c}\left(\delta\mathbf{q}_{i}(t)\cdot\nabla_{i}\mathbf{A}_{i}^{ext}\right)\mathbf{v}_{i}-e_{i}\nabla_{i}\left[\Phi_{i}^{ext}+\Phi_{c}^{int}\right]\cdot\delta\mathbf{q}_{i}(t)\right\} |_{\mathbf{q}_{j}=\mathbf{q}_{j}(t)}dt\right].
\end{aligned}
\end{equation}

Now, for an arbitrary function $f_{i}(q(t),t)$, we have the relations
\begin{equation}
\mathrm{E}\left[\int_{t_{I}}^{t_{F}}\sum_{i=1}^{N}\left[f_{i}(q(t),t)D\delta\mathbf{q}_{i}(t)\right]dt\right]=-\mathrm{E}\left[\int_{t_{I}}^{t_{F}}\sum_{i=1}^{N}\left[\delta\mathbf{q}_{i}(t)D_{*}f_{i}(q(t),t)\right]dt\right],
\end{equation}
and 
\begin{equation}
\mathrm{E}\left[\int_{t_{I}}^{t_{F}}\sum_{i=1}^{N}\left[f_{i}(q(t),t)D_{*}\delta\mathbf{q}_{i}(t)\right]dt\right]=-\mathrm{E}\left[\int_{t_{I}}^{t_{F}}\sum_{i=1}^{N}\left[\delta\mathbf{q}_{i}(t)Df_{i}(q(t),t)\right]dt\right],
\end{equation}
and 
\begin{equation}
\frac{1}{2}\left(D+D_{*}\right)f_{i}(q(t),t)=\left\{ \partial_{t}+\frac{1}{2}\left[D\mathbf{q}_{i}(t)+D_{*}\mathbf{q}_{i}(t)\right]\cdot\nabla_{i}\right\} f_{i}(q,t)|_{\mathbf{q}_{j}=\mathbf{q}_{j}(t)}.
\end{equation}
So, using Eq. (9) in section 2, the integrand of (233) becomes 
\begin{equation}
\begin{aligned}\delta J & =\mathrm{E}\left[\int_{t_{I}}^{t_{F}}\sum_{i=1}^{N}\left\{ \frac{m_{i}}{2}\left[D_{*}D+DD_{*}\right]\mathbf{q}_{i}(t)\right.\right.\\
 & \left.\left.-\frac{e_{i}}{c}\mathbf{v}_{i}\times\left(\nabla_{i}\times\mathbf{A}_{i}^{ext}\right)+\frac{e_{i}}{c}\partial_{t}\mathbf{A}_{i}^{ext}+e_{i}\nabla_{i}\left[\Phi_{i}^{ext}+\Phi_{c}^{int}\right]\right\} |_{\mathbf{q}_{j}=\mathbf{q}_{j}(t)}\delta\mathbf{q}_{i}(t)dt\right]+\vartheta(||\delta\mathbf{q}_{i}||).
\end{aligned}
\end{equation}

From the variational constraint (231-32), and using the fact that
the arbitrary variations (i.e., the virtual displacements in the generalized
coordinates) $\delta\mathbf{q}_{i}(t)$ are independent for all \emph{i}
by D'Alembert's principle \cite{Ray2006}, it follows that the first-order
variation of $J$ must be zero for each $\delta\mathbf{q}_{i}(t)$.
Moreover, since the expectation is a positive linear functional, we
will have the equations of motion 
\begin{equation}
\sum_{i=1}^{N}\frac{m_{i}}{2}\left[D_{*}D+DD_{*}\right]\mathbf{q}_{i}(t)=\sum_{i=1}^{N}e_{i}\left[-\frac{1}{c}\partial_{t}\mathbf{A}_{i}^{ext}-\nabla_{i}\left(\Phi_{i}^{ext}+\Phi_{c}^{int}\right)+\left(\frac{\mathbf{v}_{i}}{c}\right)\times\left(\nabla_{i}\times\mathbf{A}_{i}^{ext}\right)\right]|_{\mathbf{q}_{j}=\mathbf{q}_{j}(t)},
\end{equation}
and 
\begin{equation}
\frac{m_{i}}{2}\left[D_{*}D+DD_{*}\right]\mathbf{q}_{i}(t)=\left[-\frac{e_{i}}{c}\partial_{t}\mathbf{A}_{i}^{ext}-e_{i}\nabla_{i}\left(\Phi_{i}^{ext}+\Phi_{c}^{int}\right)+\frac{e_{i}}{c}\mathbf{v}_{i}\times\left(\nabla_{i}\times\mathbf{A}_{i}^{ext}\right)\right]|_{\mathbf{q}_{j}=\mathbf{q}_{j}(t)},
\end{equation}
for each time $t$ $\in$ $\left[t_{I},t_{F}\right]$ with probability
one.

\bibliographystyle{unsrt}
\bibliography{PhDthesisRefscopy}

\end{document}